\documentclass[12pt]{article}
\usepackage[normalem]{ulem}
\usepackage{tocloft}
\usepackage{amsfonts}
\usepackage{amsmath}
\usepackage{amssymb}
\usepackage{float,tikz, extarrows, tikz-cd}
\usetikzlibrary{decorations.pathmorphing,calc}
\usepackage{array}
\usepackage{bigints}
\usepackage{textcomp}
\usepackage{bm}
\usepackage{booktabs}
\usepackage[nosort]{cite}
\usepackage{color}
\usepackage{dsfont}
\usepackage{float}
\usepackage{framed}
\usepackage{graphicx}
\usepackage{tcolorbox}
\usepackage{indentfirst}
\usepackage{mathrsfs}
\usepackage{multirow}
\usepackage{pdflscape}
\usepackage{setspace}
\usepackage{titlesec}
\usepackage{wrapfig}
\usepackage{mathtools}
\usepackage[all]{xy}
\usepackage{young}
\usepackage[vcentermath]{youngtab}
\usepackage{relsize}
\usepackage{stackengine}
\usepackage{verbatim}
\usepackage{slashed}
\usepackage[compat=1.0.0]{tikz-feynman}
\usepackage{adjustbox}
\usepackage{subcaption}
\usepackage{caption}

\usepackage{hyperref}
\hypersetup{colorlinks=true}
\hypersetup{linkcolor=black}
\hypersetup{citecolor=black}
\hypersetup{urlcolor=black}

\numberwithin{equation}{section}

\usepackage[left=2.5cm,right=2.5cm,top=2.5cm,bottom=3cm]{geometry}
\fontdimen2\font=1.2\fontdimen2{\jot}{5pt}

\titleformat{\section}{\large\bfseries}{\thesection.}{4pt}{}
\titlespacing{\section}{0pt}{20pt}{6pt}

\titleformat{\subsection}{\normalfont\bfseries}{\thesubsection.}{4pt}{}
\titlespacing{\subsection}{0pt}{15pt}{6pt}

\titleformat{\subsubsection}{\normalfont\itshape}{\thesubsubsection.}{4pt}{}
\titlespacing{\subsubsection}{0pt}{15pt}{6pt}

\titleformat{\paragraph}{\normalfont\itshape}{\theparagraph.}{4pt}{}
\titlespacing{\paragraph}{0pt}{15pt}{6pt}

\def\tilde{\widetilde}

\def\bar{\overline}

\def\half{{1 \over 2}}
\def\d{\partial}

\def\1{{\mathds 1}}

\DeclareMathAlphabet{\mathbfsf}{OT1}{cmss}{bx}{n}

\newcommand{\Z}{{\mathbb Z}}

\newcommand{\R}{{\mathbb R}}

\def\CH{{\mathcal H}}

\def\CM{{\mathcal M}}

\def\CO{{\mathcal O}}

\newcommand{\beq}{\begin{equation}\begin{aligned}}
\newcommand{\eeq}{\end{aligned}\end{equation}}

\newcommand{\lam}{\lambda}

\newcommand{\al}{\alpha}
\newcommand{\be}{\beta}
\newcommand{\del}{\delta}
\newcommand{\ga}{\gamma}

\newcommand{\ov}{\over}

\newcommand{\non}{\nonumber \\}

\newcommand{\bea}{\begin{eqnarray}}
\newcommand{\eea}{\end{eqnarray}}

\newcommand{\beqa}{\begin{eqnarray}}
\newcommand{\eeqa}{\end{eqnarray}}
\newcommand{\beqar}{\begin{eqnarray*}}
\newcommand{\eeqar}{\end{eqnarray*}}

\newcommand{\reef}[1]{(\ref{#1})}
\newcommand{\eg}{{\it e.g.,}\ }
\newcommand{\ie}{{\it i.e.,}\ }

\newcommand{\mt}[1]{\textrm{\tiny #1}}

\def\({\left(} \def\){\right)}
\def\[{\left[} \def\]{\right]}

\DeclareFontShape{OT1}{cmr}{mx}{n}%
{<->cmr10}{}
\newcommand{\mytitlefont}{\fontseries{mx}\selectfont}
\DeclareMathAlphabet{\titlemath}{OT1}{cmr}{mx}{n}

\begin{document}


\begin{titlepage}

\begin{center}
			
~\\
			
{\fontsize{27pt}{0pt} \mytitlefont Thermal Order in Conformal Theories}
			
~\\

Noam Chai,$^{1}$ Soumyadeep Chaudhuri,$^{1}$ Changha Choi,$^{2}$ Zohar Komargodski,$^{2}$ Eliezer Rabinovici,$^{1}$  Michael Smolkin$^{1}$


~\\[0.1cm]
{\it $^{1}$ Racah Institute, The Hebrew University. Jerusalem 9190401, Israel}\

{\it $^{2}$ Simons Center for Geometry and Physics, SUNY, Stony Brook, NY 11794, USA}\

\end{center}

\vskip0.5cm
			
\noindent 

It is widely expected that at sufficiently high temperatures order is always lost, e.g. magnets loose their ferromagnetic properties. We pose the question of whether this is always the case in the context of quantum field theory in $d$ space dimensions. 
More concretely,  one can ask whether there exist critical points (CFTs) which break some global symmetry at arbitrary finite temperature. The most familiar CFTs do not exhibit symmetry breaking at finite temperature, and moreover,
in the context of the AdS/CFT correspondence, critical points at finite temperature are described by an uncharged black brane which obeys a no-hair theorem. 
Yet, we show that there exist CFTs which have some of their
internal symmetries broken at arbitrary finite temperature. Our main example is a vector model which we study both in the epsilon expansion and arbitrary rank as well as the large rank limit (and arbitrary dimension). The large rank limit of the vector model displays a conformal manifold, a moduli space of vacua, and a deformed moduli space of vacua at finite temperature. The appropriate Nambu-Goldstone bosons including the dilaton-like particle are identified. Using these tools we establish symmetry breaking at finite temperature for finite small $\epsilon$.  
We also prove that a large class of other fixed points, which describe some of the most common quantum magnets, indeed behave as expected and do not break any global symmetry at finite temperature. We discuss some of the consequences of finite temperature symmetry breaking for the spectrum of local operators. Finally, we propose a class of fixed points which appear to be possible candidates for finite temperature symmetry breaking in $d=2$. 

\vfill 

\end{titlepage}
	
		\setcounter{table}{1}.
	
\setcounter{tocdepth}{3}
\renewcommand{\cfttoctitlefont}{\large\bfseries}
\renewcommand{\cftsecaftersnum}{.}
\renewcommand{\cftsubsecaftersnum}{.}
\renewcommand{\cftsubsubsecaftersnum}{.}
\renewcommand{\cftdotsep}{6}
\renewcommand\contentsname{\centerline{Contents}}
	
\tableofcontents


\section{General Remarks}\label{sec:intro}

The phenomenon of Spontaneous Symmetry Breaking is commonplace in Nature. The progress in theoretical understanding of this subject was followed by searching for systems in which the symmetry was restored. The mechanism for that can be intrinsic to the system, i.e. dynamical, or it can be brought about by subjecting the system to external conditions such as varying degrees of freedom, temperature, density or large overall quantum numbers. In this work we plan to discuss only the effects of temperature. A more concise analysis of the main content of this work will be provided in \cite{Chai:2020PRL}.

We start by reviewing some aspects of Spontaneous Symmetry Breaking. For a sample of standard references on the subject, see for instance \cite{Amit:1984ms,Peskin:1995ev,Weinberg:1996kr,Parisi:1998}.

\subsection{A Review of what Symmetry Breaking is}

Let us consider a general quantum system in $d$ space dimensions with a $\Z_2$ global symmetry. We study the theory with the $d$ space dimensions being compact, denoting the space as $\CM_d$.
Let $\CO$ be a local operator which is odd under the $\Z_2$ symmetry and we consider the expectation value of $\CO$ in the thermal ensemble with inverse temperature $\be_{\text{th}}$
\begin{equation}\langle \CO\rangle^{\CM_d}_{\beta_{\text{th}} }\equiv \frac{1}{Z}Tr_{\CH_{\CM_d} } \CO  e^{-{\beta_{\text{th}}} H}~.\end{equation}
Here $Z$ is the  partition function and $\CH_{\CM_d}$ denotes the Hilbert space. We trace over the Hilbert space to obtain the expectation value of $\CO$ in the thermal ensemble. 

In quantum systems in finite volume we can always choose the energy eigenstates to be eigenstates of $\Z_2$, and for each such eigenstate $|\Psi_{n,q}\rangle$ it holds that $\langle \Psi_{n,q}|\CO|\Psi_{n,q}\rangle=0$. (Here $q\in\{0,1\}$ depending on whether the state is even or odd under $\Z_2$.)

Therefore for every compact space $\CM_d$ we have 
\begin{equation}\label{vanishingc}\langle \CO\rangle^{\CM_d}_{\beta_{\text{th}}}=0~.\end{equation}
This is the familiar statement that in compact space symmetries cannot break (whether the temperature is zero or not).\footnote{This is true as long as the number of degrees of freedom, $N$, is finite (in particle physics or condensed matter systems the number of degrees of freedom is always finite in this sense). For the case of $N=\infty$ there can be a phase transition even in compact space. Two examples where this happens are the one plaquette model \cite{Gross:1980he,Wadia:1980cp} and the four-dimensional $N=4 ~ SU(N)$ super Yang-Mills theory on a finite sphere~\cite{Witten:1998zw,Susskind:1997dr,Sundborg:1999ue,Aharony:2003sx}. Special features of the infinite $N$ limit will be important in part of this work.}

From now on, when we write $\langle \CO\rangle_{\beta_{\text{th}}}$ we mean $\langle \CO\rangle_{\beta_{\text{th}}}^{\R^d}$, i.e. the usual infinite volume limit. While in any compact space~\eqref{vanishingc} is correct, there could be a difficulty in taking the infinite volume limit while maintaining $\langle \CO\rangle_{\beta_{\text{th}}}=0$. When we are unable to maintain $\langle \CO\rangle_{\beta_{\text{th}}}=0$ in the infinite volume limit we 
say that spontaneous symmetry breaking occurs. This typically depends on ${\beta_{\text{th}}}$, in the sense that for some values of ${\beta_{\text{th}}}$ we may be able to maintain $\langle \CO\rangle_{\beta_{\text{th}}}=0$ and for some other values we may not. 

One familiar reason that the infinite volume may be singular is the following: As we increase the volume of $\CM_d$, the Hilbert space $\CH_{\CM_d}$ may develop different ``sectors'' of states which have exponentially vanishing matrix elements with states in other sectors. Then, the infinite volume limit is taken by discarding some states in the Hilbert space and it may happen that as a result we cannot diagonalize $\Z_2$ in the infinite volume limit. 

In the standard situation of symmetry breaking at low temperatures the way these sectors in the Hilbert space arise is as follows. We have two nearly degenerate eigenstates in finite volume, $|+\rangle$ and $|-\rangle$, such that they are respectively even and odd under $\Z_2$ and the energy difference is given by \begin{equation}\label{domainwallinstanton}\Delta E\sim e^{-T_{\text{w}} Vol(\CM_d)}~.\end{equation} where $T_{\text{w}}$ is a dimensionful constant known as the domain wall tension. The energy splitting is interpreted for some range of parameters as an instanton effect in quantum mechanics and therefore it is natural to define the two ``minima''
$$|VAC1\rangle=|+\rangle+|-\rangle~,\qquad |VAC2\rangle=|+\rangle-|-\rangle~.$$
These states can be thought of as being separated by a barrier that scales with the volume of space, and hence the instanton~\eqref{domainwallinstanton}. Since the barrier scales with the volume of space, the low lying states fall into two distinct sectors which do not communicate in infinite space. We can choose to be in {\it either} of $|VAC1\rangle$ or $|VAC2\rangle$ as we take the infinite volume limit. Since $|VAC1\rangle$ and $|VAC2\rangle$ are not $\Z_2$ eigenstates, the $\Z_2$ symmetry is broken spontaneously. At zero temperature as well as at sufficiently low temperatures we therefore have $\langle \CO\rangle_{\beta_{\text{th}}}\neq 0$. 

States which are obtained from $|VAC1\rangle$ with the action of only finitely many operators are nearly orthogonal to states which are obtained from $|VAC2\rangle$ by acting with finitely many operators, hence the notion of superselection sectors.
But note that for states where the energy scales with the volume and is sufficiently high, the distinction between $|VAC1\rangle$ and $|VAC2\rangle$ essentially disappears. 
For this reason, we often think that at high enough temperatures, where the typical state is a state with a larger energy density than the energy scale involved in the spontaneous symmetry breaking, the $\Z_2$ symmetry must be restored.  

Another viewpoint takes into account that at finite temperature we do not minimize the energy but instead we minimize $$F=E-{S\over {\beta_{\text{th}}}}$$ (where $S$ is the entropy), and hence at high temperature the high entropy states dominate. Since high entropy states are disordered we again expect that for high enough temperatures the symmetry will be restored. 

In this note we would like to examine the question of whether it is really the case that at high enough temperature all symmetries are restored.\footnote{In this note we only discuss ordinary, zero-form symmetries. The de-confinement transition of course famously behaves in the opposite fashion but we do not discuss higher symmetries here. Yet it is worth pointing out that, in $d=2$ space dimensions, the two questions are linked! If we have a theory $\cal{T}$ with $\mathbb{Z}_2$ global symmetry we could gauge it and obtain a new theory $\cal{T}'$ with a one-form $\mathbb{Z}_2$ symmetry instead. Then, if at finite temperature the original $\mathbb{Z}_2$ was broken in the theory $\cal T$, then in the new theory $\cal{T}'$ the one-form $\mathbb{Z}_2$ symmetry is un-broken. Therefore in $d=2$ space dimensions an example with broken ordinary symmetry at finite temperature is essentially equivalent to an example which confines at finite temperature. This relationship between ordinary symmetries and higher symmetries  was explained in~\cite{Gaiotto:2014kfa,Gaiotto:2017yup}. In the context of the AdS/CFT correspondence, the Black-Hole picture of course leads one to expect finite temperature de-confinement. For some references on this subject see the original work~\cite{Witten:1998zw} as well as some more recent developments~\cite{Gursoy:2010kw,Hofman:2017vwr} and references therein. More details about this relationship between the behavior of theory $\cal T$ and theory $\cal{T}'$ are in Appendix~\ref{FootnoteTwo}. }   There are many examples in the literature of systems that break some symmetries at intermediate temperatures, we will review some of those beautiful constructions. But our focus is on the true high temperature limit.

Using the relationship between finite temperature and a theory on a circle,  we can  conclude that  in $d=2$ only discrete symmetries can break spontaneously at finite temperature~\cite{Coleman:1973ci} and in $d=1$ no symmetries whatsoever can break at finite temperature.

\subsection{Arguments from the AdS/CFT Correspondence}

The AdS/CFT correspondence links the question of symmetry restoration at high temperatures with the no-hair ``theorem''.  According to the AdS/CFT correspondence \cite{Maldacena:1997re,Gubser:1998bc,Witten:1998qj}, a conformal theory in $\mathbb{R}^{d,1}$ is dual to the Poincar\'e patch of $AdS_{d+2}$.
Putting the field theory at finite temperature is then interpreted as a black brane in $AdS_{d+2}$ \cite{Witten:1998zw}. The statement that there is symmetry breaking in the CFT is translated to hair on the black brane\cite{Gubser:2008px,Hartnoll:2008vx,Hartnoll:2008kx}. Black branes which are charged are known to exhibit instabilities and they can develop hair through the condensation of scalar fields. But to our knowledge no such hair has been exhibited for uncharged black branes. (Equivalently, when the temperature of the black brane is much larger than the chemical potential the hair disappears.) This statement also extends to the possible condensation of scalars with deformed boundary conditions \cite{Faulkner:2010fh, Faulkner:2010gj}.\footnote{See however \cite{Buchel:2009ge, Donos:2011ut, Gursoy:2018umf, Buchel:2018bzp}. While the hairy black holes in \cite{Buchel:2009ge, Donos:2011ut,Gursoy:2018umf, Buchel:2018bzp} did not dominate the ensemble at high temperatures, their mere existence is a possible step towards a violation of the no-hair theorem for black branes. See also~\cite{Buchel:2020thm} for a construction that appeared shortly after this paper had been submitted.} The CFT constructions we present here are not at odds with the no-hair theorem for such black branes in AdS. The reason being that our models are  vector models and as such do not have standard AdS duals (rather, the dual description is via Vasiliev's equations \cite{Klebanov:2002ja}).

\subsection{Lattice Systems and the Continuum Limit} 

The notion of arbitrarily high temperature has to be clarified. Let us first examine local lattice systems with finitely many degrees of freedom per site (spin systems) and where the $\Z_2$ symmetry is realized on-site. In such systems, strictly infinite temperature corresponds to the unit density matrix, i.e. as ${\beta_{\text{th}}}\to 0$
$$ e^{-{\beta_{\text{th}}} H}\to \mathbb{I}~.$$
The state $\mathbb{I}$ makes sense in such lattice models. Let us now take some order parameter localized to a site. Since in the state $\mathbb{I}$ all sites decouple and the Hilbert space is a direct product $\CH=\otimes_{sites} \CH_i$, the expectation values of such local operators vanish since for such local operators as ${\beta_{\text{th}}}\to0$,  $\langle \CO\rangle_{\beta_{\text{th}}} \to Tr_{site} \CO=0$. Hence, for such lattice systems the symmetries must be restored at sufficiently high temperature~\cite{kliesch2014locality}.

Let us now consider the regime of QFT. If the lattice model is described by QFT at distances much larger than the lattice spacing, then we can also consider a temperature which is much larger than the inverse correlation length but much smaller than the inverse lattice spacing distance $a$
$$a\ll {\beta_{\text{th}}}\ll \xi ~.$$ This is a less trivial limit. In fact, in QFT the state $\mathbb{I}$ does not necessarily make sense\footnote{We thank A. Kapustin and D. Harlow for discussions on this topic.} and the high temperature limit in the continuum QFT sense contains potentially nontrivial physics as we will see. This is the sense in which we will find nontrivial behavior even at arbitrarily high temperature. 

In fact, a QFT does not necessarily require a lattice to be defined. It can be ultraviolet complete by itself. The short distance limit is then described by a Conformal Field Theory (CFT). The question about the behavior of the theory at very high temperatures can be then translated into a question about conformal field theory at nonzero temperature. Since there is no inherent scale in a CFT, any nonzero temperature is equivalent to any other nonzero temperature. Hence, if there is symmetry breaking in a CFT at some nonzero temperature there is symmetry breaking at all nonzero temperatures. 

\subsection{The Central Question}

\begin{tcolorbox}

Are there unitary, local, nontrivial CFTs which break a global symmetry at finite temperature?

\end{tcolorbox}

Unitarity appears to be important for the following reason: Instead of the thermal ensemble $e^{-{\beta_{\text{th}}} H}$ one could ask the same question about the high temperature behavior in the ensemble with a chemical potential 
$e^{-{\beta_{\text{th}}} H-\mu Q}$. Actually, in some situations with 't Hooft anomalies it is already known that one can guarantee symmetry breaking for any radius of the thermal circle, i.e. any $\beta_{th}$~\cite{Komargodski:2017dmc, Aitken:2017ayq, Tanizaki:2017qhf,Dunne:2018hog,Wan:2019oax} for some appropriate  imaginary values of $\mu$. A similar thing sometimes happens with random chemical potentials~\cite{Hong:2000rk}. On the other hand, for the ensemble $e^{-{\beta_{\text{th}}} H}$ no such example exists to our knowledge. The question is also interesting in systems with no translational invariance. A nice setup where one could study it is in~\cite{Dong:2012ua} and see also~\cite{Alberte:2017oqx}.

The main point in this note is the construction of conformal models in $d=3-\epsilon$ dimensions which break a symmetry at finite temperature. We will also provide a conjecture for a model in $d=2$, but since it is only conjectural at the moment, all the examples where we can rigorously establish symmetry breaking at finite temperature are in fractional dimensions and hence are not fully unitary models \cite{Hogervorst:2015akt} (and see references therein -- however, in the infinite rank limit these models may become unitary).

\subsection{Our Construction}

What we will do here is to present a construction of CFTs which have a unique gapless ground state at zero temperature and in the infinite $N$ limit some of them have flat directions in field space. At nonzero temperature, however, we will find examples that exhibit spontaneous symmetry breaking.

Our examples are in a class of conformal vector models. We first prove a no-go theorem: such symmetry breaking at finite temperature cannot occur in models with a single quadratic Casimir. This explains a posteriori why many familiar quantum magnets restore their symmetries at high temperature. But in the bi-conical class of fixed points \cite{Nelson:1974xnq,Kosterlitz:1976zza,Calabrese:2002bm,Rychkov:2018vya}, which have two quadratic Casimirs, we find examples which display symmetry breaking at any finite temperatures. 

We treat the bi-conical models both in the limit of small epsilon and in the limit of finite epsilon and large rank.  We find that the two approaches essentially overlap and agree.  These bi-conical CFTs have symmetry group $O(m)\times O(N-m)$ and the smaller group of the two breaks at finite temperature. For instance (and without loss of generality) if $m<N/2$ the unbroken symmetry group is $O(m-1)\times O(N-m)$. Therefore there is no thermal gap and instead we have Nambu-Goldstone bosons living on $S^{m-1}$. In the equal rank case $2m=N$ no symmetry breaking occurs at finite temperature!

We find some special features when studying the large rank limit of the biconical models. We find a one-dimensional conformal manifold and a moduli space of vacua though these models have no supersymmetry. In addition, the moduli space of vacua does not disappear at finite temperature, but instead, it is deformed.  Moreover, the ground state energies of the thermal effective potential depend neither on temperature nor on the expectation value of the field leading to spontaneous symmetry breaking \cite{Bardeen:1983rv,Amit:1984ri,Rabinovici:1987tf}.\footnote{This statement excludes trivial temperature dependence that is scheme dependent.} One finds a certain hyperbola in the space of fields, where all the vacua on this hyperbola are degenerate. The curvature of the hyperbola is set by the temperature. This allows us to establish that indeed symmetry breaking takes place in these models in $d<3-\epsilon$ dimensions for finite small enough $\epsilon$. For the case of equal rank $2m=N$ the hyperbola is not deformed at finite temperature and indeed symmetry breaking at finite temperature does not occur. In $d=3$ these models are free and hence trivial and at $d=2$ the Nambu-Goldstone bosons on $S^{m-1}$ are lifted by non-perturbative effects and hence, strictly speaking, no symmetry breakdown occurs. This is of course in line with the general expectation that no continuous symmetry breaking can occur at finite temperature in 2+1 dimensions.\footnote{There are known exceptions to this expected behaviour. We refer the reader to \cite{2019JSP...175..521H} for a brief review on such exceptions. See \cite{Witten:1978qu} as well for a discussion on this topic.}

It is still interesting though that the thermal gap is exponentially small for large $m$ because these non-perturbative effects occur at an exponentially small scale. This is a huge hierarchy between the thermal scale and the actual correlation length. 

A very interesting special case are the models with symmetry $O(1)\times O(N-1)$. For them we cannot straightforwardly apply the large rank methods since one of the ranks is just 1. But we can still carry out the $\epsilon$ expansion and we find that the symmetry is broken at finite temperature to $O(N-1)$, hence, there are 2 vacua. These models therefore are possible candidates for a full fledged unitary CFT in 2+1 dimensions with symmetry breaking at finite temperature. We cannot prove, though, that this indeed occurs in 2+1 dimensions and our evidence is based solely on the $\epsilon$ expansion. It is conceivable that this problem can be settled in the future. 

In summary, we report here on a construction of critical points in $d=3-\epsilon$ space dimensions which break some global symmetries at finite temperature. These models also display some other interesting features, such as moduli spaces of vacua at zero and nonzero temperature. We emphasize a special case in the above class of critical points that may break a $\mathbb{Z}_2$ symmetry at finite temperature strictly in $d=2$ space dimensions.

\subsection{Consequences for the Spectrum of Operators and the Phase Diagrams}

As we reviewed above, there is an intuitive picture of what low temperature symmetry breaking means in terms of which states survive the infinite volume limit. But imagine a CFT that at finite temperature breaks a $\Z_2$ symmetry. What does that mean for the spectrum of dimensions of local operators of the theory?

It is useful to address this question in radial quantization, where the spectrum of the theory on $S^d$ is isomorphic to the space of local operators and the energies are identified with the scaling dimensions. We study the partition function on $S^d\times S^1_{\beta_{\text{th}}}$  which is hence given by \begin{equation}Z_{S^d\times S^1_{\beta_{\text{th}}}}=\sum_{\Delta} e^{-{\beta_{\text{th}}} \Delta/R}~.  \end{equation}
where $R$ is the radius of  $S^d$. Evidently, the partition function is only a function of ${\beta_{\text{th}}}/R$ due to conformal invariance. The limit of large volume is obtained by taking $R\to\infty$ with fixed ${\beta_{\text{th}}}$. In this limit we can use effective field theory since there is an approximately local theory on $S^d$ at distances much bigger than ${\beta_{\text{th}}}$. If we assume a thermal gap and no symmetry breaking, then this effective theory on $S^d$ is obtained from a formal series expansion of local functionals of the metric in the $d$-dimensional theory
\begin{equation}\label{HighTexp}\sqrt g{\cal L} = A{\beta^{-d}_{\text{th}}} \sqrt g+B {\beta^{-d+2}_{\text{th}}} R\sqrt g+\cdots~.\end{equation}
where $A,B...$ are dimensionless, model-dependent constants. This leads to the usual expansion of the partition function at large $R$ (or, alternatively, small $\beta$)
 \begin{equation}{\beta_{\text{th}}}\to 0~,\quad \log Z_{S^d\times S^1_{\beta_{\text{th}}}}\sim {2A\pi^{d/2+1/2}\over \Gamma\left({d\over 2}+{1\over 2}\right)} {\beta^{-d} _{\text{th}}}R^d+{2Bd(d-1)\pi^{d/2+1/2}\over \Gamma\left({d\over 2}+{1\over 2}\right)}{\beta^{-d+2}_{\text{th}}}R^{d-2}+\cdots~.\end{equation}
We can then infer the density of operators at high $\Delta$ $$\log \rho= {2^{1\over d+1}A^{1\over d+1}(d+1)\pi^{1\over 2}\over d^{d\over d+1}\Gamma\left({d\over 2}+{1\over 2}\right)^{1\over d+1}}
 \Delta^{{d\over d+1}}+\cdots~.$$
(The density should be interpreted in a Tauberian sense \cite{Pappadopulo:2012jk,Qiao:2017xif,Mukhametzhanov:2019pzy}.) This is how standard CFTs, satisfying the assumptions above (i.e. a thermal gap and a unique vacuum at nonzero temperature) behave.

In the event that there is a gap but the $\Z_2$ symmetry is broken, there is a mild but important violation of~\eqref{HighTexp}. The expansion~\eqref{HighTexp} still holds in each of the two vacua but the action~\eqref{HighTexp} is missing a non-local piece due to the fact that there are two vacua. It is nonlocal in the sense that it cannot be described by a local functional of the metric.

Hence to leading order we have now \begin{equation}\label{factortwo}\log\rho(\Delta) =   {2^{1\over d+1}A^{1\over d+1}(d+1)\pi^{1\over 2}\over d^{d\over d+1}\Gamma\left({d\over 2}+{1\over 2}\right)^{1\over d+1}}
 \Delta^{{d\over d+1}}+\cdots+\log 2+\cdots ~.\end{equation} 
Say for $d=3$, the additive contribution to the density of states  $\log 2$ cannot be obtained from any local term in the action~\eqref{HighTexp}.\footnote{For even $d$ a constant piece can be obtained from a local term in~\eqref{HighTexp}. But the factor of $\log 2$ we are talking about has a truly non-local origin due to the two vacua and it would exist even on a torus where no term in~\eqref{HighTexp} would give such a contribution. Another important thing to realize is that in even $d$ there could be a dimensionless gravitational counter-term in these two vacua. The number of such counter-terms depends on the number of space-time dimensions. In the particular case of $d=3$ there are no such counter-terms and hence the $\log 2$ contribution on $S^3\times S^1$ is scheme independent. } 

The $\log2$ contribution suggests a mechanism for why the infinite volume limit fails to preserve $\langle \CO\rangle_{\beta_{\text{th}}}=0$. The factor of $\log 2$ suggests that the spectrum of {\it high dimension} operators comes in two sectors, each of which furnishes what would seem like a local theory in the thermodynamic limit. Each of these sectors consists of operators which are not $\mathbb{Z}_2$ eigenstates. In the event that the symmetry that is spontaneously broken at finite temperature is a continuous symmetry, $\log 2$ is replaced by a constant times $\log R/{\beta_{\text{th}}}$. Therefore, as we take the infinite volume limit we will find again that the space of states breaks up into sectors. But unlike in the standard, familiar, situation where this happens for the low lying states, here these are the states with fixed energy density that break up into such distinct sectors (while the low lying states do not!).  It would be nice to understand better this situation. 

The general framework for CFTs at finite temperature has been recently studied in \cite{Iliesiu:2018fao,Petkou:2018ynm,Manenti:2019wxs}. Ultimately, symmetry breaking at finite temperature in CFTs should be understood in this language.  

\begin{figure}[]
\centering
\includegraphics[scale=0.35]{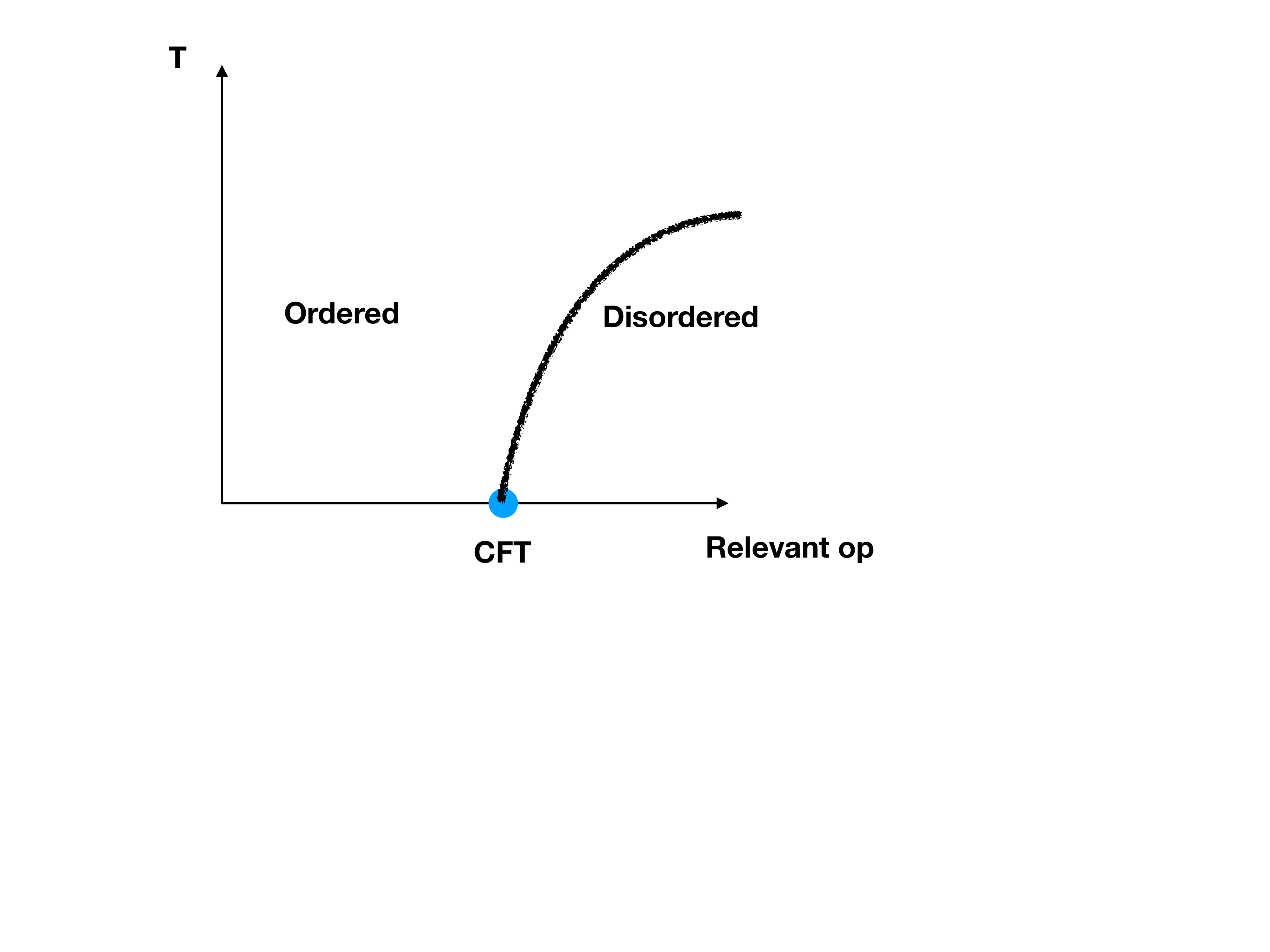}
\vspace{-6em}
\caption{A possible phase diagram in a theory where the critical point breaks a symmetry at finite temperature. In familiar systems, the black line always turns the other way.}
\label{PT}
\end{figure}

Quantum critical points with such unfamiliar behavior at finite temperature would lead to rather unfamiliar phase diagrams. Symmetry breaking in the CFT at finite temperature implies that, had we started in the ordered zero-temperature phase, the order could persist for any temperature. This is the opposite situation than what is encountered in most of the quantum critical points. Schematically, if we had just one relevant operator, one could find a phase diagram such as in figure~\ref{PT}. By contrast, in the more familiar situations, the finite temperature phase transition line bends in the other direction.

\subsection{The Outline}
The outline of this paper is as follows. In section 2 we discuss some general facts about thermal field theory. We emphasize the infrared problem, review some familiar examples, and also present the construction of intermediate-temperature symmetry breaking. We also make some general remarks about weakly coupled conformal gauge theories in 3+1 dimensions.
In section~3 we discuss our results about vector models. We prove a general theorem about single Casimir models, discuss the small epsilon and large rank limit of the bi-conical models and construct controlled examples of symmetry breaking in CFTs. We then discuss a possible candidate for finite temperature symmetry breaking in $d=2$.  Finally,  the details leading to footnote 2 are given in appendix A, and some properties of the large rank limit are discussed in appendix B.

\section{Aspects of Thermal Field Theory} \label{sec:review}
\subsection{The $\phi^4$ Model}

To introduce some of the ideas that will be crucial below, it is useful to start with the $\phi^4$ model. Let us take the potential to be 
\begin{equation}\label{Isfour}V={1\ov 2}m^2\phi^2 +{1\ov 4! }\lambda \phi^4~.\end{equation} 
We will first consider this model in $3+1$ dimensions and then discuss what happens in other space-time dimensions. Of course, the model~\eqref{Isfour} is not a UV complete QFT. But that would not be important for us yet, as we will only try to understand its behavior at intermediate temperatures, much below the Landau pole scale. 
We first set $m^2=0$. Then the model at zero temperature is at a 2nd order phase transition described at very long distance by a free field theory. But due to the coupling $\lambda\ll1$ (which is irrelevant from the point of view of the infrafred), at finite temperature one obtains an effective mass. The best way to think about this effective mass is to rotate to Euclidean signature, compactify the model on a circle of radius ${\beta_{\text{th}}}\over 2\pi$ and study the physics at distances $ x\gg {\beta_{\text{th}}}$ in the remaining $\mathbb{R}^3$. The physics at long distances on $\mathbb{R}^3$ is guaranteed to be a local QFT in 3 (Euclidean) dimensions. Expanding in modes on the circle we find fields $\phi_n$ labeled by integer $n$ such that $\phi_n=\phi_{-n}^*$. The Lagrangian in $\mathbb{R}^3$ takes the form (after canonically normalizing the fields)
\begin{equation}\label{Lagfour}{\cal L} =\int d^3x \left[\half (\partial\phi_0)^2+{\lambda{\beta^{-1}_{\text{th}}}\over 4!}\phi_0^4+\sum_{n=1}^\infty \partial\phi_n\del\bar\phi_n+\sum_{n=1}^\infty{4\pi^2n^2\over {\beta^2_{\text{th}}}}|\phi_n|^2+{\lambda{\beta^{-1}_{\text{th}}}\over 2}\phi_0^2\sum_{n=1}^\infty|\phi_n|^2\right]~. \end{equation}
(We have not included the self-interactions of the KK modes for reasons that will soon become clear.) The modes $\phi_n$ with $n\geq 1$ are massive with mass $m_n=2\pi n/{\beta_{\text{th}}}$. However the mode $\phi_0$ is massless and one should worry about it. Unlike the $\phi^4$ interaction which is infrared free in four space-time dimensions, the $\phi^4$ interaction in three space-time dimensions leads to strong coupling below the energy scale $\lambda{\beta^{-1}_{\text{th}}}$ and hence unless the mode $\phi_0$ decouples beforehand we will run into strong coupling. The strong coupling dynamics of such zero modes is a source of infrared problems in thermal field theory. Of course, there are no actual infrared problems; it is up to us whether we can or cannot solve the dynamics of the zero mode. 

Due to the last term in~\eqref{Lagfour} one may be saved from strong coupling physics since the radiative corrections from the massive particles running in the loop may induce a sufficiently large mass for $\phi_0$.
The induced mass to leading order in $\lambda$ is \cite{Weinberg:1974hy,Dolan:1973qd} \begin{equation}\label{thermal}m^2_{\text{th}}=\lambda {\beta^{-1}_{\text{th}}}\sum _n\int {d^3k\over (2\pi)^3} {1\over k^2+{(2\pi n)^2 \over {\beta^2_{\text{th}}}}}=-{\lambda\over 2} {\beta^{-2}_{\text{th}}} \sum_{n>0} n = {\lambda\over 24} {\beta^{-2}_{\text{th}}}~.  \end{equation} 
The integrals in~\eqref{thermal} are clearly divergent but we have nevertheless evaluated them using dimensional regularization. This requires some clarification. If this was purely a 3d QFT, then the mass would have been incalculable as one could add a counter-term. But since we are studying a four dimensional theory on a circle, the counter-terms must descend from local functionals in four dimensions. Four dimensional counter-terms can never lead to a dependence such as ${\beta^{-2}_{\text{th}}}$ on the circle radius. So to make the discussion~\eqref{thermal} completely rigorous we could have taken a ${\beta_{\text{th}}}$ derivative of the integrals, rendering them convergent. 

The induced thermal mass squared is positive and it is of the order of $\lambda{\beta^{-2}_{\text{th}}}$.  If the thermal mass is above the strong coupling scale then we are saved from strong coupling dynamics and the analysis is self-consistent. Indeed, the strong coupling scale is $\lambda{\beta^{-1}_{\text{th}}}$, which should be compared to the thermal mass, $\lambda^{1/2}{\beta^{-1}_{\text{th}}}$. Therefore, as long as as $\lambda\ll1$ we see that the thermal mass is far above the strong coupling scale and our results are self consistent. In this regime the mode $\phi_0$ is weakly coupled and higher-order contributions to the thermal mass are negligible. 
If we started at zero temperature in the ferromagnetic phase with $m^2<0$, our analysis shows that at temperatures of order $m/\sqrt\lambda$ the $\mathbb Z_2$ symmetry would be restored. Of course, our model is not ultraviolet complete so we cannot quite discuss extremely large temperatures. But the restoration of the symmetry at temperatures higher than $m/\sqrt\lambda$ takes place in an entirely controlled fashion. The Landau pole scale is indeed exponentially far away.\footnote{Note that near the restoration temperature, the thermal mass essentially cancels against the zero temperature mass, which means that the model is strongly coupled in that region. But we can study the model reliably away from that region.}

The restoration of the symmetry (i.e. the exit from the ferromagnetic phase) here takes place due to the fact that the thermal mass~\eqref{thermal} is positive. This drives the system at high temperatures to the unbroken (paramagnetic) phase. If the sign of the thermal mass was reversed the physics would have been completely different. The question of symmetry restoration at high temperatures is thus intimately related to the sign of the thermal mass squared for the order parameter.

The fact that we can avoid the strong coupling dynamics in the infrared is not to be taken for granted. For instance, if we consider the model~\eqref{Isfour} in 2+1 dimensions, most of the formulae go through except that now the thermal mass squared is $m_{\text{th}}^2\sim \lambda{\beta^{-1}_{\text{th}}}$ (compare with~\eqref{thermal}) and the strong coupling scale squared is likewise at $\lambda{\beta^{-1}_{\text{th}}}$ and hence (apart from possible logarithmic effects) there is no parametric separation between the thermal mass and the strong coupling scale.

\subsection{More General Scalar Models in 3+1 Dimensions} \label{sec:Rochelle}

There is no general principle that says that the one loop thermal mass squared should be positive. In this subsection we review a construction by Weinberg \cite{Weinberg:1974hy} for a model of scalar fields in four dimensions with quartic interactions and a negative thermal mass squared. 

 The degrees of freedom consist of two scalar fields $\phi_{1}$, $\phi_2$ transforming under an $O(N)\times O(N)$ symmetry in the representations $(N,1)$ and $(1,N)$, respectively. 
The most general quartic interactions preserving the global symmetry are given by
\beq\label{Rmodel}
V=\lam _{11} (\phi_1^2)^2 +2\lam_{12} (\phi_1^2)(\phi_2^2) +\lam_{22}(\phi_2^2)^2~.
\eeq
To avoid a runaway we need to impose that $\lam_{11},\lam_{22}\geq 0$ and if $\lambda_{12}<0$ we also need to impose $\lam_{12}^2 \leq \lam_{11}\lam_{22}$. There is classically a flat direction if the latter inequality is saturated.

The one-loop thermal mass for $\phi_1$ and $\phi_2$ is evaluated very similarly to our previous example
\beq \label{eq:weinbergthermalmass}
m_{\text{th};1}^2&={1\ov 3} \left ( (N+2) \lam_{11} +N\lam_{12}\right)\beta_{\text{th}}^{-2}~,
\\m_{\text{th};2}^2&={1\ov 3} \left ( (N+2) \lam_{22} +N\lam_{12}\right)\beta_{\text{th}}^{-2}~.
\eeq

We see that the presence of $O(N)\times O(N)$ symmetry group allows a regime in parameter space where the thermal corrections destabilize the origin of field space. For instance, take negative $\lambda_{12}$ such that $|\lambda_{12}|\gg \lambda_{11}$ but $\lambda_{22}\gg|\lambda_{12}|$ and also $\lambda_{22}\gg \lambda^2_{12}/\lambda_{11}$. This can be achieved while having $\lambda_{11},\lambda_{22},|\lambda_{12}|\ll 1$, i.e. we are entirely in the perturbative regime with a stable vacuum. 

Therefore if the original theory at zero temperature were in the symmetric (disordered) phase with $m_1^2,m_2^2\geq 0$, there would be a finite domain in theory space (parameterized by $(\lam_{11},\lam_{12},\lam_{22})$) where the system develops spontaneous symmetry breaking at high enough temperatures. The system is therefore in a broken phase at high temperatures but in a symmetric phase at low temperature. This is perplexing and goes against one's usual intuition about entropy effects at high temperature. One may find consolation in that the model~\eqref{Rmodel} is not ultraviolet complete; at really high temperatures the couplings $\lambda$ grow strong and the description breaks down.  

This $O(N)\times O(N)$ model of symmetry non-restoration led to many interesting ideas in the physics of early universe by recasting various important problems (such as $CP$ violation and domain-wall formations) in the light of possible symmetry non-restoration in the Standard Model \cite{Mohapatra:1979qt, Langacker:1980kd, Salomonson:1984rh,Dodelson:1989ii,Dodelson:1991iv,Dvali:1995cj}. This $O(N)\times O(N)$ model was scrutinzed in various other approaches, the majority of which supported the existence of  symmetry breaking at high temperatures \cite{Bimonte:1995xs,AmelinoCamelia:1996hw,Orloff:1996yn,Roos:1995vm,Jansen:1998rj,Bimonte:1999tw,Pinto:1999pg} albeit with some lingering debate \cite{Fujimoto:1984hr,Klimenko:1988ng,Grabowski:1990qc,Gavela:1998ux}. The phenomenon of symmetry breaking at finite temperatures with a symmetric zero temperature phase is also found in nature: the Rochelle's salt \cite{kao2004dielectric} which is a sodium potassium tartrate (KNaC$_4$H$_4$O$_6\cdot$4H$_2$O) has three crystal phases. The two transition temperatures are at -18\textdegree C and 24\textdegree C where the intermediate phase develops an orthorhombic crystal while the other two phases are monoclinic. As the orthorhombic phase is more ordered than the monoclinic crystal, the phase transition at -18\textdegree C can be regarded as a phenomenon of symmetry non-restoration. (The salt finally restores all the spontaneously broken crystal symmetries once it becomes a liquid at 55\textdegree C.) Since our QFT model~\eqref{Rmodel} is not ultraviolet complete, one should regard this construction as some intermediate symmetry non-restoration, while the fate of the system at asymptotically high temperatures remains unknown (or rather, not well defined within the QFT).

The behavior of the model~\eqref{Rmodel} is perplexing but we consoled ourselves in that it does not imply symmetry non-restoration at asymptotically high temperatures. Surprisingly, later in this paper we construct theories that are well defined at arbitrarily short distance scales and they exhibit symmetry breaking at arbitrarily high temperature. (Though, as emphasized in the introduction, all the models where we can establish symmetry non-restoration rigorously, live in non-integer dimensions.)

In the following subsection we discuss some basics of theories which include gauge fields. We discuss the infrared ``problem''  and quote the results we have found for the simplest weakly coupled conformal gauge theories. 

\subsection{Thermal Field Theory with Gauge Fields} \label{sec:IR problem}
In this subsection we make some remarks about the thermal properties of 3+1 dimensional gauge theories. This section can be skipped if one is only interested in the main results of this paper, which are in the next section about vector models. Essentially the content of this subsection is that we will cover some of the simplest weakly coupled conformal gauge theories and argue that they do not provide examples of conformal field theories that break an ordinary global symmetry at finite temperature.

Let us start from the free $U(1)$ gauge field in 3+1 dimensions at finite temperature. Reducing on a circle, $A_\mu$ breaks up into $A_0$ which is a compact scalar in 3 dimensions and a massless 3-dimensional gauge field $A_i$. The latter is also equivalent to a compact scalar through Poincar\'e's duality $F=d\varphi$. So we have two massless compact scalars in 3 dimensions at any value of the temperature. While these look like superfluid modes, they do not correspond to ordinary symmetry breaking, rather they are related to the electric and magnetic one-form symmetries of the original massless gauge theory in 3+1 dimensions. 

If we were to add some dynamical electric particles, then the compact scalar $A_0$ would obtain a mass while $\varphi$ would remain massless. This is the familiar fact that in QED the electric fields are screened in the thermal plasma while the magnetic fields are not.   

The situation becomes conceptually more complicated in non-Abelian gauge theories with (or without) matter \cite{Gross:1980br}. While the treatment of $A_0$ (which becomes an adjoint scalar field) is quite similar -- it obtains a mass of order \begin{equation} \label{electricmass}m_{el} \sim g_{YM}{\beta^{-1}_{\text{th}}}~.\end{equation}
(where $g_{YM}$ is the four-dimensional gauge coupling)
the $A_i$ components furnish a non-Abelian gauge theory in 3 dimensions. 
Such gauge theories are never infrared free, regardless of how much matter is put in, since the effective three-dimensional gauge coupling is \begin{equation}\label{massgap}g^2_{3d}\sim g^2_{YM}{\beta^{-1}_{\text{th}}}~.\end{equation}
which is always a relevant perturbation in the UV since it has mass dimension 1.
For instance, this three-dimensional sector may confine and develop a mass gap at the scale~\eqref{massgap}.

This is reminiscent of the discussion in the $\phi^4$ model, where the three-dimensional theory which is obtained at distances much larger than the circle size could be strongly coupled even if the original model is infrared free. 
However here the problem is a little more complicated. While in the $\phi^4$ model the thermal fluctuations essentially drove the zero mode away from strong coupling (and the whole theory was weakly coupled at sufficiently high temperatures\footnote{This improved perturbation theory where the thermal fluctuations are included is identical to the resummation of the so-called `daisy diagram' at each order of the perturbation theory as described in \cite{Dolan:1973qd}.}), here this will not be the case. There will be a left-over strongly interacting sector which we will have to treat carefully. A related point is the hierarchy between the scales~\eqref{electricmass} and \eqref{massgap}. More generally, there are three important scales in the problem, $O({\beta^{-1}_{\text{th}}}),O(g_{YM}{\beta^{-1}_{\text{th}}}),O(g_{YM}^2{\beta^{-1}_{\text{th}}})$ which are called `hard', `soft' and `ultrasoft', respectively. The hard scale corresponds to the energy scale of the non-zero Matsubara modes, while the soft and ultrasoft scales correspond to the energy scales of the Matsubara zero mode of the ordinary matter fields (including $A_0$) and the transverse gluon field, respectively. Symmetry breaking could take place from effects of order $g_{YM}{\beta^{-1}_{\text{th}}}$, namely from the soft scale, and then strong coupling dynamics at the ultra-soft scale would be negligible.

Now let us  provide a more concrete discussion that applies to large $N$ weakly coupled conformal gauge theories in 3+1 dimensions. These weakly coupled conformal field theories are made up of non-abelian gauge fields and some matter fields. Various coupling constants are tuned to a fixed point. Those couplings can be made arbitrarily small by adjusting  the matter content carefully. 

The study of these weakly-coupled fixed points has been an important source of insights into quantum field theory. 
The simplest model in this class consists of $SU(N_c)$ gauge fields minimally coupled to $N_f$ Dirac fermions in the fundamental representations. The presence of the non-trivial fixed point was suggested by Caswell \cite{Caswell:1974gg} and Banks and Zaks \cite{Banks:1981nn}. The two-loop beta function for the gauge coupling constant is as follows ($\al\equiv g^2/(4\pi)^2$)
\beq\label{betafun}
~&\beta(\al)_{\text{QCD}}^{2-loop}=b_0 \al^2+b_1\al^3+O(\al^4)
\\& b_0=-{11\ov 3}C_2(G)+{4\ov 3}T(R),~ b_1=-{34 \ov 3} C_2^2(G)+{20\ov 3}C_2(G) T(R)+4C_2(R)T(R)~.
\eeq
We used the quadratic Casimir $C_2(G)=N_c,~C_2(R)={(N_c^2-1)/ 2N_c}$ and Dynkin index $T(R)={N_f\ov 2}$. The positivity of $b_0$ when the number of flavors satisfies  $N_f\leq {11\ov 2}N_c$ indicates asymptotic freedom and the possibility of a UV completion by the free fixed point $g=0$. An important observation is that $b_1$ is positive as long as $34 N_c^3/(13N_c^2-3)<N_f<{11}N_c/2$ and hence one could naively expect a non-trivial unitary fixed point with coupling $\al=b_0/b_1$.
%
Such a conclusion is not necessarily correct since the truncation~\eqref{betafun} is not a priori justified. Banks and Zaks discovered that a systematic expansion is possible when one takes a limit of large $N_c$ and $N_f$ with appropriately chosen ratio $x_f=N_f/N_c$. Even though $N_f$ and $N_c$ are integers, in the limit of large $N_c$ and $N_f$, $x_f$ can be adjusted to achieve the limit $x_f={11\ov 2}-\epsilon$ with arbitrarily small $\epsilon$. This makes the coupling constant at the fixed point $\al=b_0/b_1\sim O(\epsilon/N_c)$ arbitrarily small. This should be thought of as the planar expansion with small `t Hooft coupling $\lam= N_c^2 g\ll 1$. (More precisely, this is the Veneziano limit \cite{Veneziano:1976wm} since we have fixed $x_f=N_f/N_c$.) 

The construction of similar weakly-coupled fixed points with scalar fields is richer due to the additional classically marginal interactions: scalar quartic couplings and Yukawa couplings. The simplest model is given by $SU(N)$ gauge theory with $N_f$ Dirac fermions $\psi$ and $N_s$ scalars $\phi$ in the fundamental representation.  There are two types of scalar quartic interactions which preserve the $U(N_s)$ global symmetry acting on the scalars: a single-trace interaction $hN \tilde Tr (\phi^\dagger \phi \phi^\dagger\phi)  $ and a double-trace interaction $ f Tr( \phi^\dagger \phi)Tr(\phi^\dagger \phi)$. (Here we think of the scalars as $N\times N_s$ matrices and $\phi^\dagger$ denotes the ordinary Hermitian conjugation.)

Let us make some general comments on the 't Hooft/Veneziano limit. If we have an action which is given by $S\sim NTr(\cdot)$, i.e. a single trace action  proportional to $N$, then the connected correlation function of $n$ single trace operators scales like $N^{2-n}$. The connected correlation function of  $m$ double trace operators and $n$ single trace operators scales like $N^{2-n}$ for any $m$. Therefore, if we like to add single trace deformations and double trace deformations to the action while preserving a smooth large $N$ limit we need to add the single trace operators with coefficients that scale like $N$ and the double trace operators with coefficients that scale like $O(1)$. This is why the couplings must scale like $hN \tilde Tr (\phi^\dagger \phi \phi^\dagger\phi)  $ and $ f Tr( \phi^\dagger \phi)Tr(\phi^\dagger \phi)$. 

Imagine we start from a large $N$ CFT and there are such single trace and double trace marginal deformations. 
In conformal field theory (or in conformal perturbation theory) one has to be more careful with counting the factors of $N$ since one-point functions vanish. As a result, correlation functions of two double trace  operators scale like $N^{0}$, correlation functions of one double trace operator and $2$ single trace operators scale like $N^{-2}$ (since the single trace operators are assumed to be marginal, the correlation function cannot factorize in any channel), correlation functions of two double trace operators and one single trace operator scale like $N^{-1}$, and finally, correlation functions of three double trace operators scale like $N^0$. 
 
Denoting collectively the single trace couplings by $h N$ and the double trace couplings by $f$ (such that $h,f$ are fixed in the large $N$ limit), the beta functions can now be extracted from the three-point functions of these operators as usual in conformal perturbation theory.  To leading order in $h,f$ the beta functions take the general form to leading order in $N$ \begin{equation}\label{genbeta}\beta(h) = Ah^2 ~,\qquad 
\beta(f) =Bf^2+Chf+Dh^2 ~. \end{equation}
The coefficients $A,B,C,D$ are $O(1)$ in the large $N$ limit and should be computed on a case-by-case basis. In short, the double trace operators do not backreact on the single trace couplings but the single trace couplings do affect the double trace couplings.

The structure~\eqref{genbeta} is very general. Let us now go back to the model with $N_s$ fundamental scalars and $N_f$ fundamental fermions, which has a smooth `t Hooft limit if we keep $h,f$ (as well as $g^2N_c$) fixed in the large $N_c$ limit. The existence of a nontrivial weakly coupled  fixed point depends now on $x_f=N_f/N_c$ (from which we can also infer $ x_s\equiv N_s /N_c$ since the total beta function at one loop has to be nearly vanishing).
Interestingly, one finds an upper bound $x_s< 0.84$ \cite{Benini:2019dfy}, which if violated, no controlled weakly coupled fixed point exists! In particular, the model with only scalars and non-Abelian gauge fields (i.e. $x_f=0$) does not have a controlled weakly coupled fixed point (we will soon discuss some possible consequences of that).

Let us now fix some $0\leq x_s<0.84$ and study the properties of the conformal gauge theory at finite temperature. 
After reducing on a circle one need not worry about the fermions since they all obtain a mass of order ${\beta^{-1}_{\text{th}}}$ as they have no zero modes on the circle. Below this scale we have a three-dimensional $SU(N_c)$ gauge theory with an adjoint scalar (the holonomy) and $N_s$ fundamental scalars with some quartic interactions. Both the adjoint scalar and the fundamental scalars obtain mass of order $g_{YM}{\beta^{-1}_{\text{th}}}$. Because  $0\leq x_s<0.84$ it does not even matter whether the thermal mass squared of the fundamental scalars is positive or negative. Either way, there is no spontaneous symmetry breaking (due to ``color-flavor locking''). Amusingly, for other gauge groups there are similar bounds on $x_s$ which prevent the existence of a symmetry breaking phase due to the condensation of scalar fields. 
See also~\cite{King:1987gz} for a lattice gauge theory point of view.

At the risk of deviating from the main theme of this paper, let us close this subsection with a brief discussion of the bound $0\leq x_s<0.84$ on the existence of weakly coupled Banks-Zaks fixed points. 
It is useful to consider first the case of $N_f=0$, i.e. the purely bosonic theory. Near $x_s=22$ the one-loop beta function vanishes but as we remarked above there is no weakly coupled fixed point. For $x_s>22$ the theory is infrared free, but that does not mean that it flows in the infrared to the free fixed point. Indeed, as in the Coleman-Weinberg mechanism \cite{Coleman:1973jx}, there could be a first-order transition instead. The absence of a weakly coupled fixed point for $x_s\leq22$ suggests that the same first order transition persists. The transition is between a trivial phase for $m_s^2>0$ and a phase with NGBs for $m_s^2<0$. Since the NGBs live on the group manifold ${U(N_s)\over U(N_s-N_c)\times SU(N_c)}$, for $x_s< 1$ there is no first-order transition anymore. In summary, in the model with $N_f=0$ it seems natural to conjecture no zero temperature phase transition for $x_s< 1$ and a zero-temperature first order phase transition for $x_s\geq 1$. This is in line with the general expectations for small $x_s$ laid out in~\cite{Fradkin:1978dv,Banks:1979fi} and see also~\cite{Cherman:2018jir} for some recent observations on the subject for larger values of $x_s$. For related observations about the nature of the phase diagram of the scalar model see~\cite{Bi:2019ers}.

We earnestly hope that the question of symmetry breaking in finite temperature conformal gauge theories will be clarified in the future.

\section{Vector Models}\label{sec:model}
We consider models with $N$ real scalar fields $\phi_i$, $i=1,...,N$ and potential \begin{equation}\label{potential}V= {1\over 4!}\lambda_{ijkl}^{\text{B}}\phi_i\phi_j\phi_k\phi_l~.\end{equation} in $4-\epsilon$ space-time dimensions where the superscript `B' denotes bare coupling. This class of models always admits a $\Z_2$ symmetry that flips the signs of all the fields $\phi\to-\phi$. 
These models are interacting systems for finite positive $\epsilon$. 
There are two limits in which we can carry out a perturbative study. One is when $\epsilon\ll 1$ and the other is when the number of fields $N$ is very large (in the latter case we should typically impose some additional symmetries). These two 
limits also have an overlapping regime where both $\epsilon$ is small and the rank is large. 
We will study both limits, allowing us to establish a rather coherent picture for the thermal properties of such models. 
We will start from the limit where $\epsilon\ll 1$ is the smallest parameter in the problem.

\subsection{Thermal Physics in the $\epsilon$ Expansion} \label{sec:ThermalEpsilon}

We are interested in fixed points in the $\epsilon$ expansion \cite{Wilson:1973jj}. Since in this subsection we take $\epsilon$ to be the smallest parameter in the problem we will content ourselves with a one-loop analysis of the fixed points:
The leading order beta function for the renormalized quartic coupling $\lambda_{ijkl}$ is
\begin{equation} \beta(\lam_{ijkl})=-\epsilon\lambda_{ijkl}+{1\over 16\pi^2} \left(\lambda_{ijmn}\lambda_{mnkl}+2\ {\rm permutations}\right)~.\end{equation}
It is convenient to rescale out the factors of $\epsilon$ and ${1\over 16\pi^2}$ by defining 
$\tilde \lambda = {\lambda \over 16\pi^2\epsilon}$ in terms of which the fixed point equations become 
\begin{equation}\label{FP}\tilde\lambda_{ijkl}=\tilde\lambda_{ijmn}\tilde\lambda_{mnkl}+2\ {\rm permutations}~.\end{equation}
These are rather complicated equations and the solutions are not classified. However, there are many known families of solutions and we will mention some of them below. The equations can be further simplified by imposing that the model~\eqref{potential} obeys a symmetry. 
An important observation is that as long as the fixed point equations~\eqref{FP} are satisfied the potential is bounded from below~\cite{Rychkov:2018vya}. This follows from the fixed point equation since $\tilde\lambda_{ijkl}\phi_i\phi_j\phi_k\phi_l \sim  Tr(\tilde\lambda_{ijmn}\phi_i\phi_j)^2$, where the square means the square of a matrix with the indices $mn$. The matrix $\tilde\lambda_{ijmn}\phi_i\phi_j$ could have zero eigenvalues, so there could be flat directions, as we will see. But the potential is certainly bounded from below by $V=0$. 
Many of the solutions to~\eqref{FP} correspond to fixed points which are theoretically and experimentally interesting. (An extrapolation is required to make contact
with = 1 which is the case we are ultimately interested in.)

We next turn to the study of the thermal properties of these fixed points. 
The thermal mass is of order $\epsilon$ and the corrections to the quartic potential due to thermal effects are of order $ \epsilon^2$. The zero temperature quartic potential is of order $\epsilon$ and hence we need not consider the thermal effects for the quartic interactions unless there are flat directions at zero temperature.

Therefore we focus our attention on the thermal mass. To compute it, we follow the same procedure of integrating out the non-zero Matsubara modes as in~\eqref{thermal}.  
We find that to leading order in $\epsilon$ the thermal mass  squared matrix is given by \begin{equation}\label{GenThma}{\cal M}^2_{ij} = {{\beta^{-2}_{\text{th}}}\over 24}\lambda_{ijkk}={2\over 3}\pi^2\epsilon{\beta^{-2}_{\text{th}}}\tilde\lambda_{ijkk}~. \end{equation}
We can use the fixed point equation~\eqref{FP} to write the thermal mass (up to a proportionality factor) as  
\begin{equation}\label{userel}{\cal M}^2_{ij} \sim \tilde\lambda_{ijmn}\tilde\lambda_{mnkk}+2\tilde\lambda_{ikmn}\tilde\lambda_{mnjk}~.\end{equation}
The last term is obviously positive definite. 
The first term is not necessarily positive definite. 
We should therefore embark on a search of CFTs which break some of their symmetries at finite temperature. This may not sound very promising. The Wilson-Fisher fixed points correspond (upon extrapolating to $\epsilon=1$) to critical points of various quantum magnets and it would be quite surprising to find that some of these magnets do not loose their magnetism upon heating them up. Nevertheless, we will indeed find fixed points which break their symmetries at arbitrary finite temperature.

We start with the first class of models, where the scalar potential~\eqref{potential} is invariant under some symmetry group $G\leq O(N)$, such that $G$ has only a single quadratic invariant. In other words, the only possible quadratic invariant is $\sum_i \phi_i \phi_i$, or, equivalently, the thermal mass must be proportional to $\delta_{ij}$. (This is equivalent to requiring that the $O(N)$ fundamental representation is irreducible under the symmetry group $G$ of the fixed point.) 

For such models, there must be a constant $z$ such that $\lambda_{ijkk}=z\delta_{ij}$ and hence from~\eqref{userel} we have 
$$z\delta_{ij} = z^2\delta_{ij}+2\tilde\lambda_{ikmn}\tilde\lambda_{jmnk}~.$$
Now, there must be some constant $C>0$ such that $\tilde\lambda_{ikmn}\tilde\lambda_{jmnk}=C \delta_{ij}$, as follows from the assumption of a single quadratic invariant. Its positivity follows from the positivity of $\tilde\lambda_{ikmn}\tilde\lambda_{jmnk}$. Therefore we have 
$$z\delta_{ij} = z^2\delta_{ij}+2C\delta_{ij}~.$$
This implies that $z>0$. Therefore, the thermal mass matrix is positive definite and there is no symmetry breaking at finite temperature.

The class of models with a single quadratic invariant covers several families: the $O(N)$ models, the cubic, tetrahedral, bi-fundamental, MN, tetragonal, the Michel fixed points etc. These classes include some of the most familiar quantum magnets upon extrapolating to three space-time dimensions. One can view these arguments as a retroactive  explanation for why some of the simplest critical points are disordered at finite temperature.

One interesting class of models not covered by the above analysis are the biconical models which have $O(m)\times O(N-m)$ symmetry. These models have two quadratic invariants.
We now turn to a detailed analysis of these fixed point. We have three quartic invariants, $(\phi_1^2)^2, (\phi_2^2)^2, \phi_1^2\phi_2^2$ where $\phi_1$ is a vector of length $m$ and $\phi_2$ is a vector of length $N-m$. We have therefore correspondingly 3 coefficients that need to be fixed to their fixed point values, $\alpha',\beta',\gamma'$
$$V={\alpha'\over 8}(\phi_1^2)^2+{\beta'\over 8}(\phi_2^2)^2+{\gamma'\over 4}\phi_1^2\phi_2^2~.$$

The $\phi_1$ indices are labelled with uppercase letters and the $\phi_2$ indices are labeled with lowercase letters. We have therefore 
$$\lambda_{ABCD} = {\alpha'}\left[\delta_{AB}\delta_{CD}+\delta_{AC}\delta_{BD}+\delta_{AD}\delta_{BC}\right]~,$$
$$\lambda_{abcd} = {\beta'}\left[\delta_{ab}\delta_{cd}+\delta_{ac}\delta_{bd}+\delta_{ad}\delta_{bc}\right]~,$$
$$\lambda_{ABcd} = {\gamma'}\delta_{AB}\delta_{cd}~.$$
and $\lambda_{AcBd}$, $\lambda_{AcdB}$ etc. are fixed by the total symmetry of the tensor.
%
%
%

We are now ready to write the one loop equations for $\alpha,\beta,\gamma$ (which differ from $\alpha',\beta',\gamma'$ by $16\pi^2\epsilon$, as above)
\begin{equation}\label{equationone}{\alpha}={\alpha^2}(m+8)+\gamma^2(N-m)~,\end{equation}
\begin{equation}\label{equationtwo}{\beta}={\beta^2}(N-m+8)+m\gamma^2~,\end{equation}
\begin{equation}\label{equationthree}\gamma={\alpha\gamma}(m+2)+{\beta\gamma}(N-m+2)+4\gamma^2~.\end{equation}
Since we are only interested in fixed points with $\gamma\neq 0$ (otherwise the theory reduces to two copies of a theory for which we proved a no-go theorem above) the last equation can be simplified to
\begin{equation}\label{equationthreeprime}1={\alpha}(m+2)+{\beta}(N-m+2)+4\gamma~.\end{equation} 
A quick consistency check of the above equations is that $\alpha=\beta=\gamma={1\over N+8}$ is the $O(N)$ fixed point. We will not be interested in this solution since the no-go theorem applies to it.

Now there are two quadratic invariants and the thermal mass matrix is proportional to 
$${\cal M}^2\sim\left(\begin{matrix}{\alpha}(m+2)\delta_{AB} +\gamma(N-m)\delta_{AB}& 0 \\ 0& {\beta}(N-m+2)\delta_{ab} + \gamma  m\delta_{ab} \end{matrix}\right)~.$$

Unfortunately we are not able to solve analytically the equations~\eqref{equationone},\eqref{equationtwo},\eqref{equationthreeprime}. But we will attack them instead in several steps which will be sufficient to demonstrate the main point. First we consider the simplified case of  equal rank, $2m=N$.
It follows by subtracting the equations~\eqref{equationone} and~\eqref{equationtwo} that $\alpha=\beta$ and\footnote{Let us prove that $\alpha=\beta$ is necessary. We subtract the beta functions for $\alpha$ and $\beta$ and we find (assuming that $\alpha$ and $\beta$ are different)
$$1=(\alpha+\beta)(m+8) $$
and hence $\alpha+\beta = {1\over m+8}$. Plugging this into the beta function for $\gamma$ we find $${3\over 2(m+8)}=\gamma$$ now we plug $\gamma$ into the equation for $\alpha$ and find
$$0=-\alpha+\alpha^2(m+8)+{9m\over 4(m+8)^2}$$
The discriminant is 
$$\Delta=1-{9m\over m+8} $$
and this is negative for all $m>1$. Hence the only allowed fixed points have $\alpha=\beta$.} 
$${\alpha}={\alpha^2}(m+8)+m\gamma^2~,$$
$$1=2{\alpha}(m+2)+4\gamma~.$$
%
There are two solutions. One solution is $\alpha=\gamma={1\over 2(m+4)}$ which has enhanced $O(N)$ symmetry and therefore we discard it.
The more interesting solution is
\begin{equation}\label{alphasol}\alpha={m\over 2(m^2+8)}~,\end{equation}
\begin{equation}\label{gammasol}\gamma={4-m\over 2(m^2+8)}~.\end{equation}
This solution exists for all positive $m$, and it always has $\alpha>0$. For $m>4$ we have a negative $\gamma$ but the potential is still increasing in all directions because $\gamma^2<\alpha^2$.
Finally, the thermal masses squared are  both proportional to $\alpha(m+2)+\gamma m$. It is easy to verify that the thermal masses are positive (for any positive m). In conclusion, the equal rank bi-conical critical model has no symmetry breaking at finite temperature. 

This bi-conical critical model can be contrasted with Weinberg's equal rank model that we have discussed in the previous section. We see that once we study the critical version of it slightly below 3 space dimensions, it no longer leads to symmetry breaking at finite temperature.

We now turn our attention to non-equal rank models. We cannot solve the equations analytically so instead we will resort to an approximate solution which will be however sufficient to establish the main conclusion. 
We will be staying in the regime where $\epsilon$ is the smallest parameter but we will now take large $N$. This will turn out to be a useful way to simplify the equations and attack the non-equal rank bi-conical models. In addition, this study will allow to make later on  comparisons with the large N results (those large N results are valid also at finite $\epsilon$).

To warm up, let us go back to the equal rank case and consider the large $N$ limit.
We consider the large $N$ expansion of the solutions~\eqref{alphasol} and \eqref{gammasol}. We find that (dropping terms of order $1/N^2$)
\begin{equation}\label{largerankfp}\alpha={1\over N}~,\quad \gamma={-1\over N}~. \end{equation}
In particular to this order in the $1/N$ expansion the zero temperature theory has a flat direction as the potential can be written as $V\sim (\phi_1^2- \phi_2^2)^2$ (hence there is a flat direction for $\phi_1^2=\phi_2^2$). At the origin of the flat direction there is a CFT and elsewhere the low-energy theory consists of a dilaton and Nambu-Goldstone bosons. 
We know that when the finite rank corrections are taken into account, the flat direction disappears and the origin is the only true minimum. We can also ask what happens to this flat direction in the large rank limit but at finite temperature. 
Recall the thermal masses, which in the leading large rank limit take the form \begin{equation}m^2_{thermal}\sim(\alpha+\gamma)N/2~.\end{equation}
We see that for the fixed point~\eqref{largerankfp}  the thermal mass cancels out to this order in the $1/N$ expansion. This strongly suggests that the flat direction remains at finite temperature, which is indeed true to this order in the expansion. 
In fact, in addition to this flat direction in field space, there is also a flat direction in coupling constant space (i.e. an exactly marginal operator) to this order in the $1/N$ expansion.

To see this, observe that the couplings $\alpha,\beta,\gamma$ all scale like $1/N$. To study systematically the large rank limit (keeping in mind that the smallest parameter is still $\epsilon$) we rescale the couplings accordingly. We find the set of fixed point equations for general rank (with $\tilde \alpha = N\alpha, \tilde \beta = N\beta, \tilde \gamma =N\gamma$) and to leading order in $1/N$: 

\begin{equation}\label{lrone}
{\tilde \alpha}=x{\tilde \alpha^2}+(1-x)\tilde \gamma^2~,\end{equation}
\begin{equation}\label{lrtwo}
{\tilde \beta}=(1-x){\tilde \beta^2}+x\tilde \gamma^2~,\end{equation}
\begin{equation}\label{lrthree}
1=x{\tilde \alpha}+(1-x){\tilde \beta}~,\end{equation}
where we have denoted $x=m/N$. 
The thermal mass matrix likewise simplifies in the large rank limit to
\begin{equation}\label{masslarge}{\cal M}^2\sim \left(\begin{matrix}x{\tilde\alpha}\delta_{AB} +(1-x) \tilde\gamma\delta_{AB}& 0 \\ 0& (1-x){\tilde\beta}\delta_{ab} + x\tilde\gamma  \delta_{ab} \end{matrix}\right)~.\end{equation}

\begin{figure}[]
\centering
\includegraphics[scale=0.35]{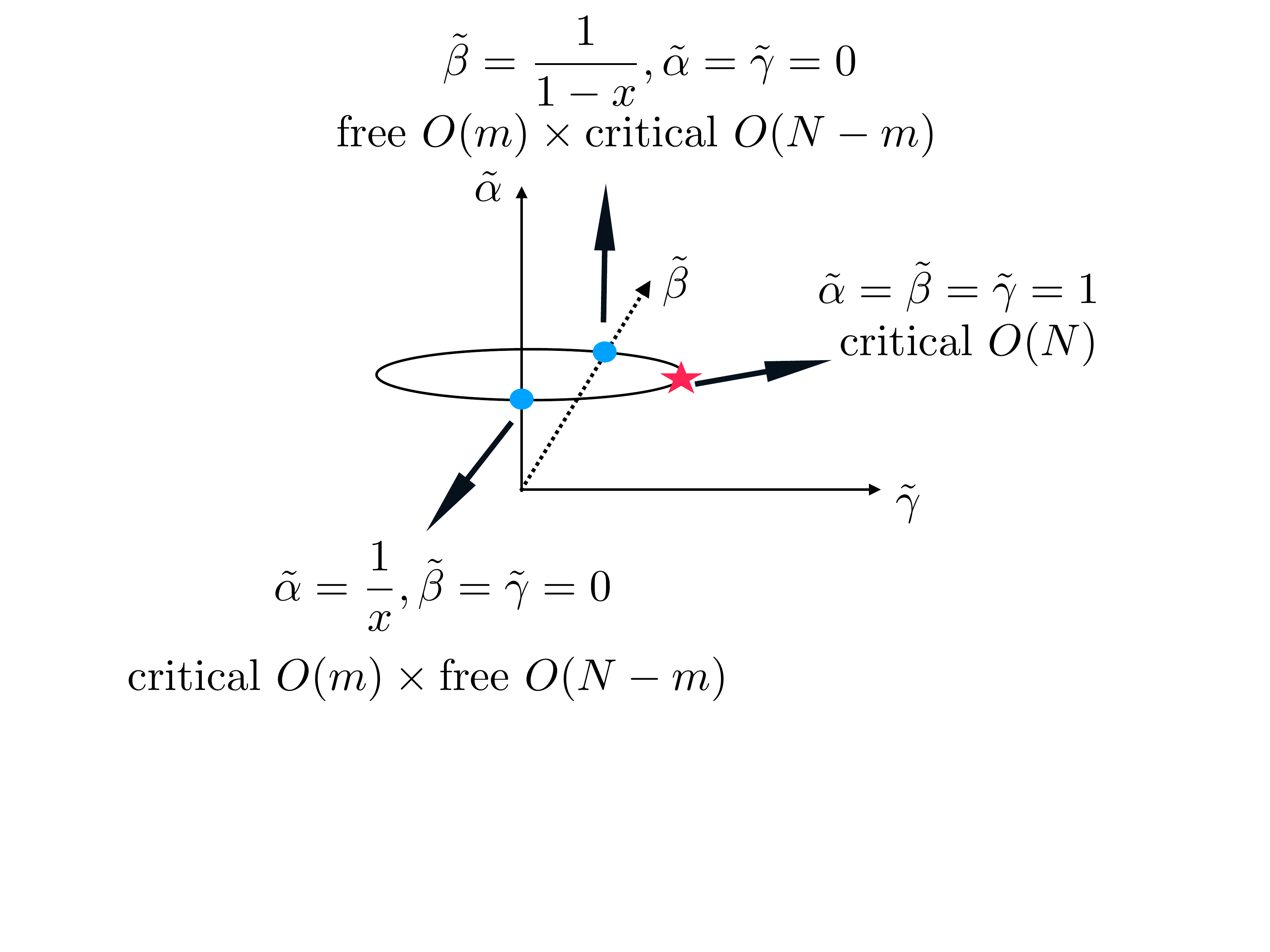}
\vspace{-6em}
\caption{A circle of fixed points in the large rank limit. The blue dots and red star surely survive the finite rank corrections, but there is another fixed point with $\gamma<0$ that likewise survives the finite rank corrections.}
\label{linefp}
\end{figure}

The three beta function equations~\eqref{lrone}, \eqref{lrtwo}, \eqref{lrthree} are in fact degenerate. There is therefore a co-dimension 2 set (a line) of fixed points in this large rank limit. Actually, there are two such sets of fixed points.
The two lines of fixed points are parameterized as follows:
\begin{equation}\label{largerankone} \tilde \alpha^{\pm} = {1\over 2x}\left(1\pm\sqrt{1-4x(1-x)\tilde\gamma^2}\right)~,\end{equation}
\begin{equation}\label{largerankonep} \tilde \beta^{\pm} = {1\over 2(1-x)}\left(1\mp\sqrt{1-4x(1-x)\tilde\gamma^2}\right)~.\end{equation}
where $\tilde\gamma$ belongs to the interval
\begin{equation}\label{tilderange}\tilde\gamma\in[-\frac1{2\sqrt{x(1-x)}}, \frac1{2\sqrt{x(1-x)}}]~.\end{equation}
The two branches of solutions \eqref{largerankone} and \eqref{largerankonep} are connected at the end points $\tilde\gamma=\pm \frac1{2\sqrt{x(1-x)}}$. So the two branches together form a closed co-dimension 2 curve (i.e. topologically a circle -- interestingly, a similar circle of fixed points  appeared in~\cite{Kiritsis:2008at}). Some particularly simple points on the circle are the $O(N)$ invariant point corresponding to $\tilde\gamma=\tilde\beta=\tilde\alpha=1$ (which is on the branch $\tilde\alpha^-,\tilde\beta^-$), $\tilde\alpha=\tilde\gamma=0$, $\tilde\beta={1\over 1-x}$ (which is on the branch $\tilde\alpha^-,\tilde\beta^-$ and corresponds to $m$ free bosons coupled to $N-m$ critical ones) and $\tilde\beta=\tilde\gamma=0$, $\tilde\alpha={1\over x}$ (which is on the branch $\tilde\alpha^+,\tilde\beta^+$ and corresponds to $N-m$ free bosons coupled to $m$ critical ones). These particular points certainly survive the finite rank corrections. One may expect that for generic points on this conformal manifold, which is topologically a circle, do not survive finite rank corrections.  See fig~\ref{linefp}.

For $x=1/2$ these two branches are one and the same (since we can interchange them by a change of variables) and the circle collapses to an interval.  $\tilde\alpha=\tilde\beta=\tilde \gamma=1$ corresponds to the $O(N)$ fixed point (where the thermal masses do not vanish also in the large rank limit) and $\tilde\alpha=\tilde\beta=-\tilde \gamma=1$ corresponds to the fixed point~\eqref{largerankfp}, where the thermal masses vanish in the large rank limit. The rest of the fixed points with $\tilde \gamma\in (-1,1)$ are large $N$ artifacts (save the one with $\tilde\gamma=0$ and either of $\tilde\alpha=0$ or $\tilde\beta=0$, which are related to each other by a change of variables and were discussed above).

It is easy to check that $\tilde\alpha^{\pm}\tilde\beta^{\pm}=\tilde\gamma^2$ for all $\tilde\gamma$. Therefore there is always a flat direction in field space at zero temperature, as long as $\tilde \gamma<0$. 
Thus, the large rank limit leads to a line of fixed points, and those with $\tilde\gamma<0$ have a flat direction in field space at zero temperature. The flat direction persists even at finite temperature! Indeed, the two thermal masses are proportional to $x\tilde\alpha+(1-x)\tilde\gamma$ and $(1-x)\tilde\beta+x\tilde\gamma$. The zero temperature flat direction is given by $\phi_1^2= \sqrt{\frac{\tilde\beta}{\tilde\alpha}} \phi_2^2$. The thermal mass term in the potential is proportional to  $(x\tilde\alpha+(1-x)\tilde\gamma)\phi_1^2+((1-x)\tilde\beta+x\tilde\gamma)\phi_2^2$. We find that it vanishes for as long as $\tilde\gamma<0$ when we plug in the flat direction:  $(x\tilde\alpha+(1-x)\tilde\gamma)\sqrt{\tilde\beta}+((1-x)\tilde\beta+x\tilde\gamma)\sqrt{\tilde\alpha} =(-x\tilde\gamma\sqrt{\tilde\alpha}+(1-x)\sqrt{\tilde\beta}\tilde\gamma)+(-(1-x)\tilde\gamma\sqrt{\tilde\beta}+x\sqrt{\tilde\alpha}\tilde\gamma)=0$.
Therefore, the moduli space of finite temperature vacua is the hyperbola
\begin{equation}\label{eq:hyp}\sqrt \alpha\phi_1^2-\sqrt\beta\phi_2^2+\frac{x\alpha+(1-x)\gamma}{12\sqrt\alpha}N{\beta^{-2}_{\text{th}}}=0~.\end{equation}

This hyperbola degenerates and touches the origin for $x\alpha+(1-x)\gamma=0$, which is one particular point on the circle in figure~\ref{linefp}. For general $x$, this may not be the physical fixed point that survives the finite rank corrections. For equal rank, $x=1/2$, it is precisely this fixed point, where the hyperbola degenerate, that survives finite rank corrections.

Suppose we knew that the theory that survives finite rank corrections has a non-degenerate hyperbola moduli space of vacua at finite temperature. That would be sufficient to imply symmetry breaking at finite temperature and finite rank. This follows from the fact that the origin is not on the hyperbola and hence, regardless of the form of the small corrections due to finite rank, the vacuum would be away from the origin. Aside from our interest in thermal physics, it is quite curious to see a model which has no supersymmetry but yet has, in the large rank approximation, a conformal manifold, a moduli space of vacua, allowing a spontaneous breaking of conformal symmetry, and even more mysteriously, a deformed moduli space of vacua upon including finite temperature corrections. This deformation of the moduli space may remind one of the deformed moduli space in some supersymmetric theories~\cite{Seiberg:1994bz}.

In the equal rank case we have found which fixed points survive the $1/N$ expansion: The interesting fixed point has $\tilde \gamma=-1$ and the thermal mass matrix vanishes in the large rank limit. The hyperbola degenerates and one cannot conclude whether the symmetry is broken at finite temperature without doing more work. (Upon computing subleading $1/N$ corrections the origin remains as the only true vacuum.)

It is interesting to understand which values of $\tilde\gamma$ correspond to fixed points that survive the expansion in $1/N$ for non-equal rank. In light of~\eqref{eq:hyp} this is a crucial question. If only the fixed point that survives the large rank expansion is not the one where the hyperbola degenerates, then the symmetry breaking would surely persist to the finite rank fixed point! In order to determine which of the fixed points survive to finite rank we can either attempt to solve the beta functions numerically, or we can include subleading corrections in the beta functions. Let us begin with the latter strategy and then we will check that it agrees with numerical solutions.

Continuing to use the rescaled couplings, the beta functions including the leading $1/N$ corrections are 
$${\tilde\alpha}={\tilde \alpha^2}(x+8/N)+\tilde\gamma^2(1-x)~,$$
$${\tilde\beta}={\tilde\beta^2}(1-x+8/N)+x\tilde\gamma^2~,$$
$$1={\tilde\alpha}(x+2/N)+{\tilde\beta}(1-x+2/N)+4\tilde\gamma/N~.$$
Let $t$ be a parameter in the range
\begin{equation}\label{tilderanget}t\in\left[-\frac1{2\sqrt{x(1-x)}}, \frac1{2\sqrt{x(1-x)}}\right]~.\end{equation} We found the leading order solution 
\begin{equation}\label{largerankonet} \tilde \alpha^{\pm}_0 = {1\over 2x}\left(1\pm\sqrt{1-4x(1-x)t^2}\right)~,\end{equation}
\begin{equation}\label{largerankoneti} \tilde \beta^{\pm}_0 = {1\over 2(1-x)}\left(1\mp\sqrt{1-4x(1-x)t^2}\right)~,\end{equation} 
\begin{equation}\label{tildegammat} \tilde \gamma_0 = t~.\end{equation}
Now we suppose a more general form for the solution, incorporating the subleading $1/N$ corrections
$$\tilde\alpha=\tilde\alpha_0+{1\over N}\delta \tilde \alpha~,\tilde\beta=\tilde\beta_0+{1\over N}\delta \tilde \beta~,\tilde\gamma=t+{1\over N}\delta \tilde \gamma~.$$
Plugging all of this back into the fixed point equations and we find
$$\delta\tilde \alpha=8\tilde\alpha_0^2+2x\tilde\alpha_0\delta\tilde \alpha+2(1-x)t\delta\tilde \gamma~,$$
$$\delta\tilde \beta=8\tilde\beta_0^2+2(1-x)\tilde\beta_0\delta\tilde \beta+2xt\delta \tilde \gamma~,$$
$$0=2\tilde\alpha_0+x\delta \tilde \alpha+2\tilde\beta_0+(1-x)\delta\tilde \beta+4t~.$$
We are trying to solve the system 
\begin{equation}\label{lineqs}\left(\begin{matrix}-1 +2x\tilde\alpha_0 & 0 & 2(1-x) t  \cr 
0 & -1 +2(1-x)\tilde\beta_0 & 2x t \cr
x & 1-x & 0\end{matrix}\right)\left(\begin{matrix}\delta\tilde\alpha\cr\delta \tilde \beta\cr\delta \tilde \gamma\end{matrix}\right)=
\left(\begin{matrix}-8\tilde\alpha_0^2\cr -8\tilde\beta_0^2\cr-2\tilde\alpha_0-2\tilde\beta_0-4t\end{matrix}\right)~.\end{equation} 
Since the matrix on the left-hand side is degenerate (this follows as the leading order solution has a zero mode) only discrete values of $t$ yield a solution. This is the mechanism by which the line of fixed points disappears at finite rank and only discrete values of $t$ yield fixed points that exist at finite rank. We must impose that the vector on the right-hand side lies in the co-dimension 1 image of the linear transformation. The image of the linear transformation is $V_{im}=Span\left\{ \left(\begin{matrix}x\tilde\alpha_0-(1-x)\tilde\beta_0\cr 0\cr x\end{matrix}\right),\left(\begin{matrix}0\cr  -x\tilde\alpha_0+(1-x)\tilde\beta_0\cr 1-x \end{matrix}\right) \right\}$. A vector that is orthogonal to this subspace is $$\left(\begin{matrix}-x\cr (1-x)\cr x\tilde\alpha_0-(1-x)\tilde\beta_0\end{matrix}\right)~.$$
We must require that the right hand side of~\eqref{lineqs} is orthogonal to this vector (which is the same as requiring that the right hand side lies in the two-dimensional subspace $V_{im}$). This leads to an algebraic equation for $t$ which determines which of the fixed points on our lines of fixed points survive to finite rank 
$$(1-2x) t^2 - 2t(x\tilde\alpha_0-(1-x)\tilde\beta_0)+3x\tilde\alpha_0^2-3(1-x)\tilde\beta_0^2=0~.$$ More explicitly 
\begin{equation}\label{quartic} 2(2x-1)t^2+{3\over 2} {1-2x\over x(1-x)} +\left({3\over 2x(1-x)}-2t\right)\sqrt{1-4x(1-x)t^2}=0~.\end{equation}

The equation~\eqref{quartic} only describes one of the two branches of  \eqref{largerankonet},\eqref{largerankoneti}. This is sufficient because $x\to 1-x$ interchanges the two branches. The radical equation \eqref{quartic} can be simplified as follows
\beq \label{cubic}
(t-1)(4x (1-x) t^3 -20  x(1-x) t^2+3 t+9)=0~.
\eeq

This equation has two real solutions $t=1,\tilde \gamma_*(x)$ and two complex solutions for $x\in (0,1/2)\cup (1/2,1)$. At $x=1/2$ the two complex solutions become degenerate real solution with $t=3$.  This additional real solution at $x=1/2$ is not physical since it makes $\al,\be$ complex after plugging back into our choice of branch in \eqref{largerankonet}. The solution $t=1$ is the $O(N)$ invariant fixed point.

As usual, there are extraneous solutions which need to be excluded when we transform the radical equation to the polynomial one. One can check from the discriminant analysis that two real solutions of \eqref{cubic} $t=1,\tilde \gamma_*(x)$ are genuine solution of \eqref{quartic} for $1/2\leq x<1$. We note that $\tilde \gamma_*(x)=\tilde \gamma_*(1-x)$in accord with the expectation following from $\mathbb Z_2$ symmetry among two branches.

When the dust settles, we obtain two physical fixed points for $0<x<1$ with the following leading large $N$ values of the couplings (excluding the theories with $\tilde\gamma=0$)
\beq \label{fps}
\text{FP}_+^{\text{bicon}}:~ &(\tilde \al,\tilde \be,\tilde \ga)=(1,1,1)~,
\\ \text{FP}_-^{\text{bicon}}:~ &(\tilde \al,\tilde \be,\tilde \ga)=\left({1+\text{sgn}(x-{1\ov 2})\sqrt{1-4x(1-x)\tilde \ga_*(x)^2}\over 2x}\right.
\\ &~~~~~~~\quad\quad\quad,\left.{1-\text{sgn}(x-{1\ov 2}) \sqrt{1-4x(1-x)\ga_*(x)^2}\over 2(1-x)},\tilde \ga_*(x)\right)~.
\eeq 

The first fixed point $\text{FP}_+^{\text{bicon}}$ is nothing but the $O(N)$ symmetric fixed point. The second fixed point $\text{FP}_-^{\text{bicon}}$ is more interesting since it turns out that one of the two thermal mass is negative for $x\neq 1/2$. This means that the moduli space of vacua in the large rank limit is a non-degenerate hyperbola~\eqref{eq:hyp}. 
A simple analytic way to show that the hyperbola does not degenerate on this point of the conformal manifold is to first observe that the cubic polynomial factor in the equation \eqref{cubic} has a positive (negative) value for t=-1 (t=-3) in the given range of $x$. This directly leads to $-3<\tilde \ga_*(x)<-1$ for $x\in (0,1/2)\cup(1/2,1)$ and hence the sum over the thermal masses
\beq\label{tmf}
~&m_1^2 \propto (1-x)\tilde \gamma_*+{1\ov 2}\left({1+\text{sgn}(x-{1\ov 2})\sqrt{1-4x(1-x)\tilde \ga_*(x)^2}}\right)~,
\\&m_2^2\propto x \tilde \gamma_*+{1\ov 2}\left({1-\text{sgn}(x-{1\ov 2}) \sqrt{1-4x(1-x) \tilde \ga_*(x)^2}}\right)~.
\eeq
becomes negative $m_1^2+m_2^2 \propto 1+\tilde \ga_*(x)<0$.
This means that the hyperbola does not degenerate and one necessarily has finite temperature symmetry breaking even at finite rank, as long as the ranks of the two symmetry groups are not equal. 

Upon taking finite rank corrections only one point on the hyperbola remains as the true vacuum. It is important to find which one it is since the symmetry breaking pattern is not the same everywhere on the hyperbola.

Without loss of generality, we consider the $1/2<x<1$ case where $m_1^2>0, m_2^2<0$. Extremization of the potential gives two possible candidates for the vacua\footnote{The origin $(\phi_1,\phi_2)=(0,0)$ cannot be a minimum since one of the thermal masses squared is negative.}
\beq \label{eq:can}
 (\phi_1^2,\phi_2^2)=\left({N(m_2^2 \tilde \gamma -m_1^2 \tilde\beta )\ov 8\pi^2 (\tilde \al \tilde \be-\tilde \gamma^2)\epsilon}, ~{N(m_1^2 \tilde \gamma -m_2^2 \tilde\alpha) \ov 8\pi^2 (\tilde \al \tilde \be-\tilde \gamma^2)\epsilon}\right) ~\text{or}~\left(0,-{N m_2^2\ov 8 \pi^2 \tilde \beta \epsilon }\right)~.
\eeq 
Using the leading-order values for the couplings leads to a singularity due to the flat direction (i.e. the hyperbola).  One must use the corrected couplings in order to find the true vacuum. So we must compute $(\delta \tilde \al,\delta \tilde \be,\delta\tilde \gamma)$. Rather than using  second-order perturbation theory to determine the $(\delta \tilde \al,\delta \tilde \be,\delta\tilde \gamma)$, there is a simple way to exclude the first solution of the equation \eqref{eq:can}. If we substitute the leading $\epsilon$ thermal masses into the numerator of $\phi_1^2$ (we can equally take $\phi_2^2$ as well), it becomes $-Nx(\tilde \al \tilde \be -\tilde \gamma^2)-2\tilde \be (\tilde \al-\tilde \gamma)+O(1/N)$. Since both $\tilde \al \tilde \be -\tilde \gamma^2$ and $\tilde \al-\tilde \gamma$ are positive quantities (the former is $O(1/N)$ because of the flat direction at $N=\infty$ and is positive because of the stability of the potential), we conclude that there is no solution with real $\phi_i$ in this case.

In summary, the second expression in \eqref{eq:can}, which represents the vertices of the hyperbola, survives as the true vacuum of the biconical model in the finite non-equal rank case. The vacua can be expressed in terms of $\tilde \gamma_*$ which solves \eqref{quartic} as
\beq
\text{VAC}^{\text{bicon}}:~ (\Phi_1^2, \Phi_2^2)=\begin{dcases}
\left({\tilde \gamma _*^2 \left(2 x-2 x^2\right)+\tilde \gamma _* \left(-2 x^2+5 x-3\right)-3 x \ov 12 (2 \tilde \gamma _*^2 (x-1) (2 x-1)+2\tilde  \gamma _* (x-1)+3)}{\beta^{-2}_{\text{th}}},0\right)& 0<x<1/2
\\ \left(0,{\tilde \gamma _*^2 \left(2 x-2 x^2\right)+\tilde \gamma _* \left(-2 x^2+5 x-3\right)-3 x \ov 12 \tilde \gamma_*(3-4x(1-x)\tilde \gamma_*)}{\beta^{-2}_{\text{th}}}\right ) & 1/2<x<1  \\ (0,0) & x=1/2~.
\end{dcases}
\eeq

We conclude that for the finite non-equal rank case, we found a critical point with symmetry breaking at arbitrary non-zero temperature and the following symmetry breaking pattern
 \beq \label{sympat}
 G_{\text{global}}: ~O(m_1)\times O(m_2)\xlongrightarrow[\beta_{th}^{-1}>0]{\text{FP}_-^{\text{bicon}}} \begin{dcases}O(m_1-1)\times O(m_2)&m_1<m_2 \\ O(m_1)\times O(m_2-1) &m_1>m_2\\ {\rm no\ breaking} &m_1=m_2\end{dcases} ~.
 \eeq
 We proved that this is all correct within the leading order $\epsilon$ expansion. More precisely, this was proven for large finite $m_1,m_2$. We will explore the case of $m_1=1$ later.

Our arguments here were somewhat formal, but since the equations are entirely algebraic~\eqref{equationone},\eqref{equationtwo},\eqref{equationthree} one can easily verify the claims numerically. 
We take $N=10^4$ and $x=0.6$, and to leading order in $\epsilon$ find the fixed point (we provide so many digits with the hope of convincing the reader that the fixed point indeed exists) $$(\tilde \al, \tilde \be, \tilde \gamma)=( 0.9176394600760599, 1.1235347774762552,  -1.0145547091210763)~. $$ Furthermore, this fixed point has the thermal masses $$(m_1^2,m_2^2)={2\ov 3}{ \pi^2 \epsilon {\beta^{-2}_{\text{th}}}}(0.1449453202892206, -0.15909420752664844896)~.$$ 
This can be plugged back into the second expression of \eqref{eq:can} and one finds the vacuum $(\phi_1^2,\phi_2^2)\sim (0,{0.0118\times 10^4})$. 

One subject we will not discuss in much detail here is the RG flow diagram between the various fixed points preserving $O(m_1)\times O(m_2)$ symmetry. Let us only say that our fixed point (at finite rank) $\text{FP}_-^{\text{bicon}}$ has 3 relevant operators-- two masses and one relevant quartic operator. Turning on the relevant quartic operator, one can flow to the decoupled critical bosons with $O(m_1)\times O(m_2)$ symmetry. Our fixed point is therefore multi-critical.

\subsection{Large-$N$ Analysis}
\label{largeN}

In this subsection we explore the large $N$ limit of the biconical model with $O(m)\times O(N-m)$ symmetry and fixed $m/N$ in $d$ spatial dimension. This limit corresponds to an opposite hierarchy with $1/N$ rather than $\epsilon=3-d$ being the smallest parameter. While small $\epsilon$ makes the model perturbatively tractable, the large $N$ techniques allow resummation of the perturbation series, and therefore some non-perturbative aspects of the model are elucidated in this limit. Therefore, this study allows to extend some of the results of the previous section to finite $\epsilon$. 

For large $N$ and fixed $m/N$ the symmetry breaking~\eqref{sympat} always leads to Nambu-Goldstone bosons and at finite temperature in 2+1 dimensions they are lifted by small non-perturbative infrared effects \cite{Coleman:1973ci}. Therefore, while many of the claims here about the large rank limit hold true also for finite small $\epsilon$, they certainly do not hold for $\epsilon=1$. In fact, we will see that some of the results may break down even before one reaches $\epsilon=1$. This requires a further analysis which we leave for the future. Our aim for now is only to show that the results about symmetry breaking at finite temperature hold for small finite $\epsilon$.

To begin, let us recall that vector models, in particular the bi-conical one, tend to be free in the large $N$ limit. Hence, the ground state approaches a Gaussian state as $N\to\infty$ \cite{Bardeen:1983st,Moshe:2003xn}, \ie up to a normalization constant it takes the following form in the space of fields
\beq
 \Psi(\phi_1, \phi_2) \propto \exp\(-{1\over 2} \sum_{i=1}^2 \int {d^{d}k\over (2\pi)^{d}} \, \omega_i(k)\, |\phi_i(k)|^2 \)~, \quad \omega_i(k)=\sqrt{k^2+m_i^2} ~.
\eeq
This functional has a well defined norm as long as $m_i^2$ are non-negative.  In position space it can be written as
\beq
 \Psi(\phi_1, \phi_2) \propto \exp\(-{1\over 4} \sum_{i=1}^2 \int d^{d}x \int d^{d}y  \big(\phi_i(x) -\sigma_i\big) D^{-1}_i(x-y) \big(\phi_i(y)-\sigma_i\big)  \)~,
 \label{trial}
\eeq
where $D^{-1}_i(x-y)$ is the Fourier transform of $2\omega_i(k)$, and two arbitrary constants $\sigma_i$ parametrize the location of the Gaussian state in the space of fields. While $m_i^2$ are singlets of the $O(m)\times O(N-m)$ group, $\sigma_1$ and $\sigma_2$ transform as vectors under $O(m)$ and $O(N-m)$ respectively. They  are associated with the order parameters in what follows. 

To determine the values of  $m_i^2$ and $\sigma_i$ for the biconical model at the fixed point, we resort to the variational principle\footnote{Note that the variational principle approach is identical to the approach using the Hubbard-Stratnovich transformation \cite{Coleman:1974jh} where the counterpart of the parameter $m_i^2$ in the former is the VEV of the corresponpding auxiliary field in the latter.}
\bea
 && \mathcal{W}= \langle \Psi| \mathcal{H} |\Psi \rangle \geq \langle 0| \mathcal{H} |0\rangle ~,
 \label{varW}
 \\
 && \mathcal{H}= {1\over2} \pi_i \pi_i +\frac{1}{2}\nabla \phi_i \nabla \phi_i +\frac{g^B_{ij}}{4N}  \,  \phi^2_i\,  \phi^2_j
\, , \quad  \quad 0\leq i,j\leq 2~.
\nonumber
\eea
where $\mathcal{W}$ is the variational functional, $|0\rangle$ is the vacuum state of the model governed by the Hamiltonian density\footnote{We use a symmetric matrix convention for the couplings $(g^B_{11}, g^B_{22}, g^B_{12})=8\pi^2\epsilon \, \mu^{3-d} (\tilde \alpha, \, \tilde \beta, \, \tilde \gamma)$, where $\mu$ is an arbitrary scale. \label{foot}} $\mathcal{H}$, and $|\Psi\rangle$ represents a family of normalized trial states \reef{trial}. As usual, the idea is to minimize the l.h.s. with respect to the variational parameters $m_i^2$ and $\sigma_i$ to find an approximation to the ground state energy. For an extremal state the inequality in \reef{varW} is saturated as $N\to\infty$. 

If the Hamiltonian is unbounded from below, then $ \mathcal{W}$ is necessarily unbounded from below too and vice versa. If the Hamiltonian is unbounded, there are states with arbitrarily negative energies, and by appropriate choice of $m_i^2\geq 0$ and $\sigma_i$ we can force $ \mathcal{W}$ to approach any negative value. 

Evaluating $ \mathcal{W}$ boils down to Gaussian integration. For instance, 
\beq
 \langle\Psi|\phi^2_j |\Psi\rangle=\int \prod_{i=1}^2 \mathcal{D}\phi_i \, \phi^2_j  |\Psi(\phi_1, \phi_2)|^2=\sigma_j^2 + N x_j D_j~,
\eeq
 where for brevity we introduced $x_1=x$ and $x_2=1-x$, whereas  $D_j$ represents an ordinary loop integral\footnote{The positive nature of $ \langle\Psi|\phi^2_j |\Psi\rangle$ is not guaranteed in dimensional regularization, but physical results are regularization independent.}
 \beq
  D_j =  \int {d^{d}k\over (2\pi)^{d}}  {1\over 2 \omega_j(k)} = 
   { \Gamma\({1-d\over 2}\) \over (4\pi)^{d+1\over 2}}(m^2_j)^{d-1\over2} ~.
 \eeq
Similarly
\beq
 \langle\Psi|\pi^2_j |\Psi\rangle=\int \prod_{i=1}^2 \mathcal{D}\phi_i \, 
 \Psi^*(\phi_1, \phi_2)\({\delta\over i \delta\phi_j}\)^2  \Psi(\phi_1, \phi_2)={N x_j\over 4}D_j^{-1}(0)~.
\eeq
It is convenient to introduce a separate notation for the kinetic energy density of each field
\beq
 K_j={1\over2 N}\langle\Psi | \big( \pi_j^2+ (\nabla \phi_j)^2\big) |\Psi\rangle=
{x_j\over 4} \int {d^{d}k\over (2\pi)^{d}} \( \omega_j(k)+ {k^2\over \omega_j(k)} \) ~.
\label{KineticE}
\eeq
Up to a mass independent constant, we have
\bea
 K_j&=& - {x_j \over 2 } \int_0^{m_j^2} dm^2\, m^2 {\d D_j \over  \d m^2}
 = { \Gamma\({3-d\over 2}\) \over (d+1)(4\pi)^{d+1\over 2}}  x_j  (m_j^2)^{d+1\over2} ~.
 \label{KineticE} 
 \eea
As a result, $ \mathcal{W}$ takes the form
\bea
 { \mathcal{W}\over N}=  \sum_i K_i
 +\sum_{i,j}\frac{g^B_{ij}}{4}  \,  \big(\sigma_i^2 + x_i D_i\big)\,  \big(\sigma_j^2 + x_j D_j\big) ~.
 \label{WzeroT}
\eea
where we rescaled $\sigma_i$'s and employed the large $N$ approximation $\langle (\phi^2_i)^2\rangle=\langle \phi^2_i\rangle^2$ to account for the contribution of the quartic potential. 

We are now in a position to be able to study the phase structure of the model starting from zero temperature. The symmetries at stake are scale invariance and the global symmetries. 

Notice that $\mathcal{W}$ is given by a sum of non-negative kinetic and potential terms, because $g^B_{ij}$ is positive semi-definite, whereas $K_i\geq 0$. Hence, the Hamiltonian of the model is bounded from below. In the large $N$ limit, the renormalized couplings $g_{ij}$ lie on a curve defined by 
\beq
  \det(g_{ij})=0~, x_1g_{11}+x_2g_{22}=8\pi^2\epsilon ~.
  \label{lineFP}
\eeq
For each set of these couplings the minimum of $\mathcal{W}$, which is obtained at $\mathcal{W}=0$, lies along  a flat direction in field space. The flat direction is given by
\beq
 m_i^2=0 ~, \quad \left(\begin{array}{c} \sigma_1^2 \\ \sigma_2^2\end{array}\right)
  = \left(\begin{array}{c} \pm\sqrt{g_{22}/ g_{11}} \\ 1\end{array}\right) \mu^{2-\epsilon} ~, \quad \text{for} \quad \text{sign}(g_{12})=\mp 1~.
  \label{broken_sol0}
\eeq
where $\mu$ is an arbitrary energy scale, and $(\sigma_1^2,\sigma_2^2)$ is aligned along the eigenvector of $g_{ij}$ with zero eigenvalue. Each field configuration along the flat direction can serve as a ground state of the theory.

Since $\sigma_i^2\geq 0$, we conclude that for $g_{12}\geq 0$ there is a unique vacuum at $\mu=0$ which respects the symmetries, whereas for $g_ {12}<0$ there is a flat direction in field space for ground states passing through the origin. 

At the origin, scale invariance, the $O(m)$, and $O(N-m)$ symmetries are all retained. At any ground state along the flat direction away from the origin in field space  $\mu$ does not vanish and thus scale invariance is spontaneously broken. This breaking leads in turn, by \reef{broken_sol0},  to the spontaneous symmetry breaking of the $O(m)$ and/or $O(N-m)$ symmetries. 
Hence, away from the origin, there are massless Nambu-Goldstone bosons and a dilaton. These massless  particles will be identified in subsection \ref{particles}.

We therefore see that for arbitrary number of space dimensions, in the strict large rank limit, there is a conformal manifold and moduli spaces of vacua for  $g_ {12}<0$. This is exactly as in the $\epsilon$ expansion but now this is valid for arbitrary $\epsilon$.
We will next see that the finite temperature corrections at leading order in the large rank expansion lead to a hyperbola, again extending a result from the $\epsilon$ expansion to arbitrary $\epsilon$.

\subsubsection{Finite ${\beta_{\text{th}}}$}

The variational functional $\mathcal{W}$ at finite ${\beta_{\text{th}}}$ is obtained by introducing a trial thermal state
\bea
 \mathcal{W}&=&\mathcal{F}_0+ \text{Tr}\big[\rho_0 (\mathcal{H} - \mathcal{H}_0) \big] \geq \mathcal{F} ~,
 \non
 \mathcal{H}_0&=& {1\over2}\sum_i\( \pi_i^2 + (\nabla \phi_i)^2  +m_i^2(\phi_i-\sigma_i)^2\) ~,
\eea
where $\mathcal{F}$ is the free energy density of the model, whereas $\mathcal{F}_0$ and $\rho_0$ denote the free energy and thermal density matrix associated with $\mathcal{H}_0$. In the limit $\beta_{th}\to \infty$ we recover the previous ansatz \reef{varW}. Furthermore, \reef{KineticE} generalizes to
\beq
 K_j={\mathcal{F}_{0j}\over N} - {1\over 2N} m_j^2 \text{Tr}\big[\rho_0\, (\phi_j-\sigma_j)^2\big] 
 = - {1\over 2N} \int_0^{m_j^2} dm^2\, m^2 {\d \over  \d m^2} \text{Tr}\big[\rho_0\, \phi_j^2\big] ~,
\eeq
where $\mathcal{F}_{0j}$ is the free energy density of the free field of mass $m_j^2$, and the second equality holds up to irrelevant constant. Substituting the thermal expectation value\footnote{We rescaled $\sigma_j\to \sqrt{N} \sigma_j$.}
\bea
&&{1\over N}\langle \phi_j^2\rangle_{\beta_{\text{th}}}={1\over N} \text{Tr}\big[\rho_0\, \phi_j^2\big] 
 \label{phi2vev}
\\
&&= \sigma_j^2 + { \Gamma\({1-d\over 2}\) \over (4\pi)^{d+1\over 2}}x_j(m_j^2)^{d-1\over2} + {2 x_j\over (4 \pi)^{d\over 2} \Gamma\({d\over 2}\)}
 \int_{|m_j|}^\infty d\omega{(\omega^2-m^2)^{d-2\over 2}\over e^{{\beta_{\text{th}}} \omega}-1}~.
 \nonumber
\eea
yields
\beq
 K_j= { \Gamma\({3-d\over 2}\) \over (d+1)(4\pi)^{d+1\over 2}}  x_j (m_j^2)^{d+1\over2}
 + {x_j\over (4 \pi)^{d\over 2} \Gamma\({d-2\over 2}\)}
 \int_0^{m_j^2} dm^2\, m^2 \int_{|m|}^\infty d\omega{(\omega^2-m^2)^{d-4\over 2}\over e^{{\beta_{\text{th}}} \omega}-1}~.
 \label{kinetic}
 \eeq
As usual, the first term is associated with zero temperature contribution, whereas the second term represents thermal fluctuations. The integral over $\omega$ cannot be evaluated in full generality, but it simplifies if the mass vanishes, \eg
\beq
 {1\over N}\langle \phi_j^2\rangle_{\beta_{\text{th}}} \Big|_{m_j^2=0} = \sigma_j^2 + {\Gamma\({d-1\over 2}\)\over 2\, \pi^{d+1 \over 2}} \zeta(d-1) ~  x_j \,{\beta^{1-d}_{\text{th}}} ~,
 \label{zeroM}
\eeq
where $\zeta(s)$ is the Riemann zeta function.

In the large $N$ limit the variational functional at finite ${\beta_{\text{th}}}$ takes the form
\beq
 \mathcal{W}=  N\sum_i K_i
 +\sum_{i,j}\frac{g^B_{ij} }{4N}  \,  \langle \phi^2_i\rangle_{\beta_{\text{th}}} \,  \langle \phi^2_j\rangle_{\beta_{\text{th}}} ~.
 \label{free}
\eeq
Note that all vevs are evaluated in the Gaussian thermal state, whereas the trial parameters, $m_i^2$, which minimize $\mathcal{W}$ represent thermal masses of the excitations. Furthermore, for large values of $m_i^2$ (or $\sigma_i$) and any given inverse temperature ${\beta_{\text{th}}}$, the variational functional approaches \reef{WzeroT} evaluated at zero temperature. This follows immediately from \reef{free} and eqs. \reef{kinetic}, \reef{phi2vev}. In particular, $ \mathcal{W}$ is bounded from below. Moreover, as shown earlier in this section, $\det(g_{ij})$ vanishes in the large $N$ limit, therefore there is always a flat direction in the space of $\sigma_i$'s determined by the eigenvector of $g_{ij}$ with zero eigenvalue. This is exactly as was found in the $\epsilon$ expansion.

\subsubsection*{\bf Phases at finite ${\beta_{\text{th}}}$}
 
For $g_{12}\geq 0$ there is a unique vacuum which respects the symmetries, and therefore we proceed to the cases with $g_ {12}<0$ where the symmetry can be broken. By construction $\mathcal{W}$ is non-negative, because $g_{ij}$ is positive semi-definite, whereas the kinetic free energy satisfies $K_i\geq 0$. Furthermore,  $\mathcal{W}=0$ at any point on the ridge
\beq
 m_1^2=m_2^2=0 ~, \quad 
 \left(\begin{array}{c} \sigma_1^2 \\ \sigma_2^2\end{array}\right)=
 \left(\begin{array}{c}  \sqrt{g_{22}/ g_{11}} \\ 1\end{array}\right) \mu^{2-\epsilon} 
 -  {{c(\epsilon)\beta^{\epsilon-2}_{\text{th}}} \over 12} 
 \left(\begin{array}{c} x_1 \\ x_2 \end{array}\right) 
 ~,
 \label{broken_sol}
\eeq
where we used \reef{zeroM} to align the order parameters $(\sigma_1^2,\sigma_2^2)$ such that $(\langle \phi^2_1\rangle_{\beta_{\text{th}}},\langle \phi^2_2\rangle_{\beta_{\text{th}}})$ is parallel to the eigenvector of $g_{ij}$ with zero eigenvalue, whereas
$\mu$ is an arbitrary scale emphasizing flatness of $\mathcal{W}$ even at finite ${\beta_{\text{th}}}$. It should be sufficiently big to ensure positive $\sigma_i^2$. The function $c(\epsilon)$ in the above expression is defined below:
\beq
c(\epsilon)\equiv \frac{6 \Gamma(\frac{2-\epsilon}{2})\zeta(2-\epsilon)}{\pi^{\frac{4-\epsilon}{2}}}.
\label{c(epsilon) def}
\eeq
Note that this function diverges in the $\epsilon\rightarrow 1$ limit, and hence restricts the validity of this analysis to $\epsilon<1$. Such divergences of thermal expectation values of the fields are consistent with the impossibility of a symmetry-broken phase in (2+1) dimensions at nonzero temperatures.

Since $\mathcal{W}\geq 0$ for all admissible masses and order parameters, we conclude that each point on the ridge \reef{broken_sol} corresponds to the global minimum of the free energy, and therefore it represents a thermodynamically stable phase in the large $N$ limit. In general, the line \reef{broken_sol} does not pass through the origin, and therefore $ O(m) \times O(N-m) $ is broken at finite ${\beta_{\text{th}}}$. The introduction of the temperature ${\beta_{\text{th}}}$ explicitly breaks scale invariance but a moduli space of vacua continues to exist.

We elaborate now on those cases where the line \reef{broken_sol} does reach the origin. For a given $x_1, x_2$ and ${\beta_{\text{th}}}$ this can occur due to \reef{broken_sol} only for that point of the curve \reef{lineFP} which satisfies in addition $\sqrt{g_{22}/ g_{11}} =x_1/x_2$. In this case the phase structure analysis follows precisely the one at ${\beta_{\text{th}}}=\infty$, the introduction of a temperature does not result in creating a horizon which prevents the field from reaching the origin. A presence of a quantum correction to the moduli space that results in forming a ``horizon" is known from some supersymmetric theories \cite{Seiberg:1994bz}.

Note that in the small $\epsilon$ regime, the admissible vacua \reef{broken_sol} are lying on the hyperbola \reef{eq:hyp}. Hence, the phase structure in the large $N$ limit (and arbitrary $\epsilon$) matches our results obtained within the $\epsilon$ expansion. 

Now comes the more difficult question regarding which of these fixed points survives at finite rank. In the $\epsilon$ expansion we provided an explicit answer which shows that indeed symmetry breaking takes place at finite, large rank. But now that $\epsilon$ is arbitrary, to find out the answer, one needs to do some sub-leading $1/N$ computations and examine how the conformal manifold and the hyperbola of vacua are lifted. We hope that this will be addressed in the future. 

In summary, we have shown that the conformal manifold and moduli spaces of vacua exist at arbitrary $d$ and $N=\infty$. The  $1/N$ corrections needed to find out the true finite temperature vacua at finite, large, rank were only found for $3-\epsilon$ dimensions with small $\epsilon$. Therefore, we can only conclude that symmetry breaking at finite temperature in the bi-conical models takes place in $3-\epsilon$ dimensions for finite small $\epsilon$. It would be interesting to complete this analysis at finite $\epsilon$.

As an aside, one might wonder if the model exhibits metastable phases. While such states necessarily decay into one of the admissible stable states, the decay rate is exponentially suppressed in the large $N$ limit, and therefore a metastable phase is a long-lived steady state as $N$ goes to infinity. To explore such a possibility we should extremize rather than minimize $\mathcal{W}$. Varying it with respect to $m_i^2$ leads to the gap equation
\beq
 {\d  \mathcal{W}\over \d m_i^2}=0 \quad \Leftrightarrow \quad m_i^2=  \sum_j g^B_{ij} 
 \langle \phi^2_j\rangle_{\be_{\text{th}}}/ N ~.
  \label{gap}
\eeq
This is simply a statement that the full two-point function in the large $N$ limit is given by the sum of all possible cactus diagrams with two external legs. 

The free energy density in the large $N$ limit is given by $ \mathcal{W}$ evaluated on the non-negative solution $m_i^2(\sigma_j)$ to the gap equations \reef{gap}. 
In particular, the order parameters $\sigma_j$ are derived by minimizing $ \mathcal{W}\big(\sigma_j, m_i^2(\sigma_j) \big)$ with respect to $\sigma_j$. For any non-zero extremum $\sigma_j\neq 0$, we always have $m_j^2=0$. Indeed
\beq
 0={\d  \mathcal{W}\over \d\sigma_i} + {\d  \mathcal{W}\over \d m_j^2} {\d m_j^2\over \d \sigma_i}
 = \sigma_i  \sum_k g^B_{ik} \langle \phi^2_k\rangle_{\be_{\text{th}}}/ N 
-{1\over 2}\sum_{j,k} {\d m_j^2\over \d \sigma_i} {\d \langle \phi_j^2\rangle_{\be_{\text{th}}} \over  \d m_j^2} \[m_j^2 
 - g^B_{jk}\big\langle \phi^2_k\rangle_{\be_{\text{th}}}/ N\] ~.
\eeq
Or equivalently, using the gap equations \reef{gap},  
\beq
 \sigma_i m_i^2=0~,  \quad \forall i~.
 \label{gap2}
\eeq
Hence, a non-zero $\sigma_i$ is necessarily linked to $m_i^2=0$ even at finite ${\beta_{\text{th}}}$. In particular, all extrema of $\mathcal{W}$ when one of the $\sigma_i$'s or both are non zero take the form \reef{broken_sol}, and we considered these cases already. They correspond to the global minima of $\mathcal{W}$ in the large $N$ limit. 
The only thing remaining is to search for the possibility of a metastable phase which respects $O(m) \times O(N-m)$.

If $ O(m) \times O(N-m) $ is unbroken, then $\sigma_1=\sigma_2=0$ and \reef{gap2} is trivially satisfied. In this case the gap equations \reef{gap} have no solution where both masses are strictly positive. Indeed, in the large $N$ limit $g_{ij}$ is a 2$\times$2 degenerate matrix. Hence, up to an overall multiplicative factor it projects $(x_1\,  \langle \phi^2_1\rangle_{\beta_{\text{th}}}, x_2 \,   \langle \phi^2_2\rangle_{\beta_{\text{th}}})$ onto the eigenvector $(-\sqrt{g_{11}/ g_{22}}, 1)$ with a non-zero eigenvalue.  As a result of an opposite sign in the entries of this eigenvactor, one of the masses is necessarily negative. The latter excludes the existence of a phase in which $ O(m) \times O(N-m) $ is unbroken and scale invariance is broken.

Finally, let us consider a symmetric phase where both $ O(m) \times O(N-m) $ and scale invariance are unbroken. Substituting $\sigma_i^2=m_i^2=0$ into \reef{gap} and \reef{gap2}, we conclude that the gap equations are satisfied provided that
 \beq
 0=  \( {{\beta^{\epsilon-2}_{\text{th}}}\over 12}  \)  \sum_j g_{ij} x_j ~.
\eeq
This constraint trivially holds at ${\beta_{\text{th}}}=\infty$, and therefore a symmetric phase minimizes $\mathcal{W}$ at zero temperature. At finite temperature, however, it is lifted relative to the solutions \reef{broken_sol},  \ie $\mathcal{W}$ is strictly positive for a symmetric configuration and the gap equations are not satisfied unless $(x_1,x_2)$ is aligned along the eigenvector of $g_{ij}$ with zero eigenvalue. Hence, we conclude that the symmetry is necessarily broken at ${\beta_{\text{th}}}\neq \infty$ if  $\sqrt{g_{22}/ g_{11}} \neq x_1/x_2$.

Since spontaneous symmetry breaking affects the spectrum of particles, it is interesting to match  different excitations of the model with the symmetry breaking patterns found above. This is the main goal of the next subsection where we analyze excitations around the large $N$ vacua of the model. We find a precise match between the symmetry breaking pattern and the particle content. In addition, we derive a composite excitation with scaling dimension 2 which is inherent to the critical vector model.

\subsection{Excitations of the Biconical Model in the large $N$ limit}
\label{particles}

The Euclidean action of the model can be written as 
\beq
 I_\mt{E} ={1\over 2} \sum_i \int \Big( \d\phi_i \d\phi_i + s_i \(\phi_i^2 - N \rho_i\) \Big)
  +{N\over 4}\sum_{ij} g^B_{ij}  \int   \rho_i\,  \rho_j 
 ~,
\eeq
where the auxiliary fields $s_i, \rho_i$ are singlets of $ O(m) \times O(N-m) $. Integrating them out leads to the standard Lagrangian of the biconical model. The integral over $s_i$ yields a delta-functional $\delta\(\phi_i^2 - N  \rho_i\)$ which simplifies the integration over $\rho_i$. The final result for the Lagrangian is ${\cal L} ={1\over 2}\d\phi_i \d\phi_i  + {g^B_{ij}\over 4N} \phi_i^2\phi_j^2 $. 

In fact, the quadratic form of $I_\mt{E}$ suggests that $\rho_i$ and $\phi_i$ can be integrated out analytically leaving us with fluctuating $s_i$ only. However, integration over $\rho_i$ should be done with caution, because $g^B_{ij}$ is degenerate in the large $N$ limit, whereas the integral over $\phi_i$ should account for the possibility of broken $ O(m) \times O(N-m)$. Hence, to identify the effective degrees of freedom of the theory, we proceed in two steps.

First we change variables $\rho_i \to M _{ij} \rho_j$ and similarly for $s_i$, where the 2$\times$2 orthogonal matrix $M_{ij}$ diagonalizes the renormalized $g_{ij}$
\beq
 M^T g M = \left(\begin{array}{cc}\text{Tr} (g) & 0 \\0 & 0\end{array}\right)~, \quad 
 M={1\over \sqrt{g_{11}+g_{22}}}
 \left(\begin{array}{cc} -\sqrt{g_{11}} & \sqrt{g_{22}} \\ \sqrt{g_{22}}& \sqrt{g_{11}}\end{array}\right) ~.
\eeq
The integral over $\rho_1$ is Gaussian, whereas integral over $\rho_2$ simply gives the $\delta$-functional $\delta(s_2)$. Hence, we get
\beq
 I_\mt{E} ={1\over 2} \sum_i \int \Big( \d\phi_i \d\phi_i +  \phi_i^2 M_{i1}s_1  \Big)
  -{N\over 2 \, \text{Tr} (g^B)}  \int   s_1^2
 ~.
  \label{eff_action0}
\eeq

Next we account for the possibility that $\phi_i$'s may develop a non-trivial expectation value. For simplicity we align  $\langle \phi_i\rangle$'s along the first components of the vector fields which are henceforth denoted by $\sigma_i$. Integrating over all other components yields\footnote{We rescale $\sigma_i\to \sqrt{N} \sigma_i$.}  
\beq
  I_\mt{E}={N\over 2} \sum_i \int  \( \d\sigma_i \d\sigma_i + \sigma_i^2 M_{i1}s_1 \)-{N\over 2 \, \text{Tr} (g^B)}  \int   s_1^2+\sum_i {x_iN-1\over 2} \text{Tr}\log(-\d^2 + M_{i1}s_1)  ~.
  \label{eff_action}
\eeq
Since $I_\mt{E}\propto N$ it follows that the large $N$ vacuum state of the model is determined by a constant solution $\bar \sigma_i, \bar s_1$ to the classical equations of motion obtained by varying $I_\mt{E}$  with respect to  $s_1$ and $\sigma_i$. In fact, after identifying $m_i^2=M_{i1} \bar s_1$ these equations become identical with \reef{gap} and \reef{gap2}. In particular, the thermal masses, $m_i^2$, of $(x_iN-1)$ fields $\phi_i$ vanish, whereas $\bar \sigma_i$ 's lie on the hyperbolic curve \reef{broken_sol}. Expanding \reef{eff_action} around $\bar \sigma_i, \bar s_1$ and keeping quadratic terms only, yields\footnote{There are no linear terms present in the action, because $\bar \sigma_i, \bar s_1$ extremize $I_\mt{E}$. The interaction terms between the fluctuating fields introduce 1/N corrections to the propagators that we discuss in what follows, and therefore we suppressed them in \reef{eff_action2}.}
\beq
  I_\mt{E}={N\over 2} \sum_i \int  \big( \d\sigma_i \d\sigma_i + (2  \bar\sigma_i M_{i1}) \sigma_i s_1 \big)-{N\over 2 \, \text{Tr} (g^B)}  \int   s_1^2 - {N\over 4}  \sum_i x_i M_{i1}^2 \text{Tr} \Big( (\d^2)^{-1} s_1 (\d^2)^{-1} s_1 \Big)~.
  \label{eff_action2}
\eeq
Finally, we perform an orthogonal transformation to disentangle the fields $\sigma_i$
\beq
 \sigma'_i= {R_{ij} \sigma_j\over \sqrt{ \det(R)} } ~, \quad 
 R= 2 \left(\begin{array}{cc} \bar\sigma_1 M_{11} &  \bar\sigma_2 M_{21} \\  - \bar\sigma_2 M_{21}&  \bar\sigma_1 M_{11}\end{array}\right) ~.
\eeq
The quadratic action at $T=0$ eventually takes the form
\bea
  {I_\mt{E}\over N}&=&{1\over 2} \int  \( \d\sigma'_1 \d\sigma'_1 +  \d\sigma'_2 \d\sigma'_2  
  +  \sqrt{ \det(R)} \sigma'_1s_1 \)
  \non
  &-&{N\over 2 \, \text{Tr} (g^B)}  \int   s_1^2
   - {\Gamma^2\({d-1\over 2}\)\over 64 \pi^{d+1}}  \sum_i x_i M_{i1}^2 \int\int {s_1(y_1) s_1(y_2)\over |y_1-y_2|^{2(d-1)}  }~.
\eea
At finite ${\be_{\text{th}}}$, we get essentially the same action, except that the last term needs to be modified on a thermal cylinder. 
\begin{figure}
\centering
\includegraphics[scale=0.53]{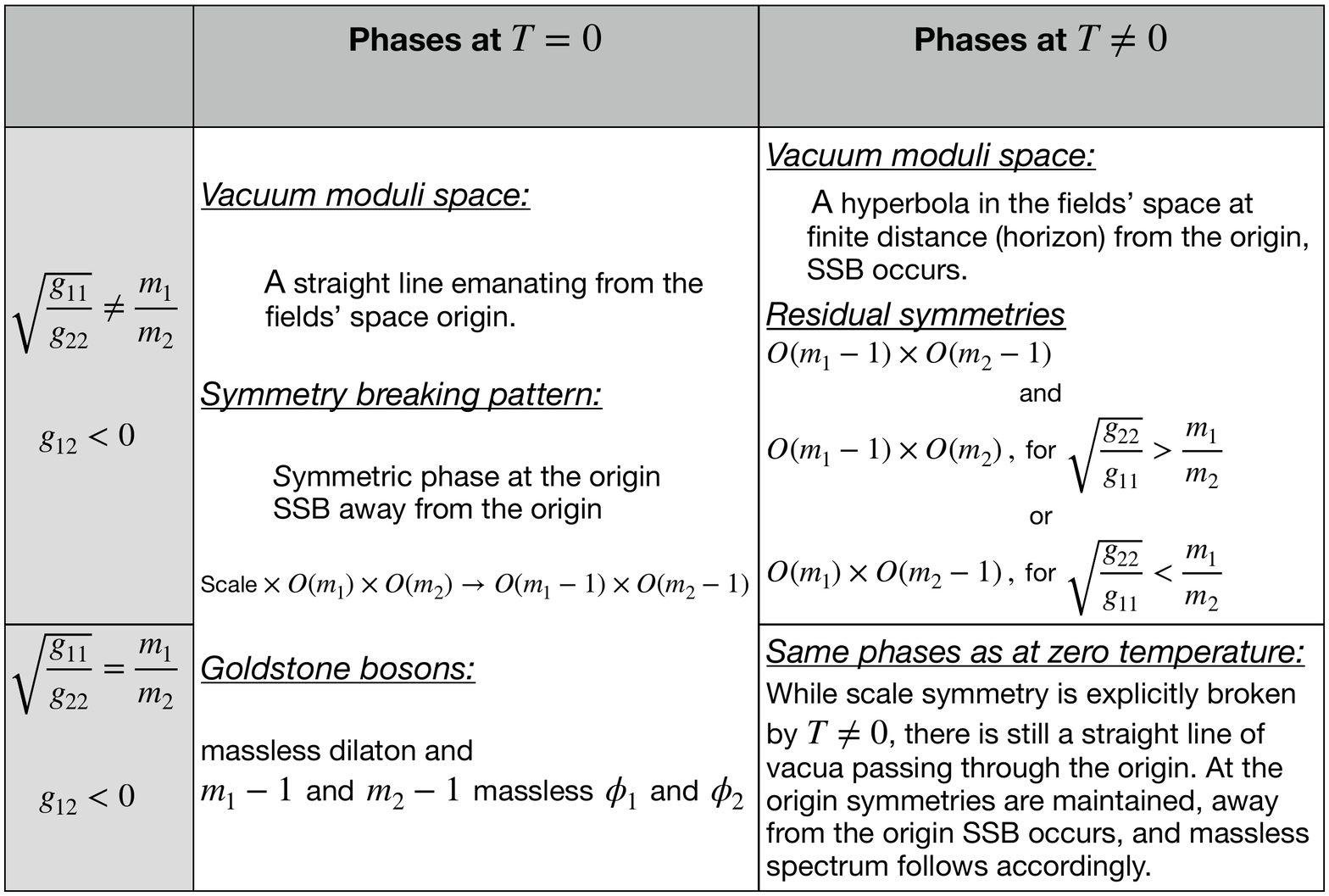}
\caption{Symmetry breaking patterns in the large rank limit.}
\label{Table_Phases}
\end{figure}

We see that the large $N$ critical biconical model has a number of massless excitations around any vacuum state \reef{broken_sol}. First, there are $m-1$ and $N-m-1$ massless modes $\phi_1$ and $\phi_2$ respectively. They are associated with the Nambu-Goldstone particles of the broken $O(m) \times O(N-m)$ symmetry.\footnote{As can be seen from \reef{eff_action0},  $m_i^2=M_{i1} \bar s_1=0$ is the mass of $\phi_i$-excitations orthogonal to the $(\sigma_1,\sigma_2)$ plane.} In addition, we have a massless mode $\sigma'_2$ which represents fluctuations along the equipotential valley \reef{broken_sol}. This excitation is a singlet of the residual symmetry group, and we interpret it as a massless dilaton associated with spontaneously broken scale invariance. We have thus accounted for $N-1$ of the particles in the original Lagrangian. A remaining degree of freedom whose fate we can follow is the massive "Higgs scalar particle". Its mass is fixed by the scale at which the symmetries were broken. In Appendix \ref{Large_N_diag} we derive these properties using the diagrammatic expansion in the large rank limit. This includes the analysis of the four and two point functions. Finally, there is an excitation $s_1$ built of the original fields $\phi_i$. The correlation function of $s_1$ in momentum space scales as $\langle s_1(p) s_1(-p)\rangle \sim p^{3-d}$. Hence, it represents a composite field with scaling dimension 2. The results are summarized in figure~\ref{Table_Phases}.

\subsection{Towards a Model in 2+1 Dimensions}

The finite temperature symmetry breaking pattern of the bi-conical model is (for $m_1<m_2$)
\begin{equation}\label{SymmB}O(m_1)\times O(m_2)\longrightarrow O(m_1-1)\times O(m_2)~.\end{equation}
This cannot hold true all the way up to $\epsilon=1$, i.e. 2+1 dimensions, due to  the Mermin-Wagner-Hohenberg-Coleman theorem \cite{PhysRevLett.17.1133, PhysRev.158.383, Coleman:1973ci} (remember that we are at finite temperature). In fact it may even break down before we reach $\epsilon=1$, as explained in the previous subsections. The only exception is $m_1=1$, in which case one can potentially have the symmetry breaking pattern~\eqref{SymmB} at finite temperature \begin{equation}\label{Pcase}\mathbb{Z}_2\times O(N)\longrightarrow O(N)~.\end{equation}
This may in principle occur at finite temperature in 2+1 dimensions and hence the case $m_1=1$ warrants some attention. 

Let us now analyze whether~\eqref{Pcase} occurs in the $\epsilon$ expansion. The beta functions in this case take the form 
\begin{equation}\label{first}{\alpha}=9{\alpha^2}+N\gamma^2~,\end{equation}
\begin{equation}\label{middle}{\beta}={\beta^2}(N+8)+\gamma^2~,\end{equation}
\begin{equation}\label{last}1=3{\alpha}+{\beta}(N+2)+4\gamma~.\end{equation} 
It is useful to take the large $N$ limit. General considerations suggest that the energy operator of the Ising model should couple to $\vec \phi^2$ of the $O(N)$ sector with strength $1/\sqrt{N}$. The quartic coupling $(\vec \phi^2)^2$ should be $O(1/N)$ as usual. Therefore we define:
$$\tilde \gamma=\sqrt N \gamma~,\quad \tilde \beta=N\beta~,\quad \tilde \alpha=\alpha~,$$
and obtain to leading order the equations
\begin{equation}\label{first}\tilde\alpha=9\tilde \alpha^2+\tilde\gamma^2~,\end{equation}
\begin{equation}\label{middle}\tilde \beta=\tilde \beta^2+\tilde \gamma^2~,\end{equation}
\begin{equation}\label{last}1=3\tilde\alpha+\tilde\beta~.\end{equation} 
Subtracting the second equation from the first, and then using the third equation, one finds $\tilde\alpha-\tilde\beta=3\tilde \alpha-\tilde\beta$ so clearly the only solution is $\tilde\alpha=\tilde\gamma=0$, and $\tilde \beta=1$ which describes the critical $O(N)$ model accompanied by a decoupled real free field.
One may thus worry that a nontrivial fixed point with symmetry $\mathbb{Z}_2\times O(N)$ may not exist. 

But it could be that the fixed point is such that the real field charged under $\mathbb{Z}_2$ does not strongly backreact on the $N$ fields transforming in the fundamental representation of $O(N)$. So we must try a new scaling for the couplings $$\tilde \gamma= N \gamma~,\quad \tilde \beta=N\beta~,\quad \tilde \alpha= N\alpha~.$$
which now leads in the large $N$ limit to the equations 
 \begin{equation}\label{first}\tilde\alpha=\tilde\gamma^2~,\end{equation}
\begin{equation}\label{middle}\tilde \beta=\tilde \beta^2~,\end{equation}
\begin{equation}\label{last}1=\tilde\beta~.\end{equation} 
Clearly then $\tilde \beta=1$ and $\tilde\alpha=\tilde\gamma^2$, which again parameterizes a one-dimensional conformal manifold, except that now it is unbounded and looks like a parabola. In addition, for $\tilde \gamma<0$ there is a moduli space of vacua which intersects the origin.
These theories describe a free field in an $O(N)$ bath -- the backreaction of the free field sector on the $O(N)$ model is very small.
It is crucial to find which of the fixed points on the conformal manifold correspond to fixed points which exist also for finite rank. Following the same strategy as before one finds the following equation
$$(\tilde\gamma-1)(\tilde\gamma+3)=0~.$$
One quick way to obtain this equation is by taking the $x\to1$ limit carefully in~\eqref{quartic}. Of course $\tilde \gamma=1$ is the $O(N)$ invariant fixed point while $\tilde\gamma=-3$ is the new, more interesting, fixed point. To leading order in the large rank expansion, the thermal masses at this new fixed point are ${2 \pi^2 \epsilon \over 3 \be_{th}^2}(-3,1)$. Therefore the scalar potential at finite temperature at leading order in the large $N$ expansion is 
$$V={2 \pi^2 \epsilon \over 6 \be_{th}^2} (\mu \be_{th})^\epsilon\left(-3\Psi^2+\vec\phi^2\right)+{2\pi^2\epsilon \over N} \, \mu^{\epsilon}\left(3\Psi^2-\vec \phi^2\right)^2~.$$
This leads to a hyperbola of vacua 
\beq
3\Psi^2-\vec\phi^2={N \over 12 \be_{th}^{2-\epsilon}}~.
\label{Z2hyper}
\eeq
Following a similar analysis to what we have done in the fixed $x$ limit, one can further show that upon including finite rank corrections the only true vacuum that remains is the one where $\Psi$ obtains a VEV ($\left\langle \Psi^2\right\rangle = {N\over 36 \beta^{2-\epsilon}_{th}}$) and $\vec \phi $ does not. Therefore the $\mathbb{Z}_2$ symmetry at finite temperature is certainly broken at large enough finite $N$. We have therefore found that~\eqref{Pcase} indeed takes place in the $\epsilon$ expansion.\footnote{As an example, here are the numerical, high-precision, values of the coupling constants and thermal masses for $N=10^4$: $$(\alpha,\beta,\gamma)=(0.0008914755083784347,
0.00009984152941453665,
 -0.0002973905499790778)~,$$ $$(m^2_1,m_2^2)={2 \pi^2 \epsilon \over 3 \be_{th}^2}(-2.9709336827156636, 0.9994004793411305)~.$$}

Since this model exhibits $\mathbb{Z}_2$ symmetry breaking at finite temperature, it is possible in principle that it continues to hold true not just for small $\epsilon$ but also for $\epsilon=1$, namely, in $2+1$ dimensions. From this perspective it is instructive to explore this model in the large $N$ limit along the guidelines of section \ref{largeN}. The kinetic free energy of the field $\Psi$ is suppressed in the large $N$ limit, and therefore the variational functional simplifies\footnote{See footnote \ref{foot} for the relation between $g_{ij}^B$ and the couplings $\tilde \gamma= N \gamma~,\quad \tilde \beta=N\beta~,\quad \tilde \alpha= N\alpha$. }
\beq
 \mathcal{W}= - {1\over 2} \int_0^{m_\phi^2} dm^2\, m^2 {\d \over  \d m^2} \langle \vec \phi^2\rangle_{\beta_{\text{th}}} 
 +{g^B_{11}\over 4N} \langle\Psi\rangle_{\be_{\text{th}}}^4 
 + {g^B_{22}\over 4N} \langle \vec\phi^2\rangle_{\beta_{\text{th}}}^2
 +{g^B_{12}\over 2N}  \langle\Psi\rangle_{\be_{\text{th}}}^2 \langle \vec\phi^2\rangle_{\beta_{\text{th}}} ~,
\eeq
where $ \langle \vec\phi^2\rangle_{\beta_{\text{th}}}$ is given by \reef{phi2vev}, and $\langle\Psi\rangle_{\be_{\text{th}}}$ is the thermal expectation value determined by minimizing $ \mathcal{W}$. In the large $N$ limit this is the only remnant of $\Psi$. 

Extremizing  $ \mathcal{W}$ with respect to $\langle\Psi\rangle_{\be_{\text{th}}}$, yields
\beq
 \langle\Psi\rangle_{\be_{\text{th}}}\( \langle\Psi\rangle_{\be_{\text{th}}}^2 + {g_{12}^B\over g_{11}^B}  \langle \vec\phi^2\rangle_{\beta_{\text{th}}}\)=0~.
\eeq
Two additional constraints are obtained by varying $\mathcal{W}$ with respect to $m_\phi^2$ and $\sigma_\phi$ respectively,
\beq
 m_\phi^2= {g_{22}^B\over N} \langle \vec\phi^2\rangle_{\beta_{\text{th}}} + {g_{12}^B\over N} \langle\Psi\rangle_{\be_{\text{th}}}^2 ~, \quad \sigma_\phi \, m_\phi^2=0~.
 \label{Z2gap}
\eeq

Now, let us study a possible phase with broken $\Z_2$ symmetry ($\langle\Psi\rangle_{\be_{\text{th}}}\neq 0$). It satisfies
\beq
 \langle\Psi\rangle_{\be_{\text{th}}}^2 + {g_{12}^B\over g_{11}^B}  \langle \vec\phi^2\rangle_{\beta_{\text{th}}}=0~.
 \label{Z2break}
\eeq
This equation has no solution unless $g_{12}^B<0$. As was argued in the beginning of this subsection, the latter inequality holds at the fixed point in $d=3-\epsilon$. In fact, for $\epsilon\ll 1$ the relation \reef{Z2break} is a hallmark of the global minimum of the variational functional ($\mathcal{W}=0$). To see it, we notice that $\mathcal{W}$ is given by the sum of positive kinetic and potential terms.  The kinetic term equals zero at $m_\phi^2=0$, whereas the potential vanishes provided that the couplings are tuned to the fixed point values  \reef{first}-\reef{last} and \reef{Z2break} is satisfied. Furthermore, substituting \reef{first}-\reef{last} into \reef{Z2break} and using \reef{zeroM}, we recover  \reef{Z2hyper}. The upshot is that $\epsilon$ expansion and the large $N$ approaches agree. In particular, the $\Z_2$ symmetry is necessarily broken at finite ${\beta_{\text{th}}}$, whereas at zero temperature there is a moduli space of vacua which intersects the origin.

It is particularly interesting to find out whether our conclusions survive all the way to $\epsilon=1$. We leave this question for the future.  
 
\section*{Acknowledgments}

The authors would like to thank O. Aharony, D. Harlow, D. Jafferis, A. Kapustin, M. Metlitski, R. Sinha, S. Yankielowicz, and Y. Zheng for useful discussions. E. Rabinovici would like to thank the IHES in Bures sur Yvette, the NHETC at Rutgers Physics Department and CCPP at NYU for hospitality and support. Z.K and C.C are supported in part by the Simons Foundation grant 488657 (Simons Collaboration on the Non-Perturbative Bootstrap) and the BSF grant no. 2018204.  The Hebrew University group is partially supported by the Binational Science Foundation (grant No. 2016186), the Israeli Science Foundation Center of Excellence (grant No. 2289/18) and by the Quantum Universe I-CORE program of the Israel Planning and Budgeting Committee (grant No. 1937/12).

\appendix

\section{More Details on Footnote 2}
\label{FootnoteTwo}

Our first comments are about 1+1 dimensional massive QFT with $\mathbb{Z}_2$ symmetry. 
We will assume that the $\mathbb{Z}_2$ is non-anomalous. To understand what precisely ``non-anomalous'' means see~\cite{Lin:2019kpn} for a recent discussion. 
Now let us couple the theory to a background $\mathbb{Z}_2$ gauge field $a$. Since the symmetry is non-anomalous we may gauge the $\mathbb{Z}_2$, which means that we can sum over $a$. The theory obtained in this way automatically has a bonus $\mathbb{Z}_2$ symmetry, and we can in turn couple it to a gauge field $b$ by adding the phase $e^{ \pi i a b}$ to the action. (The $a b$ product is just the cup product.) Since we sum over $a$, this gives a functional of $b$. 

A familiar claim is that if the original $\mathbb{Z}_2$ is broken then the dual is not and vice versa. This claim is important, for instance, in the Kramers–Wannier duality. 

Let us review the general proof.
We have to understand what does it mean for the original $\mathbb{Z}_2$ symmetry to be broken. Take a torus with sides $R,T$. Take them both to be very large compared to any mass scale of the infinite volume theory. 
Then, if the symmetry is broken the partition function with nontrivial $a$ background is always exponentially smaller than the partition function without $a$ insertions. This is clear for instance in the interpretation that $a$ is along the time direction. Indeed, on the circle the two lowest states mix with energy difference 
$\Delta E = e^{-mR} $
($m$ is the domain wall tension) and putting $a=1$ along the time direction the partition function becomes
$e^{E_1T}-e^{E_2T}$ which is exponentially smaller than each of the terms due to the small energy difference (the minus sign is due to the fact that the two eigenstates have different $\mathbb{Z}_2$ charges). The same is true for any other cycle $a$ wraps. 
Therefore if we sum over $a$ without additional phases the result is dominated by the contribution from the torus without any $a$ insertions. Hence, the partition function is essentially $\frac12\left(e^{E_1T}+e^{E_2T}\right)$.
But now we can activate $b$ in the time direction. Since it does not affect the partition function in the sector with $a=0$, we see that the partition function with $b$ turned on in the time direction is approximately $\frac12\left(e^{E_1T}+e^{E_2T}\right)$, and in particular, it is not exponentially smaller. 

That the partition function with $b$ in the time direction is not exponentially smaller means that the bonus $\mathbb{Z}_2$ is unbroken. The converse argument works identically. 

Now we have to explain the relationship of these observations to order at finite temperature in 2+1 dimensions.  Consider a theory $\cal T$ in 2+1 dimensions with $\mathbb{Z}_2$ symmetry. We take space to be a cylinder with radius $\beta/2\pi$. At long distances, i.e. distances much longer than $\beta$ or any other scale in the problem, the theory is assumed to be gapped and it should be thought of as a 1+1 dimensional theory. Now consider in parallel the theory $\cal{T}'$ obtained by gauging the $\mathbb{Z}_2$ symmetry of $\cal T$. $\cal{T}'$ has a one-form symmetry in 2+1 dimensions~\cite{Gaiotto:2014kfa}. If we put the theory on the same cylinder and take the long distance limit, from the point of view of 1+1 dimensions, it has an ordinary $\mathbb{Z}_2$ symmetry that can be interpreted precisely as the bonus symmetry in our previous discussion. (The 1+1 dimensional theory obtained in this way also has a one-form symmetry which we will ignore.) This bonus $\mathbb{Z}_2$ symmetry is precisely the symmetry acting on the confinement/deconfinement order parameter of $\cal{T}'$ introduced by Polyakov~\cite{Polyakov:1978vu} in the context of gauge theories. Hence, breaking the ordinary $\mathbb{Z}_2$ symmetry in the theory $\cal T$ at finite temperature is equivalent to a finite temperature deconfined phase of $\cal{T}'$.

\section{4-point Correlators in the Large $N$ Biconical Model}
\label{Large_N_diag}
In this appendix, we  will discuss the 4-point correlators of the biconical model in the large N limit.  In  the strictly $N\rightarrow \infty$ limit, the dynamics of the model is essentially Gaussian. Therefore, all the connected 4-point correlators are suppressed by powers of $\frac{1}{N}$. We will restrict our attention to the correlators that are nonvanishing at $O(\frac{1}{N})$. 

This analysis will shed light on the effective interactions between the particles in the model at different temperatures. We will see that there are interesting differences in the behaviour of the correlators in a ground state and in a thermal state. These differences arise essentially from an interaction vertex which vanishes in a ground state but is nonzero for a thermal state. We will show that the presence of this vertex leads to new  poles in certain thermal correlators. In addition, it also leads to some thermal correlators to be nonzero while their vacuum counterparts vanish at $O(\frac{1}{N})$.

In this analysis, we will include  correlators with insertions  of the Goldstone bosons. Thus, in contrast to the main text, we will not integrate out these modes. Rather, we will  derive an effective  action which includes these modes. 

\subsection{Effective Action with an Auxiliary Field}

We remind the reader that the Euclidean action of the model is given by
\begin{equation}
 I_E =\frac{1}{2} \sum_i \int d^{d+1}x \Big( \partial_\mu\overrightarrow{\phi}_i \cdot\partial^\mu\overrightarrow{\phi}_i + s_i (\overrightarrow{\phi}_i^2 - N \rho_i) \Big)
  +\frac{N}{4}\sum_{ij} g^B_{ij}  \int d^{d+1}x  \rho_i\,  \rho_j ,
\end{equation}
Here the fields $s_i$ and $\rho_i$ are auxiliary degrees of freedom introduced to simplify the analysis.

We will consider perturbations of the fields $\phi_i$ about the expectation values $\sigma_i$ which are aligned along some particular direction (with the unit vector $\widehat{n}_i$) in the $(x_iN)$-dimensional space. Therefore, we take
\begin{equation}
\overrightarrow{\phi}_i=(\sigma_i+\eta_i)\widehat{n}_i+\overrightarrow{\theta}_i,
\end{equation}
where $\eta_i$ is the fluctuation along  the direction $\widehat{n}_i$, whereas $\overrightarrow{\theta}_i$ is the fluctuation transverse to this direction.
In terms of these degrees of freedom, the action is given by
\begin{equation}
\begin{split}
 I_E =&\frac{1}{2} \sum_i \int d^{d+1}x \Big( \partial_\mu\overrightarrow{\theta}_i \partial^\mu\overrightarrow{\theta}_i +\partial_\mu\eta_i \partial^\mu\eta_i + s_i (\overrightarrow{\theta}_i^{2}+\eta_i^2+2\eta_i\sigma_i+\sigma_i^2) \Big)\\
 &-\frac{1}{2} \sum_i \int d^{d+1}xN \rho_is_i
  +\frac{N}{4}\sum_{ij} g_{ij}^B  \int d^{d+1}x  \rho_i\,  \rho_j.
 \end{split}
\end{equation}

Now, we will integrate out the fields $\rho_1$ and $\rho_2$. For this, let us first introduce the following variables:
\begin{equation}
\begin{split}
&\rho_1^\prime=\frac{1}{\sqrt{g_{11}^B+g_{22}^B}}\Big(\sqrt{g_{11}^B}\rho_1-\sqrt{g_{22}^B}\rho_2\Big),\ \rho_2^\prime=\frac{1}{\sqrt{g_{11}^B+g_{22}^B}}\Big(\sqrt{g_{22}^B}\rho_1+\sqrt{g_{11}^B}\rho_2\Big),\\
&s_1^\prime=\frac{1}{\sqrt{g_{11}^B+g_{22}^B}}\Big(\sqrt{g_{11}^B}s_1-\sqrt{g_{22}^B}s_2\Big),\ s_2^\prime=\frac{1}{\sqrt{g_{11}^B+g_{22}^B}}\Big(\sqrt{g_{22}^B}s_1+\sqrt{g_{11}^B}s_2\Big).\\
\end{split}
\end{equation}
The action then takes the  form
\begin{equation}
\begin{split}
 I_E =&\frac{1}{2} \sum_i \int d^{d+1}x \Big( \partial_\mu\overrightarrow{\theta}_i \partial^\mu\overrightarrow{\theta}_i +\partial_\mu\eta_i \partial^\mu\eta_i + s_i (\overrightarrow{\theta}_i^{2}+\eta_i^2+2\eta_i\sigma_i+\sigma_i^2) \Big)\\
 &-\frac{1}{2} \sum_i \int d^{d+1}xN \rho_i^\prime s_i^\prime
  +\frac{N}{4} (g_{11}^B+g_{22}^B) \int d^{d+1}x  \rho_1^{\prime 2}.
 \end{split}
\end{equation}
Integrating out the field  $\rho_2^\prime$, we get a delta function in the path integral which imposes the constraint 
\begin{equation}
\begin{split}
s_2^\prime=0\implies s_2=-\sqrt{\frac{g_{22}^B}{g_{11}^B}}s_1.
 \end{split}
\end{equation}
Integrating out the field $\rho_1^\prime$ contributes the following term to the action

\begin{equation}
\begin{split}
-\frac{N}{4 (g_{11}^B+g_{22}^B)} \int d^{d+1}x \ s_1^{\prime2}
&=-\frac{N}{4g_{11}^B} \int d^{d+1}x \ s_1^2.
  \end{split}
\end{equation}
Finally, integrating out the field $s_2$, we get the following action which is a functional of the remaining fields:
\begin{equation}
\begin{split}
  \widetilde{I}_E =&\Bigg[\frac{1}{2} \sum_i \int d^{d+1}x \Big( \partial_\mu\overrightarrow{\theta}_i \partial^\mu\overrightarrow{\theta}_i +\partial_\mu\eta_i \partial^\mu\eta_i  \Big)-\frac{N}{4g_{11}^B} \int d^{d+1}x \ s_1^2\\
 &+\frac{1}{2} \int d^{d+1}x s_1 (\overrightarrow{\theta}_1^{2}+\eta_1^2+2\eta_1\sigma_1+\sigma_1^2) \\
 &-\frac{1}{2}\sqrt{\frac{g_{22}^B}{g_{11}^B}} \int d^{d+1}x s_1 (\overrightarrow{\theta}_2^{2}+\eta_2^2+2\eta_2\sigma_2+\sigma_2^2) \Bigg].
 \end{split}
\end{equation}

Let us now define
\begin{equation}
\begin{split}
s_0\equiv-i\sqrt{\frac{N}{2g_{11}^B}}s_1.
 \end{split}
 \label{s1-s0 relation}
\end{equation}
Then the action is given by
\begin{equation}
\begin{split}
  \widetilde{I}_E =&\int d^{d+1}x\Bigg[\frac{1}{2} \sum_i  \Big( \partial_\mu\overrightarrow{\theta}_i \partial^\mu\overrightarrow{\theta}_i +\partial_\mu\eta_i \partial^\mu\eta_i  \Big)+\frac{1}{2} s_0^2+is_0\Big(\sqrt{\frac{g_{11}^B}{2N}} \sigma_1^2-\sqrt{\frac{g_{22}^B}{2N}}\sigma_2^2 \Big)\\
  &\qquad\qquad+i\sqrt{\frac{g_{11}^B}{2N}} \ s_0 (\overrightarrow{\theta}_1^{2}+\eta_1^2+2\eta_1\sigma_1) -i\sqrt{\frac{g_{22}^B}{2N}}  \ s_0(\overrightarrow{\theta}_2^{2}+\eta_2^2+2\eta_2\sigma_2)\Bigg].
 \end{split}
\end{equation}

At a temperature $T=\frac{1}{\beta_{\text{th}}}$, the expectation values $\sigma_1$ and $\sigma_2$  lie on a moduli space defined by the following equation:
\begin{equation}
\begin{split}
&\sqrt{\frac{g_{11}^B}{2N}}\sigma_1^2-\sqrt{\frac{g_{22}^B}{2N}}\sigma_2^2=\frac{\sqrt{2N}c(\epsilon)\beta_{\text{th}}^{-2+\epsilon}}{24}(\sqrt{g_{22}^B}x_2-\sqrt{g_{11}^B}x_1).\\
  \end{split}
  \label{thermal VEV relation}
 \end{equation}
 where $c(\epsilon)$ is the function defined in \eqref{c(epsilon) def}. Therefore, the above action reduces to
\begin{equation}
\begin{split}
  \widetilde{I}_E =&\int d^{d+1}x\Bigg[\frac{1}{2} \sum_i  \Big( \partial_\mu\overrightarrow{\theta}_i \partial^\mu\overrightarrow{\theta}_i +\partial_\mu\eta_i \partial^\mu\eta_i  \Big)+\frac{1}{2} s_0^2+i s_0 \Bigg(\sqrt{\frac{2g_{11}^B}{N}} \sigma_1\eta_1-\sqrt{\frac{2g_{22}^B}{N}} \sigma_2\eta_2\Bigg)\\
  &\qquad\quad-i\frac{\sqrt{2N}c(\epsilon)\beta_{\text{th}}^{-2+\epsilon}}{24}(\sqrt{g_{11}^B}x_1-\sqrt{g_{22}^B}x_2)s_0\\
  &\qquad\quad+i\sqrt{\frac{g_{11}^B}{2N}} \ s_0 (\overrightarrow{\theta}_1^{2}+\eta_1^2) -i\sqrt{\frac{g_{22}^B}{2N}}  \ s_0(\overrightarrow{\theta}_2^{2}+\eta_2^2)\Bigg].
 \end{split}
\end{equation}

From the above expression of the action, we can see that only a linear combination of the fields $\eta_1$ and $\eta_2$ couples to $s_0$ at the quadratic level. This implies that this mode picks up a mass by the Higgs mechanism. The mode orthogonal to this combination remains massless. As  discussed in the main text, this massless boson (the dilaton) arises due to the spontaneous breaking of scale invariance. We provide the forms of these modes corresponding to the massive boson ($\eta_{-}$) and the dilaton ($\eta_{+}$) below:
\begin{equation}
\begin{split}
\eta_{-}\equiv \frac{\sqrt{g_{11}^B}\sigma_1 \eta_1-\sqrt{g_{22}^B}\sigma_2 \eta_2}{\sqrt{g_{11}^B\sigma_1^2+g_{22}^B\sigma_2^2}},\ \eta_{+}\equiv \frac{\sqrt{g_{11}^B}\sigma_1 \eta_2+\sqrt{g_{22}^B}\sigma_1 \eta_2}{\sqrt{g_{11}^B\sigma_1^2+g_{22}^B\sigma_2^2}}.
\end{split}
\label{thermal: def. higgs, dilaton}
\end{equation}
As we will soon see, the mass of the $\eta_{-}$ field is given by 
\begin{equation}
\begin{split}
\widetilde{\sigma}\equiv\sqrt{\frac{2}{N}}\sqrt{g_{11}^B\sigma_1^2+g_{22}^B\sigma_2^2}.
\end{split}
\label{higgs mass}
\end{equation}
Note that this mass depends on the expectation values $\sigma_1$ and $\sigma_2$. Since these expectation values are constrained to lie on a hyperbola determined by the temperature (see \eqref{thermal VEV relation}), therefore the mass is not completely independent of the temperature. However, at any given temperature, it is not uniquely determined as there is a moduli space of vacua and each of these vacua  gives a different value of the mass. For instance, at zero temperature, this mass is given by
\begin{equation}
\begin{split}
\lim_{\beta_{\text{th}}\rightarrow\infty} \widetilde{\sigma}= \sqrt{\frac{2(g_{11}^B+\sqrt{g_{11}^Bg_{22}^B})}{N}}\sigma_1,
\end{split}
\label{higgs mass zero temp}
\end{equation}
where $\sigma_1$ parametrises the different points on the moduli space of vacua.

In terms of the quantities introduced above, the action takes the following form:
\begin{equation}
\begin{split}
  \widetilde{I}_E =&\int d^{d+1}x\Bigg[\frac{1}{2} \Big(\sum_i   \partial_\mu\overrightarrow{\theta}_i \partial^\mu\overrightarrow{\theta}_i +\partial_\mu\eta_{+} \partial^\mu\eta_{+}+\partial_\mu\eta_{-} \partial^\mu\eta_{-}\Big)  +\frac{1}{2} s_0^2+i \widetilde{\sigma} s_0 \eta_{-}\\
  &\qquad\quad-i\frac{\sqrt{2N}c(\epsilon)\beta_{\text{th}}^{-2+\epsilon}}{24}(\sqrt{g_{11}^B}x_1-\sqrt{g_{22}^B}x_2)s_0\\
  &\qquad\quad+i\sqrt{\frac{g_{11}^B}{2N}} \ s_0 \overrightarrow{\theta}_1^{2} -i\sqrt{\frac{g_{22}^B}{2N}}  \ s_0\overrightarrow{\theta}_2^{2}\\
  &\qquad\quad+\frac{i}{2\sqrt{N}} A_{_{--}} s_0\eta_{-}^2+\frac{i}{\sqrt{N}} A_{_{+-}}s_0\eta_{-}\eta_{+} +\frac{i}{2\sqrt{N}} A_{_{++}} s_0\eta_{+}^2\Bigg].
 \end{split}
 \label{thermal:action}
\end{equation}
where
\begin{equation}
\begin{split}
&A_{_{--}}=\sqrt{2}\Bigg(\frac{(g_{11}^B)^{\frac{3}{2}}\sigma_1^2-(g_{22}^B)^{\frac{3}{2}}\sigma_2^2}{(g_{11}^B\sigma_1^2+g_{22}^B\sigma_2^2)} \Bigg),\\
&A_{_{+-}}=\sqrt{2}\Bigg(\frac{\sqrt{g_{11}^B g_{22}^B}(\sqrt{g_{11}^B}+\sqrt{g_{22}^B})\sigma_1\sigma_2}{g_{11}^B\sigma_1^2+g_{22}^B\sigma_2^2}\Bigg),\\
&A_{_{++}}=-\sqrt{2}\Big(\frac{\sqrt{g_{11}^B g_{22}^B}}{g_{11}^B\sigma_1^2+g_{22}^B\sigma_2^2}\Big)\Big(\sqrt{g_{11}^B}\sigma_1^2-\sqrt{g_{22}^B}\sigma_2^2\Big).\\
 \end{split}
\end{equation}

\paragraph{Essential difference between ground states and thermal states:}\mbox{}
Note that using equation \eqref{thermal VEV relation} and the definition of $\widetilde{\sigma}$ given in \eqref{higgs mass}, we  can get
\begin{equation}
\begin{split}
&A_{_{++}}
=-\Big(\frac{\sqrt{g_{11}^Bg_{22}^B}}{3\sqrt{2}\widetilde{\sigma}^2}\Big)c(\epsilon)\beta_{\text{th}}^{-2+\epsilon}(\sqrt{g_{22}^B}x_2-\sqrt{g_{11}^B}x_1).\\
 \end{split}
 \label{A++ expression}
\end{equation}
Therefore, this coefficient vanishes as the temperature goes to zero i.e. when $\beta_{\text{th}}\rightarrow\infty$. The absence of this vertex at zero temperature leads to the vanishing of certain Feynman diagrams. We will show that as a consequence, certain 4-point correlators with insertions of the dilatons are nonzero only in a thermal state, and vanish as the temperature is taken to zero.

\subsection{Feynman Diagrammatics}
From the Euclidean action given in \eqref{thermal:action}, we can  derive all the ingredients for drawing Feynman diagrams in this theory. We enumerate all the propagators and the interaction vertices appearing in such Feynman diagrams below. In a thermal state with temperature $T=\frac{1}{\beta_{\text{th}}}$, the zeroth components of the momenta in the propagators are quantised in units of $\frac{2\pi}{\beta_{\text{th}}}$. From these propagators and vertices we can compute the thermal correlators in momentum space. To get these correlators, one would have to multiply a factor of $\beta_{\text{th}}\delta_{\sum_i p_i^0,0}(2\pi)^d\delta^{d}(\sum_i \overrightarrow{p}_i)$ to the contributions of the Feynman diagrams. At zero temperature, this multiplicative factor has to be replaced as follows:
\begin{equation*}
\beta_{\text{th}}\delta_{\sum_i p_i^0,0}(2\pi)^d\delta^{d}(\sum_i \overrightarrow{p}_i)\rightarrow (2\pi)^{d+1}\delta^{d+1}(\sum_i \overrightarrow{p}_i).
\end{equation*}
\subsubsection{Propagators}\mbox{}
\begin{equation*}
\raisebox{-0.1 cm}{\scalebox{0.7}{\begin{tikzpicture}
\draw[color=black] (-1.4,0)node{$\theta_1^\alpha$};
\draw[color=black] (1.4,0)node{$\theta_1^\alpha$};
\draw[color=black] (0,0.6)node{$\overrightarrow{k}$};
\draw[thick,color=yellow] (-1.0,0) -- (1.0,0);
\draw[thick,color=black,->] (-0.4,0.3) -- (0.4,0.3);
\end{tikzpicture}}}
=\frac{1}{k^2},\ 
\raisebox{-0.1 cm}{\scalebox{0.7}{\begin{tikzpicture}
\draw[color=black] (-1.4,0)node{$\theta_2^\alpha$};
\draw[color=black] (1.4,0)node{$\theta_2^\alpha$};
\draw[color=black] (0,0.6)node{$\overrightarrow{k}$};
\draw[thick,color=orange] (-1.0,0) -- (1.0,0);
\draw[thick,color=black,->] (-0.4,0.3) -- (0.4,0.3);
\end{tikzpicture}}}
=\frac{1}{k^2},\ 
\raisebox{-0.1 cm}{\scalebox{0.7}{\begin{tikzpicture}
\draw[color=black] (-1.4,0)node{$\eta_{+}$};
\draw[color=black] (1.4,0)node{$\eta_{+}$};
\draw[color=black] (0,0.6)node{$\overrightarrow{k}$};
\draw[thick,color=blue] (-1.0,0) -- (1.0,0);
\draw[thick,color=black,->] (-0.4,0.3) -- (0.4,0.3);
\end{tikzpicture}}}
=\frac{1}{k^2},
\end{equation*}

\begin{equation*}
\raisebox{-0.1 cm}{\scalebox{0.7}{\begin{tikzpicture}
\draw[color=black] (-1.4,0)node{$\eta_{-}$};
\draw[color=black] (1.4,0)node{$\eta_{-}$};
\draw[color=black] (0,0.6)node{$\overrightarrow{k}$};
\draw[thick,color=red] (-1.0,0) -- (1.0,0);
\draw[thick,color=black,->] (-0.4,0.3) -- (0.4,0.3);
\end{tikzpicture}}}
=\frac{1}{k^2+\widetilde{\sigma}^2},\ 
\raisebox{-0.1 cm}{\scalebox{0.7}{\begin{tikzpicture}
\draw[color=black] (-1.4,0)node{$\eta_{-}$};
\draw[color=black] (1.4,0)node{$s_0$};
\draw[color=black] (0,0.6)node{$\overrightarrow{k}$};
\draw[thick,color=red] (-1.0,0) -- (0,0);
\draw[thick,color=green,dashed] (0,0) -- (1.0,0);
\draw[thick,color=black,->] (-0.4,0.3) -- (0.4,0.3);
\end{tikzpicture}}}
=-\frac{i\widetilde{\sigma}}{k^2+\widetilde{\sigma}^2},\ 
\raisebox{-0.1 cm}{\scalebox{0.7}{\begin{tikzpicture}
\draw[color=black] (-1.4,0)node{$s_0$};
\draw[color=black] (1.4,0)node{$s_0$};
\draw[color=black] (0,0.6)node{$\overrightarrow{k}$};
\draw[thick,color=green,dashed] (-1.0,0) -- (1.0,0);
\draw[thick,color=black,->] (-0.4,0.3) -- (0.4,0.3);
\end{tikzpicture}}}
=\frac{k^2}{k^2+\widetilde{\sigma}^2}.
\end{equation*}

From the form of the $(\eta_{-}-\eta_{-})$ propagator given above, one can easily see that $\widetilde{\sigma}$ is the mass of the  $\eta_{-}$ field.

\subsubsection{Vertices}\mbox{}

\begin{equation*}
\raisebox{-0.2 cm}{\scalebox{0.7}{\begin{tikzpicture}
\draw[thick,color=green,dashed] (-1,0) -- (1.0,0);
\draw[color=black] (1.4,0)node{$s_0$};
\draw[thick,color=green,dashed] (-1,0)node{$\times$};
\end{tikzpicture}}}
= i\frac{\sqrt{2N} c(\epsilon)\beta_{\text{th}}^{-2+\epsilon}}{24}(\sqrt{g_{11}^B}x_1-\sqrt{g_{22}^B}x_2),
\end{equation*}

\begin{equation*}
\raisebox{-0.9 cm}{\scalebox{0.7}{\begin{tikzpicture}
\draw[color=black] (-1.3,1.1)node{$\theta_1^\alpha$};
\draw[color=black] (-1.3,-1.1)node{$\theta_1^\alpha$};
\draw[color=black] (1.4,0)node{$s_0$};
\draw[thick,color=yellow] (-1.0,1.0) -- (0,0);
\draw[thick,color=yellow] (-1.0,-1.0) -- (0,0);
\draw[thick,color=green,dashed] (0,0) -- (1.0,0);
\end{tikzpicture}}}
=-i\sqrt{\frac{2g_{11}^B}{N}},\ 
\raisebox{-0.9 cm}{\scalebox{0.7}{\begin{tikzpicture}
\draw[color=black] (-1.3,1.1)node{$\theta_2^\alpha$};
\draw[color=black] (-1.3,-1.1)node{$\theta_2^\alpha$};
\draw[color=black] (1.4,0)node{$s_0$};
\draw[thick,color=orange] (-1.0,1.0) -- (0,0);
\draw[thick,color=orange] (-1.0,-1.0) -- (0,0);
\draw[thick,color=green,dashed] (0,0) -- (1.0,0);
\end{tikzpicture}}}
= i\sqrt{\frac{2g_{22}^B}{N}},
\end{equation*}

\begin{equation*}
\raisebox{-0.9 cm}{\scalebox{0.7}{\begin{tikzpicture}
\draw[color=black] (-1.3,1.1)node{$\eta_{-}$};
\draw[color=black] (-1.3,-1.1)node{$\eta_{-}$};
\draw[color=black] (1.4,0)node{$s_0$};
\draw[thick,color=red] (-1.0,1.0) -- (0,0);
\draw[thick,color=red] (-1.0,-1.0) -- (0,0);
\draw[thick,color=green,dashed] (0,0) -- (1.0,0);
\end{tikzpicture}}}
=-\frac{i}{\sqrt{N}}A_{_{--}},\ 
\raisebox{-0.9 cm}{\scalebox{0.7}{\begin{tikzpicture}
\draw[color=black] (-1.3,1.1)node{$\eta_{-}$};
\draw[color=black] (-1.3,-1.1)node{$\eta_{+}$};
\draw[color=black] (1.4,0)node{$s_0$};
\draw[thick,color=red] (-1.0,1.0) -- (0,0);
\draw[thick,color=blue] (-1.0,-1.0) -- (0,0);
\draw[thick,color=green,dashed] (0,0) -- (1.0,0);
\end{tikzpicture}}}
=-\frac{i}{\sqrt{N}}A_{_{+-}},\ 
\raisebox{-0.9 cm}{\scalebox{0.7}{\begin{tikzpicture}
\draw[color=black] (-1.3,1.1)node{$\eta_{+}$};
\draw[color=black] (-1.3,-1.1)node{$\eta_{+}$};
\draw[color=black] (1.4,0)node{$s_0$};
\draw[thick,color=blue] (-1.0,1.0) -- (0,0);
\draw[thick,color=blue] (-1.0,-1.0) -- (0,0);
\draw[thick,color=green,dashed] (0,0) -- (1.0,0);
\end{tikzpicture}}}
=-\frac{i}{\sqrt{N}}A_{_{++}}.\ 
\end{equation*}

The 1-point vertex with the field $s_0$ ensures that the thermal expectation of the field $s_0$ is zero (upto the leading order in $\frac{1}{N}$). Its contribution to this expectation value cancels the contributions of tadpole diagrams involving loops of Goldstone bosons. We will not prove this explicitly here. However, note that in the main text we have already shown that the saddle point value of the field $s_1$ must be zero in the strictly $N\rightarrow\infty$ limit. Since the field $s_0$  is related to $s_1$ by the equation \eqref{s1-s0 relation}, its expectation value also must vanish upto leading order in $\frac{1}{N}$.

From the expression of the 1-point vertex, we can see that it vanishes when the temperature goes to zero, i.e. when $\beta_{\text{th}}\rightarrow \infty$. The tadpole diagrams, whose contributions it cancelled at nonzero temperatures, also vanish in this limit.\footnote{See \cite{kleinert2001critical} for a proof of the vanishing of such tadpole diagrams with loops of massless  propagators in the vacuum.} Therefore, the expectation value of the field $s_0$ remains zero at zero temperature.

\subsubsection{Correction to the $s_0-s_0$ Propagator due to Loops of Goldstone Bosons}

In the large N limit, the $s_0-s_0$ propagator receives corrections from loops of the Goldstone bosons as shown below:
\begin{equation}
\begin{split}
&\raisebox{-0.5 cm}{\begin{tikzpicture}
\draw[thick,color=green,dashed] (-1.5,0) -- (-0.5,0);
\draw[thick,color=green,dashed] (0.5,0) -- (1.5,0);
\draw[pattern=north east lines, pattern color=black] (0,0) circle (0.5);
\end{tikzpicture}}
=\raisebox{-0 cm}{\begin{tikzpicture}
\draw[thick,color=green,dashed] (-1,0) -- (-0,0);
\draw[thick,color=green,dashed] (0,0) -- (1,0);
\end{tikzpicture}}
+\raisebox{-0.5 cm}{\begin{tikzpicture}
\draw[thick,color=green,dashed] (-1.5,0) -- (-0.5,0);
\draw[thick,color=green,dashed] (0.5,0) -- (1.5,0);
\draw[thick,color=yellow] (0.5,0 cm) arc (0:180:0.5);
\draw[thick,color=yellow] (-0.5,0 cm) arc (-180:0:0.5);
\end{tikzpicture}}
+\raisebox{-0.5 cm}{\begin{tikzpicture}
\draw[thick,color=green,dashed] (-1.5,0) -- (-0.5,0);
\draw[thick,color=green,dashed] (0.5,0) -- (1.5,0);
\draw[thick,color=orange] (0.5,0 cm) arc (0:180:0.5);
\draw[thick,color=orange] (-0.5,0 cm) arc (-180:0:0.5);
\end{tikzpicture}}+\cdots\\
\end{split}
\end{equation}
The dots in the  above expression represent diagrams with iterations of the same loops that are shown explicitly. Each of these loops comes with two vertices contributing a factor which is $O(\frac{1}{N})$. On the other  hand, since there are $O(N)$ number of Goldstone bosons, the overall contribution of these loops is O(1).

\subsection{Connected 4-point Correlators}
Let us now discuss the connected 4-point correlators of the different modes. Some of these correlators are nonzero in both ground states and thermal states. We will first consider these correlators, and then turn our attention to the ones that are nonzero only in a thermal state.

\subsubsection{Correlators which are Nonzero in Both Ground States and Thermal States}
We enumerate all the 4-point correlators that are nonzero (in both ground states and thermal states) at $O(\frac{1}{N})$ in table \ref{table: nonvanishing correlators: both states}. We also show the kinds of Feynman diagrams that contribute to these correlators at $O(\frac{1}{N})$.

\begin{table}[H]
\caption{Correlators which are nonzero in both ground states and thermal states}
\label{table: nonvanishing correlators: both states}
\begin{center}
\scalebox{0.8}{
\begin{tabular}{ |c| c|  }
\hline
Correlator & Diagrams\\
\hline
4 Goldstone bosons & \raisebox{-0.6 cm}{\scalebox{0.6}{\begin{tikzpicture}
\draw[thick,color=yellow] (-2.5,1.0) -- (-1.5,0);
\draw[thick,color=yellow] (-2.5,-1.0) -- (-1.5,0);
\draw[thick,color=yellow] (2.5,1.0) -- (1.5,0);
\draw[thick,color=yellow] (2.5,-1.0) -- (1.5,0);
\draw[thick,color=green,dashed] (-1.5,0) -- (-0.5,0);
\draw[thick,color=green,dashed] (0.5,0) -- (1.5,0);
\draw[pattern=north east lines, pattern color=black] (0,0) circle (0.5);
\draw[color=black] (-2.6,1.1)node{$\alpha$};
\draw[color=black] (-2.6,-1.1)node{$\alpha$};
\draw[color=black] (2.6,1.1)node{$\beta$};
\draw[color=black] (2.6,-1.1)node{$\beta$};
\end{tikzpicture}}},\ 
\raisebox{-0.6 cm}{\scalebox{0.6}{\begin{tikzpicture}
\draw[thick,color=orange] (-2.5,1.0) -- (-1.5,0);
\draw[thick,color=orange] (-2.5,-1.0) -- (-1.5,0);
\draw[thick,color=orange] (2.5,1.0) -- (1.5,0);
\draw[thick,color=orange] (2.5,-1.0) -- (1.5,0);
\draw[thick,color=green,dashed] (-1.5,0) -- (-0.5,0);
\draw[thick,color=green,dashed] (0.5,0) -- (1.5,0);
\draw[pattern=north east lines, pattern color=black] (0,0) circle (0.5);
\draw[color=black] (-2.6,1.1)node{$\alpha$};
\draw[color=black] (-2.6,-1.1)node{$\alpha$};
\draw[color=black] (2.6,1.1)node{$\beta$};
\draw[color=black] (2.6,-1.1)node{$\beta$};
\end{tikzpicture}}},\ 
\raisebox{-0.6 cm}{\scalebox{0.6}{\begin{tikzpicture}
\draw[thick,color=yellow] (-2.5,1.0) -- (-1.5,0);
\draw[thick,color=yellow] (-2.5,-1.0) -- (-1.5,0);
\draw[thick,color=orange] (2.5,1.0) -- (1.5,0);
\draw[thick,color=orange] (2.5,-1.0) -- (1.5,0);
\draw[thick,color=green,dashed] (-1.5,0) -- (-0.5,0);
\draw[thick,color=green,dashed] (0.5,0) -- (1.5,0);
\draw[pattern=north east lines, pattern color=black] (0,0) circle (0.5);
\draw[color=black] (-2.6,1.1)node{$\alpha$};
\draw[color=black] (-2.6,-1.1)node{$\alpha$};
\draw[color=black] (2.6,1.1)node{$\beta$};
\draw[color=black] (2.6,-1.1)node{$\beta$};
\end{tikzpicture}}}\\
\hline
4 massive bosons & \raisebox{-0.6 cm}{\scalebox{0.6}{\begin{tikzpicture}
\draw[thick,color=red] (-2.5,1.0) -- (-1.5,0);
\draw[thick,color=red] (-2.5,-1.0) -- (-1.5,0);
\draw[thick,color=red] (2.5,1.0) -- (1.5,0);
\draw[thick,color=red] (2.5,-1.0) -- (1.5,0);
\draw[thick,color=green,dashed] (-1.5,0) -- (-0.5,0);
\draw[thick,color=green,dashed] (0.5,0) -- (1.5,0);
\draw[pattern=north east lines, pattern color=black] (0,0) circle (0.5);
\end{tikzpicture}}},\ 
\raisebox{-0.6 cm}{\scalebox{0.6}{\begin{tikzpicture}
\draw[thick,color=red] (-2.0,1.0) -- (-1.5,0.5);
\draw[thick,color=green,dashed] (-1.5,0.5) -- (-1,0);
\draw[thick,color=red] (-2.0,-1.0) -- (-1,0);
\draw[thick,color=red] (2.0,1.0) -- (1,0);
\draw[thick,color=red] (2.0,-1.0) -- (1,0);
\draw[thick,color=red] (-1.0,0) -- (0,0);
\draw[thick,color=green,dashed] (0,0) -- (1.0,0);
\end{tikzpicture}}},\ 
\raisebox{-0.6 cm}{\scalebox{0.6}{\begin{tikzpicture}
\draw[thick,color=red] (-2.0,1.0) -- (-1.5,0.5);
\draw[thick,color=green,dashed] (-1.5,0.5) -- (-1,0);
\draw[thick,color=red] (-2.0,-1.0) -- (-1,0);
\draw[thick,color=red] (2.0,1.0) -- (1.5,0.5);
\draw[thick,color=green,dashed] (1.5,0.5) -- (1,0);
\draw[thick,color=red] (2.0,-1.0) -- (1,0);
\draw[thick,color=red] (-1.0,0) -- (1,0);
\end{tikzpicture}}},\ 
\raisebox{-0.6 cm}{\scalebox{0.6}{\begin{tikzpicture}
\draw[thick,color=red] (-2.0,1.0) -- (-1.5,0.5);
\draw[thick,color=green,dashed] (-1.5,0.5) -- (-1,0);
\draw[thick,color=red] (-2.0,-1.0) -- (-1,0);
\draw[thick,color=red] (2.0,1.0) -- (1.5,0.5);
\draw[thick,color=green,dashed] (1.5,0.5) -- (1,0);
\draw[thick,color=red] (2.0,-1.0) -- (1,0);
\draw[thick,color=blue] (-1.0,0) -- (1,0);
\end{tikzpicture}}}\\
\hline
2 massive bosons, &\raisebox{-0.6 cm}{\scalebox{0.6}{\begin{tikzpicture}
\draw[thick,color=red] (-2.5,1.0) -- (-1.5,0);
\draw[thick,color=blue] (-2.5,-1.0) -- (-1.5,0);
\draw[thick,color=red] (2.5,1.0) -- (1.5,0);
\draw[thick,color=blue] (2.5,-1.0) -- (1.5,0);
\draw[thick,color=green,dashed] (-1.5,0) -- (-0.5,0);
\draw[thick,color=green,dashed] (0.5,0) -- (1.5,0);
\draw[pattern=north east lines, pattern color=black] (0,0) circle (0.5);
\end{tikzpicture}}},\ 
\raisebox{-0.6 cm}{\scalebox{0.6}{\begin{tikzpicture}
\draw[thick,color=red] (-2.0,1.0) -- (-1.5,0.5);
\draw[thick,color=green,dashed] (-1.5,0.5) -- (-1,0);
\draw[thick,color=blue] (-2.0,-1.0) -- (-1,0);
\draw[thick,color=red] (2.0,1.0) -- (1,0);
\draw[thick,color=blue] (2.0,-1.0) -- (1,0);
\draw[thick,color=red] (-1.0,0) -- (0,0);
\draw[thick,color=green,dashed] (0,0) -- (1.0,0);
\end{tikzpicture}}},\ 
\raisebox{-0.6 cm}{\scalebox{0.6}{\begin{tikzpicture}
\draw[thick,color=red] (-2.0,1.0) -- (-1.5,0.5);
\draw[thick,color=green,dashed] (-1.5,0.5) -- (-1,0);
\draw[thick,color=blue] (-2.0,-1.0) -- (-1,0);
\draw[thick,color=red] (2.0,1.0) -- (1.5,0.5);
\draw[thick,color=green,dashed] (1.5,0.5) -- (1,0);
\draw[thick,color=blue] (2.0,-1.0) -- (1,0);
\draw[thick,color=red] (-1.0,0) -- (1,0);
\end{tikzpicture}}},\ 
\raisebox{-0.6 cm}{\scalebox{0.6}{\begin{tikzpicture}
\draw[thick,color=red] (-2.0,1.0) -- (-1.5,0.5);
\draw[thick,color=green,dashed] (-1.5,0.5) -- (-1,0);
\draw[thick,color=blue] (-2.0,-1.0) -- (-1,0);
\draw[thick,color=red] (2.0,1.0) -- (1.5,0.5);
\draw[thick,color=green,dashed] (1.5,0.5) -- (1,0);
\draw[thick,color=blue] (2.0,-1.0) -- (1,0);
\draw[thick,color=blue] (-1.0,0) -- (1,0);
\end{tikzpicture}}},\\
 2 dilatons & 
 \raisebox{-0.6 cm}{\scalebox{0.6}{\begin{tikzpicture}
\draw[thick,color=red] (-2.5,1.0) -- (-1.5,0);
\draw[thick,color=red] (-2.5,-1.0) -- (-1.5,0);
\draw[thick,color=blue] (2.5,1.0) -- (1.5,0);
\draw[thick,color=blue] (2.5,-1.0) -- (1.5,0);
\draw[thick,color=green,dashed] (-1.5,0) -- (-0.5,0);
\draw[thick,color=green,dashed] (0.5,0) -- (1.5,0);
\draw[pattern=north east lines, pattern color=black] (0,0) circle (0.5);
\end{tikzpicture}}},
 \raisebox{-0.6 cm}{\scalebox{0.6}{\begin{tikzpicture}
\draw[thick,color=red] (-2.0,1.0) -- (-1.5,0.5);
\draw[thick,color=green,dashed] (-1.5,0.5) -- (-1,0);
\draw[thick,color=red] (-2.0,-1.0) -- (-1,0);
\draw[thick,color=blue] (2.0,1.0) -- (1.5,0.5);
\draw[thick,color=blue] (1.5,0.5) -- (1,0);
\draw[thick,color=blue] (2.0,-1.0) -- (1,0);
\draw[thick,color=red] (-1.0,0) -- (0,0);
\draw[thick,color=green,dashed] (0,0) -- (1,0);
\end{tikzpicture}}}\\
\hline
3 massive bosons,& \raisebox{-0.6 cm}{\scalebox{0.6}{\begin{tikzpicture}
\draw[thick,color=red] (-2.5,1.0) -- (-1.5,0);
\draw[thick,color=blue] (-2.5,-1.0) -- (-1.5,0);
\draw[thick,color=red] (2.5,1.0) -- (1.5,0);
\draw[thick,color=red] (2.5,-1.0) -- (1.5,0);
\draw[thick,color=green,dashed] (-1.5,0) -- (-0.5,0);
\draw[thick,color=green,dashed] (0.5,0) -- (1.5,0);
\draw[pattern=north east lines, pattern color=black] (0,0) circle (0.5);
\end{tikzpicture}}},\ 
\raisebox{-0.6 cm}{\scalebox{0.6}{\begin{tikzpicture}
\draw[thick,color=red] (-2.0,1.0) -- (-1.5,0.5);
\draw[thick,color=green,dashed] (-1.5,0.5) -- (-1,0);
\draw[thick,color=blue] (-2.0,-1.0) -- (-1,0);
\draw[thick,color=red] (2.0,1.0) -- (1,0);
\draw[thick,color=red] (2.0,-1.0) -- (1,0);
\draw[thick,color=red] (-1.0,0) -- (0,0);
\draw[thick,color=green,dashed] (0,0) -- (1.0,0);
\end{tikzpicture}}},\ 
\raisebox{-0.6 cm}{\scalebox{0.6}{\begin{tikzpicture}
\draw[thick,color=red] (-2.0,1.0) -- (-1,0);
\draw[thick,color=blue] (-2.0,-1.0) -- (-1,0);
\draw[thick,color=red] (2.0,1.0) -- (1.5,0.5);
\draw[thick,color=green,dashed] (1.5,0.5) -- (1,0);
\draw[thick,color=red] (2.0,-1.0) -- (1,0);
\draw[thick,color=green,dashed] (-1.0,0) -- (0,0);
\draw[thick,color=red] (0,0) -- (1.0,0);
\end{tikzpicture}}},\ 
\raisebox{-0.6 cm}{\scalebox{0.6}{\begin{tikzpicture}
\draw[thick,color=red] (-2.0,1.0) -- (-1.5,0.5);
\draw[thick,color=green,dashed] (-1.5,0.5) -- (-1,0);
\draw[thick,color=blue] (-2.0,-1.0) -- (-1,0);
\draw[thick,color=red] (2.0,1.0) -- (1.5,0.5);
\draw[thick,color=green,dashed] (1.5,0.5) -- (1,0);
\draw[thick,color=red] (2.0,-1.0) -- (1,0);
\draw[thick,color=red] (-1.0,0) -- (1,0);
\end{tikzpicture}}},\ 
\raisebox{-0.6 cm}{\scalebox{0.6}{\begin{tikzpicture}
\draw[thick,color=red] (-2.0,1.0) -- (-1.5,0.5);
\draw[thick,color=green,dashed] (-1.5,0.5) -- (-1,0);
\draw[thick,color=blue] (-2.0,-1.0) -- (-1,0);
\draw[thick,color=red] (2.0,1.0) -- (1.5,0.5);
\draw[thick,color=green,dashed] (1.5,0.5) -- (1,0);
\draw[thick,color=red] (2.0,-1.0) -- (1,0);
\draw[thick,color=blue] (-1.0,0) -- (1,0);
\end{tikzpicture}}}\\
1 dilaton &\\
\hline
2 massive bosons,  &\raisebox{-0.6 cm}{\scalebox{0.6}{\begin{tikzpicture}
\draw[thick,color=red] (-2.5,1.0) -- (-1.5,0);
\draw[thick,color=red] (-2.5,-1.0) -- (-1.5,0);
\draw[thick,color=yellow] (2.5,1.0) -- (1.5,0);
\draw[thick,color=yellow] (2.5,-1.0) -- (1.5,0);
\draw[thick,color=green,dashed] (-1.5,0) -- (-0.5,0);
\draw[thick,color=green,dashed] (0.5,0) -- (1.5,0);
\draw[pattern=north east lines, pattern color=black] (0,0) circle (0.5);
\draw[color=black] (2.6,1.2)node{$\alpha$};
\draw[color=black] (2.6,-1.2)node{$\alpha$};
\end{tikzpicture}}},\ 
\raisebox{-0.6 cm}{\scalebox{0.6}{\begin{tikzpicture}
\draw[thick,color=red] (-2.0,1.0) -- (-1.5,0.5);
\draw[thick,color=green,dashed] (-1.5,0.5) -- (-1,0);
\draw[thick,color=red] (-2.0,-1.0) -- (-1,0);
\draw[thick,color=yellow] (2.0,1.0) -- (1,0);
\draw[thick,color=yellow] (2.0,-1.0) -- (1,0);
\draw[thick,color=red] (-1.0,0) -- (0,0);
\draw[thick,color=green,dashed] (0,0) -- (1.0,0);
\draw[color=black] (2.2,1.2)node{$\alpha$};
\draw[color=black] (2.2,-1.2)node{$\alpha$};
\end{tikzpicture}}},\ 
\raisebox{-0.6 cm}{\scalebox{0.6}{\begin{tikzpicture}
\draw[thick,color=red] (-2.5,1.0) -- (-1.5,0);
\draw[thick,color=red] (-2.5,-1.0) -- (-1.5,0);
\draw[thick,color=orange] (2.5,1.0) -- (1.5,0);
\draw[thick,color=orange] (2.5,-1.0) -- (1.5,0);
\draw[thick,color=green,dashed] (-1.5,0) -- (-0.5,0);
\draw[thick,color=green,dashed] (0.5,0) -- (1.5,0);
\draw[pattern=north east lines, pattern color=black] (0,0) circle (0.5);
\draw[color=black] (2.6,1.2)node{$\alpha$};
\draw[color=black] (2.6,-1.2)node{$\alpha$};
\end{tikzpicture}}},\ 
\raisebox{-0.6 cm}{\scalebox{0.6}{\begin{tikzpicture}
\draw[thick,color=red] (-2.0,1.0) -- (-1.5,0.5);
\draw[thick,color=green,dashed] (-1.5,0.5) -- (-1,0);
\draw[thick,color=red] (-2.0,-1.0) -- (-1,0);
\draw[thick,color=orange] (2.0,1.0) -- (1,0);
\draw[thick,color=orange] (2.0,-1.0) -- (1,0);
\draw[thick,color=red] (-1.0,0) -- (0,0);
\draw[thick,color=green,dashed] (0,0) -- (1.0,0);
\draw[color=black] (2.2,1.2)node{$\alpha$};
\draw[color=black] (2.2,-1.2)node{$\alpha$};
\end{tikzpicture}}}\\
2 Goldstone bosons &  \\
\hline
1 massive boson, 1 dilaton,  &\raisebox{-0.6 cm}{\scalebox{0.6}{\begin{tikzpicture}
\draw[thick,color=red] (-2.5,1.0) -- (-1.5,0);
\draw[thick,color=blue] (-2.5,-1.0) -- (-1.5,0);
\draw[thick,color=yellow] (2.5,1.0) -- (1.5,0);
\draw[thick,color=yellow] (2.5,-1.0) -- (1.5,0);
\draw[thick,color=green,dashed] (-1.5,0) -- (-0.5,0);
\draw[thick,color=green,dashed] (0.5,0) -- (1.5,0);
\draw[pattern=north east lines, pattern color=black] (0,0) circle (0.5);
\draw[color=black] (2.6,1.2)node{$\alpha$};
\draw[color=black] (2.6,-1.2)node{$\alpha$};
\end{tikzpicture}}},\ 
\raisebox{-0.6 cm}{\scalebox{0.6}{\begin{tikzpicture}
\draw[thick,color=red] (-2.0,1.0) -- (-1.5,0.5);
\draw[thick,color=green,dashed] (-1.5,0.5) -- (-1,0);
\draw[thick,color=blue] (-2.0,-1.0) -- (-1,0);
\draw[thick,color=yellow] (2.0,1.0) -- (1,0);
\draw[thick,color=yellow] (2.0,-1.0) -- (1,0);
\draw[thick,color=red] (-1.0,0) -- (0,0);
\draw[thick,color=green,dashed] (0,0) -- (1.0,0);
\draw[color=black] (2.2,1.2)node{$\alpha$};
\draw[color=black] (2.2,-1.2)node{$\alpha$};
\end{tikzpicture}}},\ 
\raisebox{-0.6 cm}{\scalebox{0.6}{\begin{tikzpicture}
\draw[thick,color=red] (-2.5,1.0) -- (-1.5,0);
\draw[thick,color=blue] (-2.5,-1.0) -- (-1.5,0);
\draw[thick,color=orange] (2.5,1.0) -- (1.5,0);
\draw[thick,color=orange] (2.5,-1.0) -- (1.5,0);
\draw[thick,color=green,dashed] (-1.5,0) -- (-0.5,0);
\draw[thick,color=green,dashed] (0.5,0) -- (1.5,0);
\draw[pattern=north east lines, pattern color=black] (0,0) circle (0.5);
\draw[color=black] (2.6,1.2)node{$\alpha$};
\draw[color=black] (2.6,-1.2)node{$\alpha$};
\end{tikzpicture}}},\ 
\raisebox{-0.6 cm}{\scalebox{0.6}{\begin{tikzpicture}
\draw[thick,color=red] (-2.0,1.0) -- (-1.5,0.5);
\draw[thick,color=green,dashed] (-1.5,0.5) -- (-1,0);
\draw[thick,color=blue] (-2.0,-1.0) -- (-1,0);
\draw[thick,color=orange] (2.0,1.0) -- (1,0);
\draw[thick,color=orange] (2.0,-1.0) -- (1,0);
\draw[thick,color=red] (-1.0,0) -- (0,0);
\draw[thick,color=green,dashed] (0,0) -- (1.0,0);
\draw[color=black] (2.2,1.2)node{$\alpha$};
\draw[color=black] (2.2,-1.2)node{$\alpha$};
\end{tikzpicture}}}\\
2 Goldstone bosons &  \\
\hline
\end{tabular}}
\end{center}
\end{table}

\paragraph{Poles at the zeros of the Mandelstam variables}\mbox{}

Notice that some of the correlators in the above table have diagrams in which a dilaton propagates as an intermediate particle. Since the dilaton is massless, such a diagram leads to a pole at the zero of a Mandelstam variable-$s$, $t$ or $u$, depending on the channel to which the diagram belongs. These correlators and the corresponding diagrams are given below.

\begin{itemize}
\item \underline{4 massive bosons:}
In this case the relevant diagram is of the following form:
\begin{center}
\raisebox{-0.6 cm}{\scalebox{0.7}{\begin{tikzpicture}
\draw[thick,color=red] (-2.0,1.0) -- (-1.5,0.5);
\draw[thick,color=green,dashed] (-1.5,0.5) -- (-1,0);
\draw[thick,color=red] (-2.0,-1.0) -- (-1,0);
\draw[thick,color=red] (2.0,1.0) -- (1.5,0.5);
\draw[thick,color=green,dashed] (1.5,0.5) -- (1,0);
\draw[thick,color=red] (2.0,-1.0) -- (1,0);
\draw[thick,color=blue] (-1.0,0) -- (1,0);
\end{tikzpicture}}}.
\end{center}
Notice that this diagram is nonvanishing in both ground states and thermal states. Therefore, the corresponding poles in this correlator are present at all temperatures (including zero).

\item \underline{2 massive bosons and 2 dilatons:}
In this case the relevant diagram is of the following form:
\begin{center}
\raisebox{-0.6 cm}{\scalebox{0.7}{\begin{tikzpicture}
\draw[thick,color=red] (-2.0,1.0) -- (-1.5,0.5);
\draw[thick,color=green,dashed] (-1.5,0.5) -- (-1,0);
\draw[thick,color=blue] (-2.0,-1.0) -- (-1,0);
\draw[thick,color=red] (2.0,1.0) -- (1.5,0.5);
\draw[thick,color=green,dashed] (1.5,0.5) -- (1,0);
\draw[thick,color=blue] (2.0,-1.0) -- (1,0);
\draw[thick,color=blue] (-1.0,0) -- (1,0);
\end{tikzpicture}}}.
\end{center}
Notice that this diagram has the vertex which couples the auxiliary field $s_0$ to two  dilatons. As we showed earlier, this vertex vanishes at zero temperature. Therefore, the contribution of this diagram and the corresponding poles in the correlator are present only at nonzero temperatures.

\item \underline{3 massive bosons and 1 dilaton:}
As in the previous case, here the relevant diagram (given below) has the vertex which vanishes at zero temperature:
\begin{center}
\raisebox{-0.6 cm}{\scalebox{0.7}{\begin{tikzpicture}
\draw[thick,color=red] (-2.0,1.0) -- (-1.5,0.5);
\draw[thick,color=green,dashed] (-1.5,0.5) -- (-1,0);
\draw[thick,color=blue] (-2.0,-1.0) -- (-1,0);
\draw[thick,color=red] (2.0,1.0) -- (1.5,0.5);
\draw[thick,color=green,dashed] (1.5,0.5) -- (1,0);
\draw[thick,color=red] (2.0,-1.0) -- (1,0);
\draw[thick,color=blue] (-1.0,0) -- (1,0);
\end{tikzpicture}}}.
\end{center}
Therefore, the contribution of this diagram and the corresponding poles in the correlator are also present only at nonzero temperatures.
\end{itemize}

\paragraph{Some examples:}\mbox{}

To illustrate the existence of the poles mentioned above, we provide the explicit forms of some correlators  below. In what follows, we will use a subscript `c' to indicate the connected piece of a correlator. In the expressions of these correlators, we will denote the contribution of the ($s_0-s_0$) propagator with momentum $\overrightarrow{p}$ by $G_{s_0}(\overrightarrow{p},\beta_{\text{th}})$. The Mandelstam variables in these expressions are defined as follows:
\begin{equation}
\begin{split}
s\equiv (\overrightarrow{p}_1+\overrightarrow{p}_2)^2,\ t\equiv (\overrightarrow{p}_1+\overrightarrow{p}_3)^2,\ u\equiv (\overrightarrow{p}_1+\overrightarrow{p}_4)^2.
\end{split}
\end{equation}

Now that we have defined all the quantities appearing in the correlators, let us provide the explicit forms of these correlators:
\begin{equation}
\begin{split}
&\langle \eta_{-}(\overrightarrow{p}_1)\eta_{-}(\overrightarrow{p}_2)\eta_{-}(\overrightarrow{p}_3)\eta_{-}(\overrightarrow{p}_4))\rangle_c\\
&=\beta_{\text{th}}\delta_{\sum_i p_i^0,0}(2\pi)^d\delta^{d}(\sum_i \overrightarrow{p}_i)\Bigg[\prod_{i=1}^4\frac{1}{p_i^2+\widetilde{\sigma}^2}\Bigg]\\
&\qquad\frac{1}{N}\Bigg[A_{_{--}}^2\Bigg\{-\Big(G_{s_0}(\overrightarrow{p}_1+\overrightarrow{p}_2,\beta_{\text{th}})+G_{s_0}(\overrightarrow{p}_1+\overrightarrow{p}_3,\beta_{\text{th}})+G_{s_0}(\overrightarrow{p}_1+\overrightarrow{p}_4,\beta_{\text{th}})\Big)\\
&\qquad\qquad\qquad+8\widetilde{\sigma}^2\Big(\frac{1}{s+\widetilde{\sigma}^2}+\frac{1}{t+\widetilde{\sigma}^2}+\frac{1}{u+\widetilde{\sigma}^2}\Big)\Bigg\}+4A_{_{+-}}^2\widetilde{\sigma}^2\Big(\frac{1}{s}+\frac{1}{t}+\frac{1}{u}\Big)\Bigg],
\end{split}
\end{equation}

\begin{equation}
\begin{split}
&\langle \eta_{-}(\overrightarrow{p}_1)\eta_{+}(\overrightarrow{p}_2)\eta_{-}(\overrightarrow{p}_3)\eta_{+}(\overrightarrow{p}_4))\rangle_c\\
&=\beta_{\text{th}}\delta_{\sum_i p_i^0,0}(2\pi)^d\delta^{d}(\sum_i \overrightarrow{p}_i)\Bigg[\frac{1}{p_1^2+\widetilde{\sigma}^2}\frac{1}{p_3^2+\widetilde{\sigma}^2}\frac{1}{p_2^2}\frac{1}{p_4^2}\Bigg]\\
&\qquad\frac{1}{N}\Bigg[A_{_{+-}}^2\Bigg\{-\Big(G_{s_0}(\overrightarrow{p}_1+\overrightarrow{p}_2,\beta_{\text{th}})+G_{s_0}(\overrightarrow{p}_1+\overrightarrow{p}_4,\beta_{\text{th}})\Big)+3\widetilde{\sigma}^2\Big(\frac{1}{s+\widetilde{\sigma}^2}+\frac{1}{u+\widetilde{\sigma}^2}\Big)\Bigg\}\\
&\qquad\qquad+A_{_{--}}A_{_{++}}\Bigg\{-G_{s_0}(\overrightarrow{p}_1+\overrightarrow{p}_3,\beta_{\text{th}})+\frac{2\widetilde{\sigma}^2}{t+\widetilde{\sigma}^2}\Bigg\}+A_{_{++}}^2\widetilde{\sigma}^2\Big(\frac{1}{s}+\frac{1}{u}\Big)\Bigg],
\end{split}
\end{equation}

\begin{equation}
\begin{split}
&\langle \eta_{-}(\overrightarrow{p}_1)\eta_{+}(\overrightarrow{p}_2)\eta_{-}(\overrightarrow{p}_3)\eta_{-}(\overrightarrow{p}_4))\rangle_c\\
&=\beta_{\text{th}}\delta_{\sum_i p_i^0,0}(2\pi)^d\delta^{d}(\sum_i \overrightarrow{p}_i)\Bigg[\frac{1}{p_2^2}\prod_{i\neq2}\frac{1}{p_i^2+\widetilde{\sigma}^2}\Bigg]\\
&\qquad\frac{1}{N}\Bigg[A_{_{--}}A_{_{+-}}\Bigg\{-\Big(G_{s_0}(\overrightarrow{p}_1+\overrightarrow{p}_2,\beta_{\text{th}})+G_{s_0}(\overrightarrow{p}_1+\overrightarrow{p}_3,\beta_{\text{th}})+G_{s_0}(\overrightarrow{p}_1+\overrightarrow{p}_4,\beta_{\text{th}})\Big)\\
&\qquad\qquad\qquad\qquad+5\widetilde{\sigma}^2\Big(\frac{1}{s+\widetilde{\sigma}^2}+\frac{1}{t+\widetilde{\sigma}^2}+\frac{1}{u+\widetilde{\sigma}^2}\Big)\Bigg\}+2A_{_{++}}A_{_{+-}}\widetilde{\sigma}^2\Big(\frac{1}{s}+\frac{1}{t}+\frac{1}{u}\Big)\Bigg].
\end{split}
\end{equation}
Note that the poles at the zeros of the Mandelstam variables in the last two correlators have the factor $A_{_{++}}$ in their coefficient. Therefore, these poles vanish as the temperature goes to zero. 

\subsubsection{Correlators which are Nonzero only in Thermal States}
Now let us look at the correlators that are nonzero at $O(\frac{1}{N})$ only in a thermal state. We provide the list of these correlators and the corresponding diagrams in table \ref{table: nonvanishing correlators: only thermal states}.
\begin{table}[H]
\caption{Correlators which are nonzero only in thermal states}
\label{table: nonvanishing correlators: only thermal states}
\begin{center}
\scalebox{0.8}{
\begin{tabular}{ |c| c|  }
\hline
Correlator & Diagrams\\
\hline
1 massive boson , 3 dilatons & \raisebox{-0.6 cm}{\scalebox{0.6}{\begin{tikzpicture}
\draw[thick,color=red] (-2.5,1.0) -- (-1.5,0);
\draw[thick,color=blue] (-2.5,-1.0) -- (-1.5,0);
\draw[thick,color=blue] (2.5,1.0) -- (1.5,0);
\draw[thick,color=blue] (2.5,-1.0) -- (1.5,0);
\draw[thick,color=green,dashed] (-1.5,0) -- (-0.5,0);
\draw[thick,color=green,dashed] (0.5,0) -- (1.5,0);
\draw[pattern=north east lines, pattern color=black] (0,0) circle (0.5);
\end{tikzpicture}}},
\raisebox{-0.6 cm}{\scalebox{0.6}{\begin{tikzpicture}
\draw[thick,color=red] (-2.0,1.0) -- (-1.5,0.5);
\draw[thick,color=green,dashed] (-1.5,0.5) -- (-1,0);
\draw[thick,color=blue] (-2.0,-1.0) -- (-1,0);
\draw[thick,color=blue] (2.0,1.0) -- (1,0);
\draw[thick,color=blue] (2.0,-1.0) -- (1,0);
\draw[thick,color=red] (-1.0,0) -- (0,0);
\draw[thick,color=green,dashed] (0,0) -- (1.0,0);
\end{tikzpicture}}}\\
\\
\hline
4 dilatons &\raisebox{-0.6 cm}{\scalebox{0.6}{\begin{tikzpicture}
\draw[thick,color=blue] (-2.5,1.0) -- (-1.5,0);
\draw[thick,color=blue] (-2.5,-1.0) -- (-1.5,0);
\draw[thick,color=blue] (2.5,1.0) -- (1.5,0);
\draw[thick,color=blue] (2.5,-1.0) -- (1.5,0);
\draw[thick,color=green,dashed] (-1.5,0) -- (-0.5,0);
\draw[thick,color=green,dashed] (0.5,0) -- (1.5,0);
\draw[pattern=north east lines, pattern color=black] (0,0) circle (0.5);
\end{tikzpicture}}}\\
\hline
2 dilatons, 2 Goldstone bosons & \raisebox{-0.6 cm}{\scalebox{0.6}{\begin{tikzpicture}
\draw[thick,color=blue] (-2.5,1.0) -- (-1.5,0);
\draw[thick,color=blue] (-2.5,-1.0) -- (-1.5,0);
\draw[thick,color=yellow] (2.5,1.0) -- (1.5,0);
\draw[thick,color=yellow] (2.5,-1.0) -- (1.5,0);
\draw[thick,color=green,dashed] (-1.5,0) -- (-0.5,0);
\draw[thick,color=green,dashed] (0.5,0) -- (1.5,0);
\draw[pattern=north east lines, pattern color=black] (0,0) circle (0.5);
\draw[color=black] (2.6,1.2)node{$\alpha$};
\draw[color=black] (2.6,-1.2)node{$\alpha$};
\end{tikzpicture}}},\ 
\raisebox{-0.6 cm}{\scalebox{0.6}{\begin{tikzpicture}
\draw[thick,color=blue] (-2.5,1.0) -- (-1.5,0);
\draw[thick,color=blue] (-2.5,-1.0) -- (-1.5,0);
\draw[thick,color=orange] (2.5,1.0) -- (1.5,0);
\draw[thick,color=orange] (2.5,-1.0) -- (1.5,0);
\draw[thick,color=green,dashed] (-1.5,0) -- (-0.5,0);
\draw[thick,color=green,dashed] (0.5,0) -- (1.5,0);
\draw[pattern=north east lines, pattern color=black] (0,0) circle (0.5);
\draw[color=black] (2.6,1.2)node{$\alpha$};
\draw[color=black] (2.6,-1.2)node{$\alpha$};
\end{tikzpicture}}}\\
\hline
\end{tabular}}
\end{center}
\end{table}
Notice that in all these diagrams, there is a vertex  with the coefficient $A_{_{++}}$ which vanishes at zero temperature. Hence, these correlators all vanish up to $O(\frac{1}{N})$ at zero temperature.

\subsection{Summary}

We studied the forms of all the connected 4-point correlators in the biconical model which are nonzero at $O(\frac{1}{N})$. From the corresponding Feynman diagrams, we saw that there are some essential differences between the correlators at zero temperature and those at nonzero temperatures. These differences are as follows:
\begin{enumerate}
\item In a thermal state, there are additional poles at the zeros of the Mandelstam variables for the following correlators:
\begin{itemize}
\item 2 massive bosons and 2 dilatons, 
\item 3 massive bosons and 1 dilaton.
\end{itemize}

\item The following correlators vanish in the ground state, but are nonzero in a thermal state:
\begin{itemize}
\item 1 massive boson and 3 dilatons, 
\item 4 dilatons,
\item 2 dilatons and 2 Goldstone bosons.
\end{itemize} 

\end{enumerate}
We saw that these differences arise due to the vanishing of an interaction vertex coupling the auxiliary field $s_0$ to two dilatons when the temperature is taken to zero.

\paragraph{Comment on the fixed point where $\sqrt{\frac{g_{11}^B}{g_{22}^B}}=\sqrt{\frac{g_{11}}{g_{22}}}=\frac{x_2}{x_1}$:} \mbox{}

As we discussed in the main text, at leading order in the large N expansion, there is a line of fixed points of the RG flow of the couplings. A special point on this line is where $\sqrt{\frac{g_{11}}{g_{22}}}=\frac{x_2}{x_1}$. This is the point at which the moduli space of vacua passes through the origin of the field space even at nonzero temperatures. From \eqref{A++ expression}, we can see that at this point, the vertex factor $-i A_{_{++}}=0$. Hence, the essential differences that we mentioned between the vacuum and thermal correlators disappear at this point.

\newpage


\bigskip

\bigskip

\renewcommand\refname{\bfseries\large\centering References\\ \vspace{-0.4cm}
\addcontentsline{toc}{section}{References}}

\bibliographystyle{utphys.bst}
\bibliography{NonRe.bib}
	
\end{document}